\documentclass[longauth]{aa}  

\usepackage{etoolbox}
\usepackage{graphicx}
\usepackage[dvipsnames]{xcolor}
\usepackage{txfonts}
\usepackage{orcidlink}

\begin{document} 

   \title{ALMAGAL} \subtitle{V. Relations between the core populations and the parent clump physical properties}

\author{D.~Elia\inst{\ref{iaps}}\orcidlink{0000-0002-9120-5890}
\and
A.~Coletta\inst{\ref{iaps},\ref{sapienza}}\orcidlink{0000-0001-8239-8304}\and
S.~Molinari\inst{\ref{iaps}}\orcidlink{0000-0002-9826-7525}\and
E.~Schisano\inst{\ref{iaps}}\orcidlink{0000-0003-1560-3958}\and
M.~Benedettini\inst{\ref{iaps}}\orcidlink{0000-0002-3597-7263} \and
\'A.~S\'anchez-Monge\inst{\ref{icecsic}, \ref{ieec}}\orcidlink{0000-0002-3078-9482} \and
A.~Traficante\inst{\ref{iaps}}\orcidlink{0000-0003-1665-6402}\and
C.~Mininni\inst{\ref{iaps}}\orcidlink{0000-0002-2974-4703}\and
A.~Nucara\inst{\ref{iaps},\ref{torverg}}\orcidlink{0009-0005-9192-5491}\and
S.~Pezzuto\inst{\ref{iaps}}\orcidlink{0000-0001-7852-1971}\and
P.~Schilke\inst{\ref{u-koln}}\orcidlink{0000-0003-2141-5689}\and
J.~D.~Soler\inst{\ref{iaps}}\orcidlink{0000-0002-0294-4465} \and
A.~Avison\inst{\ref{skao}}\orcidlink{0000-0002-2562-8609}\and
M.~T.~Beltr\'an\inst{\ref{arcetri}}\orcidlink{0000-0003-3315-5626} \and
H.~Beuther\inst{\ref{mpia}}\orcidlink{0000-0002-1700-090X} \and
S.~Clarke\inst{\ref{u-koln},\ref{asiaa}}\orcidlink{0000-0001-9751-4603} \and
G.~A.~Fuller\inst{\ref{u-man},\ref{u-koln}}\orcidlink{0000-0001-8509-1818} \and
R.~S.~Klessen\inst{\ref{u-hei},\ref{ralf-2},\ref{cfa},\ref{cfa-rad}}\orcidlink{0000-0002-0560-3172}\and
R.~Kuiper\inst{\ref{u-duisb}}\orcidlink{0000-0003-2309-8963}\and
U.~Lebreuilly\inst{\ref{saclay}}\orcidlink{0000-0001-8060-1890}\and
D.~C.~Lis\inst{\ref{caltech}}\orcidlink{0000-0002-0500-4700}\and
T.~M\"{o}ller \inst{\ref{u-koln}}\orcidlink{0000-0002-9277-8025}\and
L.~Moscadelli\inst{\ref{arcetri}}\orcidlink{0000-0002-8517-8881}\and
A.~J.~Rigby\inst{\ref{leeds}}\orcidlink{0000-0002-3351-2200}\and
P.~Sanhueza\inst{\ref{u-tokyo}}\orcidlink{0000-0002-7125-7685}\and 
F.~van der Tak\inst{\ref{sron}, \ref{u-gron}}\orcidlink{0000-0002-8942-1594}\and
Q.~Zhang\inst{\ref{cfa}}\orcidlink{0000-0003-2384-6589}\and
K.~L.~J.~Rygl\inst{\ref{ira}}\orcidlink{0000-0003-4146-9043}\and
M.~Merello\inst{\ref{astroing},\ref{mpir}}\orcidlink{0000-0003-0709-708X}\and
C.~Battersby\inst{\ref{u-conn}}\orcidlink{0000-0002-6073-9320}\and
P.~T.~P.~Ho\inst{\ref{asiaa},\ref{hawaii}}\orcidlink{0000-0002-3412-4306}\and
P.~D.~Klaassen\inst{\ref{roe}}\orcidlink{0000-0001-9443-0463}\and
P.~M.~Koch\inst{\ref{asiaa}}\orcidlink{0000-0003-2777-5861}\and
J.~Allande\inst{\ref{arcetri},\ref{unifi}}\orcidlink{0009-0007-4060-0560}\and
L.~Bronfman\inst{\ref{u-chile}}\orcidlink{0000-0002-9574-8454}\and
F.~Fontani\inst{\ref{arcetri},\ref{mpe},\ref{lux}}\orcidlink{0000-0003-0348-3418}\and 
P.~Hennebelle\inst{\ref{saclay}}\orcidlink{0000-0002-0472-7202}\and
B.~Jones\inst{\ref{u-koln}}\orcidlink{0000-0002-0675-0078}\and
T.~Liu\inst{\ref{shanghai}}\orcidlink{0000-0002-5286-2564}\and
G.~Stroud\inst{\ref{u-man}}\orcidlink{0000-0002-4935-2416}\and
M.~R.~A.~Wells\inst{\ref{mpia}}\orcidlink{0000-0002-3643-5554}\and
A.~Ahmadi\inst{\ref{leiden}}\orcidlink{0000-0003-4037-5248}\and
C.~L.~Brogan\inst{\ref{nraoCH}}\orcidlink{0000-0002-6558-7653}\and
F.~De~Angelis\inst{\ref{iaps}}\orcidlink{0009-0002-6765-7413}\and
T.~R.~Hunter\inst{\ref{nraoCH}}\orcidlink{0000-0001-6492-0090}\and
K.~G. Johnston\inst{\ref{lincoln}}\orcidlink{0000-0003-4509-1180}\and
C.~Y.~Law\inst{\ref{arcetri}}\and
S.~J.~Liu\inst{\ref{iaps}}\orcidlink{0000-0001-7680-2139}\and
S.-Y.~Liu\inst{\ref{asiaa}}\orcidlink{0000-0003-4603-7119}\and
Y.~Maruccia\inst{\ref{napoli}}\orcidlink{0000-0003-1975-6310}\and
V.~-M.~Pelkonen\inst{\ref{iaps}}\orcidlink{0000-0002-8898-1047}\and
Y.-N.~Su\inst{\ref{asiaa}}\orcidlink{0000-0002-0675-276X}\and
Y. Tang\inst{\ref{asiaa}}\orcidlink{0000-0002-0675-276X}\and
L. Testi\inst{\ref{u-bo}}\orcidlink{0000-0003-1859-3070}\and
S.~Walch\inst{\ref{u-koln},\ref{datacologne}}\orcidlink{0000-0001-6941-7638}\and
T.~Zhang\inst{\ref{zhejiang},\ref{u-koln}}\orcidlink{0000-0002-1466-3484}\and
H.~Zinnecker\inst{\ref{u-autono}}\orcidlink{0000-0003-0504-3539}
       }

   \institute{INAF - IAPS, via Fosso del Cavaliere, 100, I-00133 Roma, Italy\\
              \email{davide.elia@inaf.it}
Istituto Nazionale di Astrofisica (INAF)-Istituto di Astrofisica e Planetologia Spaziale, Via Fosso del Cavaliere 100, I-00133 Roma, Italy \label{iaps} 
\and
Dipartimento di Fisica, Sapienza Universit\`a di Roma, Piazzale Aldo Moro 2, I-00185, Roma, Italy \label{sapienza}
\and
Institut de Ci\`encies de l'Espai (ICE, CSIC), Campus UAB, Carrer de Can Magrans s/n, E-08193, Bellaterra (Barcelona), Spain\label{icecsic} 
\and
Institut d'Estudis Espacials de Catalunya (IEEC), E-08860, Castelldefels (Barcelona), Spain\label{ieec} 
\and
Dipartimento di Fisica, Università di Roma Tor Vergata, Via della Ricerca Scientifica 1, I-00133 Roma, Italy\label{torverg}
\and
Physikalisches Institut der Universit\"at zu K\"oln, Z\"ulpicher Str. 77, D-50937 K\"oln, Germany\label{u-koln}
\and
SKA Observatory, Jodrell Bank, Lower Withington, Macclesfield, SK11 9FT, UK\label{skao}
\and
Istituto Nazionale di Astrofisica (INAF), Osservatorio Astrofisico di Arcetri, Largo E. Fermi 5, Firenze, Italy \label{arcetri} 
\and
Max Planck Institute for Astronomy, K\"onigstuhl 17, 69117 Heidelberg, Germany\label{mpia}
\and
Institute of Astronomy and Astrophysics, Academia Sinica, 11F of ASMAB, AS/NTU No.\ 1, Sec.\ 4, Roosevelt Road, Taipei 10617, Taiwan \label{asiaa}
\and
Jodrell Bank Centre for Astrophysics, Department of Physics and Astronomy, The University of Manchester, Oxford Road, Manchester, M13 9PL, UK\label{u-man}
\and
Universit\"{a}t Heidelberg, Zentrum f|"{u}r Astronomie, Institut f\"{u}r Theoretische Astrophysik, Albert-Ueberle-Str. 2, 69120 Heidelberg, Germany \label{u-hei} 
\and
Universit\"{a}t Heidelberg, Interdisziplin\"{a}res Zentrum f\"{u}r Wissenschaftliches Rechnen, Im Neuenheimer Feld 205, 69120 Heidelberg, Germany\label{ralf-2} 
\and
Harvard-Smithsonian Center for Astrophysics, 60 Garden Street, Cambridge, MA 02138, U.S.A.\label{cfa} 
\and
Elizabeth S. and Richard M. Cashin Fellow at Radcliffe Institute for Advanced Studies at Harvard University, 10 Garden Street, Cambridge, MA 02138, U.S.A. \label{cfa-rad}
\and
Faculty of Physics, University of Duisburg-Essen, Lotharstra{\ss}e 1, D-47057 Duisburg, Germany\label{u-duisb}
\and
Universit\'e Paris-Saclay, Universit\'e Paris-Cit\'e, CEA, CNRS, AIM, 91191 Gif-sur-Yvette, France\label{saclay}
\and
Jet Propulsion Laboratory, California Institute of Technology, 4800 Oak Grove Drive, Pasadena, CA 91109, USA\label{caltech}
\and
School of Physics and Astronomy, University of Leeds, Leeds LS2 9JT, UK\label{leeds}
\and
Department of Astronomy, School of Science, The University of Tokyo, 7-3-1 Hongo, Bunkyo, Tokyo 113-0033, Japan\label{u-tokyo}
\and
SRON Netherlands Institute for Space Research\label{sron}
\and
Kapteyn Astronomical Institute, University of Groningen, Landleven 12, 9747 AD Groningen, The Netherlands\label{u-gron}\and
INAF-Istituto di Radioastronomia \& Italian ALMA Regional Centre, Via P. Gobetti 101, I-40129 Bologna, Italy
\label{ira}\and
Centro de Astro-Ingeniería UC, Instituto de Astrofísica, Pontificia Universidad Católica de Chile, Avda Vicuña Mackenna 4860, Macul, Santiago, Chile\label{astroing}
\and
Max-Planck-Institut für Radioastronomie, Auf dem Hügel 69, D-53121 Bonn, Germany\label{mpir}
\and
University of Connecticut, Department of Physics, 196A Auditorium Road, Unit 3046, Storrs, CT~06269, USA\label{u-conn}
\and 
East Asian Observatory, 660 N.\ A'ohoku, Hilo, Hawaii, HI 96720, USA \label{hawaii}
\and
UK Astronomy Technology Centre, Royal Observatory Edinburgh, Blackford Hill, Edinburgh EH9 3HJ, UK\label{roe}
\and
Dipartimento di Fisica e Astronomia, Università degli Studi di Firenze, Via G. Sansone 1, I-50019 Sesto Fiorentino, Firenze, Italy\label{unifi}
\and
\label{u-chile}Departamento de Astronomía, Universidad de Chile, Casilla 36-D, Santiago, Chile
\and
Max-Planck-Institute for Extraterrestrial Physics (MPE), Garching bei M\"unchen, Germany \label{mpe} 
\and
LUX, Observatoire de Paris, Université PSL, Sorbonne Université, CNRS, 75014, Paris, France \label{lux}
\and
Shanghai Astronomical Observatory, Chinese Academy of Sciences, 80 Nandan Road, Shanghai 200030, China\label{shanghai}
\and
Leiden Observatory, Leiden University, PO Box 9513, 2300 RA Leiden, The Netherlands\label{leiden}
\and
National Radio Astronomy Observatory, 520 Edgemont Road, Charlottesville VA 22903, USA\label{nraoCH}
\and 
School of Engineering and Physical Sciences, The University of Lincoln, Brayford Way, Lincoln, LN6 7TS, United Kingdom\label{lincoln}
\and
INAF - Astronomical Observatory of Capodimonte, Via Moiariello 16, I-80131 Napoli, Italy \label{napoli}
\and
Dipartimento di Fisica e Astronomia, Alma Mater Studiorum - Universit\`a di Bologna\label{u-bo}
\and
Center for Data and Simulation Science, University of Cologne, Germany\label{datacologne}
\and
Research Center for Astronomical computing, Zhejiang Laboratory, Hangzhou, China\label{zhejiang}
\and
Universidad Autonoma de Chile, Av Pedro de Valdivia 425, Providencia, Santiago de Chile, Chile\label{u-autono}
       }

   \date{Received -------, 2024; accepted -------}

\newcommand{\nnear}{476}
\newcommand{\nfar}{531}
\newcommand{\ncoletta}{6348}
\newcommand{\cornishmatches}{24}
\newcommand{\cornishsouthmatches}{50}
\newcommand{\cornishcontamination}{110}
\newcommand{\cornishcoverage}{866}
\newcommand{\scornishcoverage}{526}
\newcommand{\ncornishcoverage}{340}
\newcommand{\nocornishcoverage}{141}
\newcommand{\cornishlmten}{83}
\newcommand{\cornishlmthresh}{68}
\newcommand{\medfblue}{32.1}
\newcommand{\medfbluez}{7.8}
\newcommand{\medfbluey}{48.2}
\newcommand{\medfblued}{24.3}
\newcommand{\medfred}{39.5}
\newcommand{\medfredz}{12.0}
\newcommand{\medfredy}{59.3}
\newcommand{\medfredd}{27.5}
\newcommand{\medfPSW}{69.4}
\newcommand{\medfPSWz}{29.1}
\newcommand{\medfPSWy}{88.7}
\newcommand{\medfPSWd}{40.3}
\newcommand{\medfPMW}{39.2}
\newcommand{\medfPMWz}{22.0}
\newcommand{\medfPMWy}{48.6}
\newcommand{\medfPMWd}{17.2}
\newcommand{\medfPLW}{24.6}
\newcommand{\medfPLWz}{14.9}
\newcommand{\medfPLWy}{28.5}
\newcommand{\medfPLWd}{9.7}
\newcommand{\medtdust}{14.5}
\newcommand{\medtdustz}{12.1}
\newcommand{\medtdusty}{15.6}
\newcommand{\medmass}{805}
\newcommand{\medmassz}{729}
\newcommand{\medmassy}{825}
\newcommand{\medlum}{1538}
\newcommand{\medlumz}{423}
\newcommand{\medlumy}{2154}
\newcommand{\medsurfd}{0.6}
\newcommand{\medsurfdz}{0.3}
\newcommand{\medsurfdy}{0.8}
\newcommand{\medvold}{11.4}
\newcommand{\medvoldz}{4.5}
\newcommand{\medvoldy}{13.5}
\newcommand{\medlm}{1.7}
\newcommand{\medlmz}{0.7}
\newcommand{\medlmy}{2.4}
\newcommand{\medtbol}{38.0}
\newcommand{\medtbolz}{32.4}
\newcommand{\medtboly}{39.6}
\newcommand{\medncorepro}{4}
\newcommand{\medncorepre}{2}
\newcommand{\medncorehii}{14}
\newcommand{\medncoreprop}{5}
\newcommand{\medncoreprep}{3}
\newcommand{\medncorehiip}{14}
\newcommand{\percncorepro}{1 ; 8}
\newcommand{\percncorepre}{0 ; 4}
\newcommand{\percncorehii}{6 ; 20}
\newcommand{\percncoreprop}{2 ; 8}
\newcommand{\percncoreprep}{2 ; 4}
\newcommand{\percncorehiip}{6 ; 20}
\newcommand{\ksncore}{0.31}
\newcommand{\ksncorep}{0.31}
\newcommand{\kscriticalncore}{0.19}
\newcommand{\kscriticalthreesigm}{0.22}
\newcommand{\nnce}{116}
\newcommand{\nnco}{817}
\newcommand{\nnch}{74}
\newcommand{\nncep}{83}
\newcommand{\nncop}{684}
\newcommand{\nnchp}{73}
\newcommand{\nncen}{33}
\newcommand{\nncon}{133}
\newcommand{\nnchn}{1}
\newcommand{\nncenperc}{28}
\newcommand{\nnconperc}{16}
\newcommand{\nnchnperc}{1}
\newcommand{\nmmmce}{83}
\newcommand{\nmmmco}{684}
\newcommand{\nmmmch}{73}
\newcommand{\medmmmcpro}{1.5}
\newcommand{\medmmmcpre}{0.9}
\newcommand{\medmmmchii}{5.1}
\newcommand{\percmmmcpro}{0.6 ; 3.9}
\newcommand{\percmmmcpre}{0.4 ; 2.2}
\newcommand{\percmmmchii}{1.6 ; 13.8}
\newcommand{\nlowsurfd}{86}
\newcommand{\ncfee}{83}
\newcommand{\ncfeo}{684}
\newcommand{\ncfeh}{73}
\newcommand{\medcfepro}{0.4}
\newcommand{\medcfepre}{0.2}
\newcommand{\medcfehii}{1.3}
\newcommand{\perccfepro}{0.1 ; 1.1}
\newcommand{\perccfepre}{0.1 ; 0.4}
\newcommand{\perccfehii}{0.6 ; 2.7}
\newcommand{\cfemax}{15}
\newcommand{\cfemin}{0.01}
\newcommand{\cfemed}{0.4}
\newcommand{\percsigmalow}{2}
\newcommand{\percsamesigma}{88}
\newcommand{\ashesincommon}{32}
\newcommand{\massivesurf}{0.3}
\newcommand{\slopemaxmasssurfd}{0.75}
\newcommand{\slopemaxmasssurfdave}{0.95}
\newcommand{\eslopemaxmasssurfd}{0.06}
\newcommand{\eslopemaxmasssurfdave}{0.08}
\newcommand{\maxmasspreave}{1.8}
\newcommand{\maxmasshiiave}{17.4}
\newcommand{\ndataxurange}{142}
\newcommand{\spmaxmass}{0.34}
\newcommand{\fmmcave}{0.4}
\newcommand{\fmmcmed}{0.1}
\newcommand{\npro}{891}
\newcommand{\npre}{116}
\newcommand{\slopencorepro}{-0.051}
\newcommand{\slopencorepre}{-0.056}
\newcommand{\errslopencorepro}{0.003}
\newcommand{\errslopencorepre}{0.013}
\newcommand{\slopecfelm}{0.36}
\newcommand{\factorcfelm}{-2.54}
\newcommand{\eslopecfelm}{0.02}
\newcommand{\efactorcfelm}{0.02}
\newcommand{\slopecfet}{2.5}
\newcommand{\factorcfet}{-5.42}
\newcommand{\eslopecfet}{0.1}
\newcommand{\efactorcfet}{0.13}
\newcommand{\lummmcslope}{0.8}
\newcommand{\lummmcspear}{0.47}
\newcommand{\lumdmmcdslope}{0.7}
\newcommand{\lumdmmcdspear}{0.45}
\newcommand{\nok}{840}
\newcommand{\minspearman}{-0.25}
\newcommand{\maxspearman}{0.62}
\newcommand{\spearcfedist}{-0.16}
\newcommand{\spearcfedistrestr}{-0.24}
\newcommand{\spearcfesurfd}{0.29}
\newcommand{\spearmmmcsurfd}{0.40}
\newcommand{\ncslopemaxmasssurfd}{0.64}
\newcommand{\nceslopemaxmasssurfd}{0.07}
\newcommand{\ncslopecfelm}{0.37}
\newcommand{\nceslopecfelm}{0.02}
\newcommand{\ncslopecfet}{2.6}
\newcommand{\nceslopecfet}{0.1}
\newcommand{\nmassive}{31}
\newcommand{\nmassivenocorn}{18}
\newcommand{\nmassivelmone}{30}
\newcommand{\nmassivelmten}{24}
\newcommand{\nmassivelmonesigmhalf}{27}
\newcommand{\nmassivelmonesigmone}{22}
\newcommand{\sddratio}{0.56}
\newcommand{\sddstdev}{0.18}

  \abstract 
   {The fragmentation of massive molecular clumps into smaller, potentially star-forming cores plays a key role in the processes of high-mass star formation. The ALMAGAL project, using the Atacama Large Millimeter/submillimeter Array (ALMA), offers high-resolution data to investigate these processes across various evolutionary stages in the Galactic plane.}
   {This study aims at correlating the fragmentation properties of massive clumps, obtained from ALMA observations, with their global physical parameters (e.g., mass, surface density, and temperature) and evolutionary indicators (such as luminosity-to-mass ratio and bolometric temperature) obtained from \textit{Herschel} observations. It seeks to assess whether the cores  evolve in number and mass in tandem with their host clumps, and to determine the possible factors influencing the formation of massive cores ($M > 24 \mathrm{M}_\odot$).}
   {We analyzed the masses of \ncoletta\ fragments, estimated from 1.4~mm continuum data for 1007 ALMAGAL clumps. Leveraging this unprecedentedly large data set, we evaluated statistical relationships between clump parameters, estimated over $\sim 0.1$~pc scales, and fragment properties, corresponding to scales of a few 1000~au, while accounting for potential biases related to distance and observational resolution. Our results were further compared with predictions from numerical simulations.}
   {The fragmentation level correlates preferentially with clump surface density, supporting a scenario of density-driven fragmentation, whereas it does not show any clear dependence on total clump mass. Both the mass of the most massive core and the core formation efficiency show a broad range and increase on average by an order of magnitude in the intervals spanned by evolutionary indicators such as clump dust temperature and the luminosity-to-mass ratio. This suggests that core growth continues throughout the clump evolution, favoring clump-fed over core-fed theoretical scenarios. However, significant scatter in these relationships indicates that multiple factors, including magnetic fields, turbulence, and stellar feedback, not quantifiable with continuum data, influence fragmentation, as also suggested by comparison with numerical simulations.}
   {}


   \maketitle

\section{Introduction}
Understanding the processes that drive the formation of massive stars still represents a key challenge in astronomy, with several critical issues still under debate \citep{tan14,mot18}.
For instance, how does mass accumulate to form a high-mass star? Are specific physical conditions required for this process? Do high-mass stars form in conjunction with massive clusters, and if so, how do they influence the subsequent formation of surrounding stars? 

The fragmentation of massive condensations \citep[clumps, with diameters ranging from $\sim 0.3$ and 3~pc, e.g.,][]{ber07} within molecular clouds plays a pivotal role in the formation of stellar clusters and massive stars, also shaping the morphology of their host clouds. This process is not regulated solely by gravity, as it has been demonstrated that turbulence \citep{kle01a,elm02,fum05,zha09,wan14,dun16}, magnetic fields \citep{bas09,hen11,fon16,bel19,ane20,pal21,beu24}, and protostellar feedback \citep{kru08,kru10,gus17,men24,san24} all play an important role in counteracting gravity and shaping fragmentation.

Recent advancements in millimeter interferometry have enabled high-resolution imaging to spatially resolve the internal structure of clumps that are either candidates for being massive star formation precursors or active sites of massive star formation. These observations allow direct comparisons with theoretical models. Specifically, dust continuum observations provide valuable insights into clump fragmentation and the masses of resulting fragments. The sizes of these smaller-scale condensations, as resolved by modern interferometers, generally correspond to those of the cores \citep[0.03-0.2~pc, see][]{ber07}.

Fragmentation is a dynamic, time-evolving process, so it is essential to characterize observed targets based on their evolutionary stage.
Particular attention has been given to objects selected as representative of the earliest phases of star formation or those still in a quiescent state. These have been observed using the Submillimeter Array \citep[SMA,][]{zha09,wan14,san17}, the Atacama Compact Array \citep[ACA,][]{cse17}, the Atacama Large Millimeter Array \citep[ALMA,][]{san19,svo19,and21,mor23}, and the Northern Extended Millimeter Array \citep{rig24}.

Other studies have focused on star-forming clumps in more advanced stages. Using ALMA, \citet{liu20,liu24} and \citet{xu24quarks} surveyed more than 100 candidate ultra-compact (UC) H\textsc{ii~} regions, while \citet{xu24assemble} observed 11 massive protoclusters.

\begin{table*}
\caption{Summary of recent studies on clump fragmentation carried out on interferometric data of continuum emission.}       
\label{table:1}      
\centering           
\begin{tabular}{lccccl} 
\hline\hline             
Survey & \# Targets & Facility & Wavelength\tablefootmark{a} & Best angular resolution & Reference \\ 
Name & & & [mm] &  [\arcsec] & \\
\hline                   
-    & 4   & PdBI  & 1.3  &      0.4  & \citet{pal13}\\
CORE     & 20   & NOEMA & 1.3  &      0.4  & \citet{beu18b}\\
ASHES    & 39   & ALMA  & 1.3  &      1.2  & \citet{san19,mor23}\\
   -     &  12  & ALMA  & 1.3  & 0.8       & \citet{svo19} \\
ATOMS    & 146  & ALMA  &   3  & 1.2 & \citet{liu20}\\
   -     &   6  & ALMA  &   3  & 2.8 & \citet{and21} \\
ALMA-IMF & 59   & ALMA  & 1.3; 3 & 0.3 &\citet{mot22} \\
SQUALO   & 13   & ALMA  & 1.3; 3 & 1.1 & \cite{tra23} \\
TEMPO    & 38   & ALMA  & 1.3  & 0.8   &\citet{avi23}\\
-        & 59   & ALMA  & 1.3  & 1 &\citet{olm23} \\
QUARKS   & 139  & ALMA  & 1.3  & 0.25 & \citet{liu24,xu24quarks}\\
ASSEMBLE & 11   & ALMA  & 0.87 & 0.8 & \citet{xu24assemble}\\
   -     & 16   &  SMA  & 0.87 & 0.6      & \citet{pan24} \\
   -     &  13  & NOEMA, ALMA  & 3    &   5.5  &  \citet{rig24}\\
INFANT   &  8   & ALMA  & 1.3  & 0.6      & \citet{che24} \\
DIHCA    & 30   & ALMA & 1.3 & 0.06 & \citet{ish24}\\ 
\hline
ALMAGAL  & 1013 & ALMA  & 1.3 & 0.15 - 0.3\tablefootmark{b} & \citet{mol25}\\
\hline                                   
\end{tabular}
\tablefoot{
\tablefoottext{a}{The quoted wavelength is only indicative (e.g., 1.3 and 3 mm generically indicate ALMA Band 6 and 3, respectively); its precise value varies from case to case depending on the adopted spectral window(s).}\tablefoottext{b}{The double value arises from the use of two interferometer configurations in ALMAGAL.} 
}
\end{table*}

Several studies have aimed to cover a broad range of evolutionary stages. These include work by \citet{pal13} with the Plateau de Bure Interferometer (PdBI), \citet{beu18b} with NOEMA, \citep{mot22,olm23,tra23,avi23,che24,ish24} with ALMA, and \citet{pan24} with SMA. These observing programs, based on target samples ranging from a few to 146 sources, are summarized in  Table~\ref{table:1}.

A major step toward a statistical approach to clump fragmentation, covering a large variety of physical conditions and Galactic locations, is represented by ALMAGAL (ALMA Cycle 7 large project; P.I.: S. Molinari, P. Schilke, C. Battersby, and P. Ho). This survey observed both in the continuum and line emission, down to a scale of 1000~au, a sample of 1017 clumps in the Galactic plane \citep{mol25,san25}, selected from the catalogs of the \textit{Herschel}\footnote{\textit{Herschel} \citep{pil10} is an ESA space observatory and with important participation from NASA. On-board cameras for photometric observations were PACS \citep{pog10} and SPIRE \citep{gri10}.}
Far Infrared Galactic Plane Survey \citep[Hi-GAL][]{mol10a,eli17,eli21} and Red MSX Source \citep[RMS,][]{lum13} far-infrared continuum surveys. 
The catalog by \citet{col25} reports the cores extracted from ALMAGAL continuum maps and now represents an important data set for studying the fragmentation in those clumps.

In this paper, we investigate correlations between the key physical properties describing the core populations in ALMAGAL clumps, particularly those quantifying the fragmentation level and the potential for massive star formation, and the clump's large-scale photometric and physical parameters. The analysis of the mass distribution of cores and their spatial arrangement lies beyond the scope of this study and is addressed in the works by \citet{col25} and \citet{sch25}, respectively.

To interpret our results, we compare our observations with leading high-mass star formation theories, broadly categorized into core-fed and clump-fed scenarios, based on the mechanisms responsible for mass accumulation.
The \textit{core-fed} model posits that massive stars originate from individual, pre-existing gas cores, with masses already established at the early fragmentation \citep[e.g.,][]{mck02,mck03}. In this scenario, a massive core accumulates material in a quasi-static manner, gradually accreting onto a single protostar. 
In contrast, the \textit{clump-fed} scenario envisions star formation in a dynamic, clustered environment where gas is accreted from a larger-scale clump. Models such as competitive accretion \citep{bon01}, global hierarchical collapse \citep[GHC,][]{vaz19}, and inertial flow \citep{pad20} fall under this category. In this scenario, gas flows into the protostellar vicinity through accretion streams rather than from a single core, enabling continuous mass accretion that supports growth beyond the initial core mass \citep[e.g.,][]{kle01a}. 
In the current paper, we critically compare model predictions regarding fragments' number and mass distribution as functions of environmental density and star formation age with our observational findings.
As it is instructive to compare observations with numerical simulations of collapsing clumps \citep[e.g.,][]{fon18}, we incorporate state-of-the-art simulations by \citet{leb25} into our analysis, focusing on the case of a $500 \mathrm{M}_\odot$ magnetized clump, and comparing the number and total mass of the resulting sink particles during its evolution to the fragmentation level and formation efficiency observed in ALMAGAL clumps, respectively.

The structure of this paper is as follows: Sect.~\ref{parameters} outlines the set of parameters used in the analysis. In Sections~\ref{ncore}, \ref{mmcmass}, and \ref{cfesect}, we compare the degree of fragmentation in terms of the number of detected cores, the mass of the most massive core, and the core formation efficiency, respectively, against clump properties.
In Sect.~\ref{discussion}, we show further combinations of parameters to discuss the possible interplay of physical conditions to give rise to the observed fragmentation properties. Finally, in Sect.~\ref{summary} we draw our conclusions.

\section{A premise on clump and core properties}\label{parameters}
\subsection{Clump physical parameters}\label{clumpparameters}

Of the 1017 targets initially present in the ALMAGAL observing plan, 1007 are analyzed in this work. Indeed, \citet{mol25} ruled out four fields due to incorrect target selection or duplication of observations. Additionally, they identified six clumps for which determining physical properties from Hi-GAL data (see below) is not feasible, so that these targets cannot be included in the present analysis.

\begin{figure*}[h!]
   \centering
   \includegraphics[width=\textwidth]{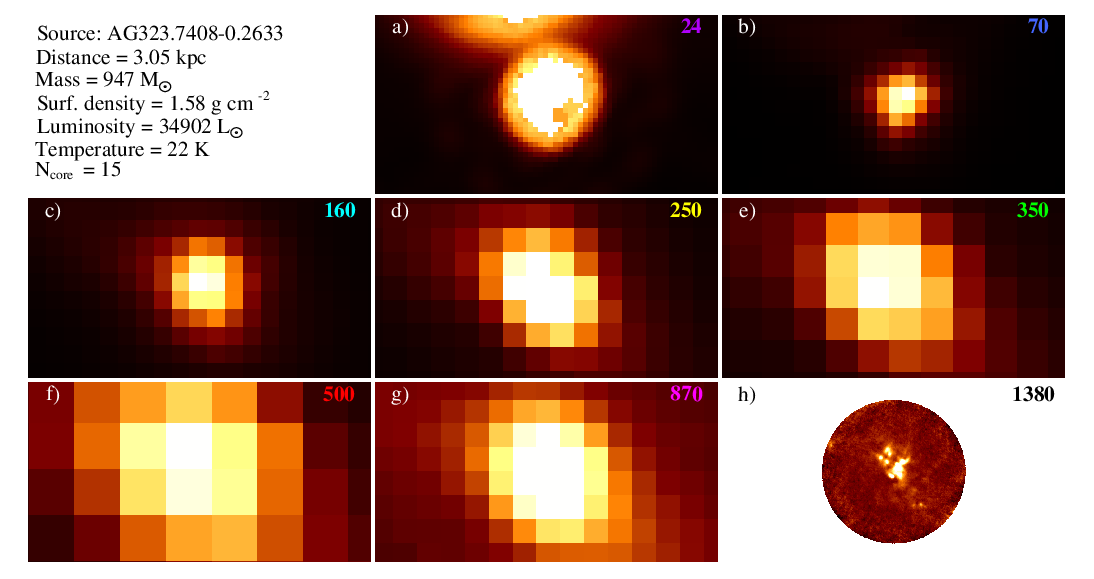}
   \caption{Example of ALMAGAL target observed from mid-infrared to millimeter. The ALMAGAL ID (AG323.7408-0.2633), its physical parameters \citep{mol25}, and the number of cores detected in ALMAGAL continuum observations \citep{col25} are reported in the top-left corner of the figure. Panels $a$-$h$ contain the source images ($\sim 80\arcsec \times 45\arcsec$) at wavelengths increasing from 24~$\mu$m to 1.38~mm (the wavelength in $\mu$m is reported in the top-right corner of each panel). Specifically, in panel $a$ the \textit{Spitzer}-MIPS image of the source (saturated in the center) is shown; in panels $b$-$c$ the \textit{Herschel}-PACS images; in panels $d$-$f$ the \textit{Herschel}-SPIRE images; in panel $g$ the ATLASGAL image; finally, in panel $h$ the new ALMAGAL image, in which 15 fragments were detected by \citet{col25}. Angular resolutions in different panels are the following: $a$: 6\arcsec; $b$: 10.2\arcsec \citep[for this panel and the following four, the resolution value is not the nominal one for \textit{Herschel}, but rather the one directly measured by][on Hi-GAL maps, and circularised here]{tra11}; $c$: 13.5\arcsec; $d$: 23\arcsec; $e$: 30\arcsec; $f$: 42\arcsec; $g$: 19\arcsec; $h$: 0.5\arcsec. Being detected at 70~$\mu$m (and also at 24~$\mu$m, in this case), in accordance with the Hi-GAL catalog criteria this source is classified as star-forming.}
   \label{targetexample}
\end{figure*}

For these 1007 objects, physical properties are calculated from Hi-GAL photometry (an example is shown in Figure~\ref{targetexample}). For the subsample of sources selected directly from the Hi-GAL catalog, data are taken from \citet{eli21}. 
Based on the physical sizes obtained using the distances estimated by \citet{meg21} for $\sim 80\%$ of the entries in the catalog of \citet{eli21}, it can be affirmed that most of these sources can be classified as clumps. 

In \citet{eli17, eli21} Hi-GAL clump photometry at 70, 160, 250, 350, and 500~$\mu$m was supplemented, where possible, with flux densities at 21, 22, 24, 870, and 1100~$\mu$m from MSX \citep{ega03}, WISE \citep{wri10}, MIPSGAL \citep{gut15}, ATLASGAL \citep{sch09}, and BGPS \citep{ros10,agu11} surveys, respectively. Sources detected at 70~$\mu$m were classified as ``star-forming'' (as the example shown in Figure~\ref{targetexample}). In contrast, those undetected at 70~$\mu$m-dark were classified as quiescent. The ALMAGAL sample is divided in \npro\ star-forming and \npre\ quiescent sources, respectively.

A modified black body \citep[e.g.,][]{eli16} with a mass opacity coefficient (hereafter opacity) of $\kappa_{300}=10$~cm$^2$~g$^{-1}$ at 300~$\mu$m equivalent to 0.1~cm$^2$~g$^{-1}$ after including a gas-to-dust ratio of 100 \cite{hil83}, and a spectral index $\beta=2$ was fitted to the portion of the SED at wavelengths $\lambda \geq 160~\mu$m, to derive temperatures and, where heliocentric distances were available, to estimate masses as well. 
In this respect, the cataloged temperature represents an average value, dominated by the colder dust present in the outer regions of the clump, and only partially reflecting any potential star formation activity in its inner regions. Correspondingly, the derived mass represents the total mass of the clump, which is predominantly contained within its large-scale envelope. As a result, the surface density, estimated as the ratio of clump mass to area, provides an average value that does not account for the highly inhomogeneous internal structure expected within a clump. It is noteworthy that in \citet{eli17,eli21}, the surface density was calculated using the beam-deconvolved, circularized size of the source at 250~$\mu$m. In some cases, this method results in extremely high surface densities ($\Sigma \gg 10$~g~cm$^{-2}$), which are not comparable with similar ALMA studies that use non-deconvolved sizes \citep[e.g.,][]{mor23}. For this reason, in this work, we have chosen to base our discussion of clump surface densities on non-deconvolved sizes. These densities are, by definition, higher than those derived using deconvolved sizes, with the average ratio of the two being \sddratio, and a standard deviation \sddstdev.

The clump bolometric luminosity was obtained as the sum of the integral under the observed SED at $\lambda \leq 160~\mu$m (if available) and the integral under the best-fitting modified black body at $\lambda > 160~\mu$m. The ratio of bolometric luminosity and mass, which is distance-independent, is generally considered a suitable descriptor of the clump evolution \citep[e.g.,][]{mol08,mol16,smi14,eli17}, and will be used in the current paper as such. 
At first glance, including both the dust temperature $T$ and the $L/M$ parameter in our analysis may seem redundant, as the two show a good degree of correlation \citep{eli17,urq18}. However, they are estimated from different data sets, as $T$ is derived from fluxes at $\lambda \geq 160~\mu$m, while $L$ derives from the integral of the SED extended to shorter wavelengths, and this accounts for the spread observed in the aforementioned relation between these two quantities. For this reason, and to facilitate future comparisons with the outputs of possible numerical simulations, we prefer to retain both quantities in our analysis.

We also make use of the bolometric temperature, defined by \citet{mye93} as a weighted average frequency (converted into a temperature) of the observed SED and estimated as described in \citet{eli17}, as a powerful evolutionary tool.

Once ALMAGAL spectroscopic data became available, it was possible to reconsider the radial velocities quoted by \citet{meg21} to ALMAGAL targets, so that heliocentric distances were updated by Benedettini et al. (in prep.) using radial velocities directly extracted from ALMAGAL line observations to obtain an updated and self-consistent set of parameters. Hence, re-scaled distance-dependent quantities were listed by \citet{mol25}.

Furthermore, also for the part of the ALMAGAL target list directly selected from the RMS survey \citep{lum13} instead of from Hi-GAL, \citet{mol25} made an effort to identify possible Hi-GAL counterparts and to derive their physical parameters. We refer the reader to \citet{mol25} for further details on this recovery procedure. As mentioned above, this procedure was not applicable to six targets, excluded for this reason from the present analysis.

To enrich the evolutionary characterization of ALMAGAL targets, we searched for 
possible compact or ultra-compact H\textsc{ii} (UCH\textsc{ii}) region counterparts in the source catalogs of the radio surveys CORNISH \citep[][obtained for the coordinate range $10^\circ < \ell < 65^\circ$, $|b|<1^\circ$]{pur13} and CORNISH South \citep[][$295^\circ < \ell < 350^\circ$, $|b|<1^\circ$]{ira23}. A total of \ncornishcoverage\ and \scornishcoverage\ ALMAGAL targets fall within the coverage of the former and latter surveys, respectively. Adopting a searching radius of 10\arcsec, \cornishmatches\ and \cornishsouthmatches\ matches with ALMAGAL were found in the former and the latter, respectively. As expected, all matched sources are detected at 70-$\mu$m, confirming that they constitute a subsample of the star-forming clump class. In the following, UCH\textsc{ii} region counterparts will be treated as a separate class with respect to that of the generic star-forming clumps.

\begin{figure}[h!]
   \centering
   \includegraphics[width=0.49\textwidth]{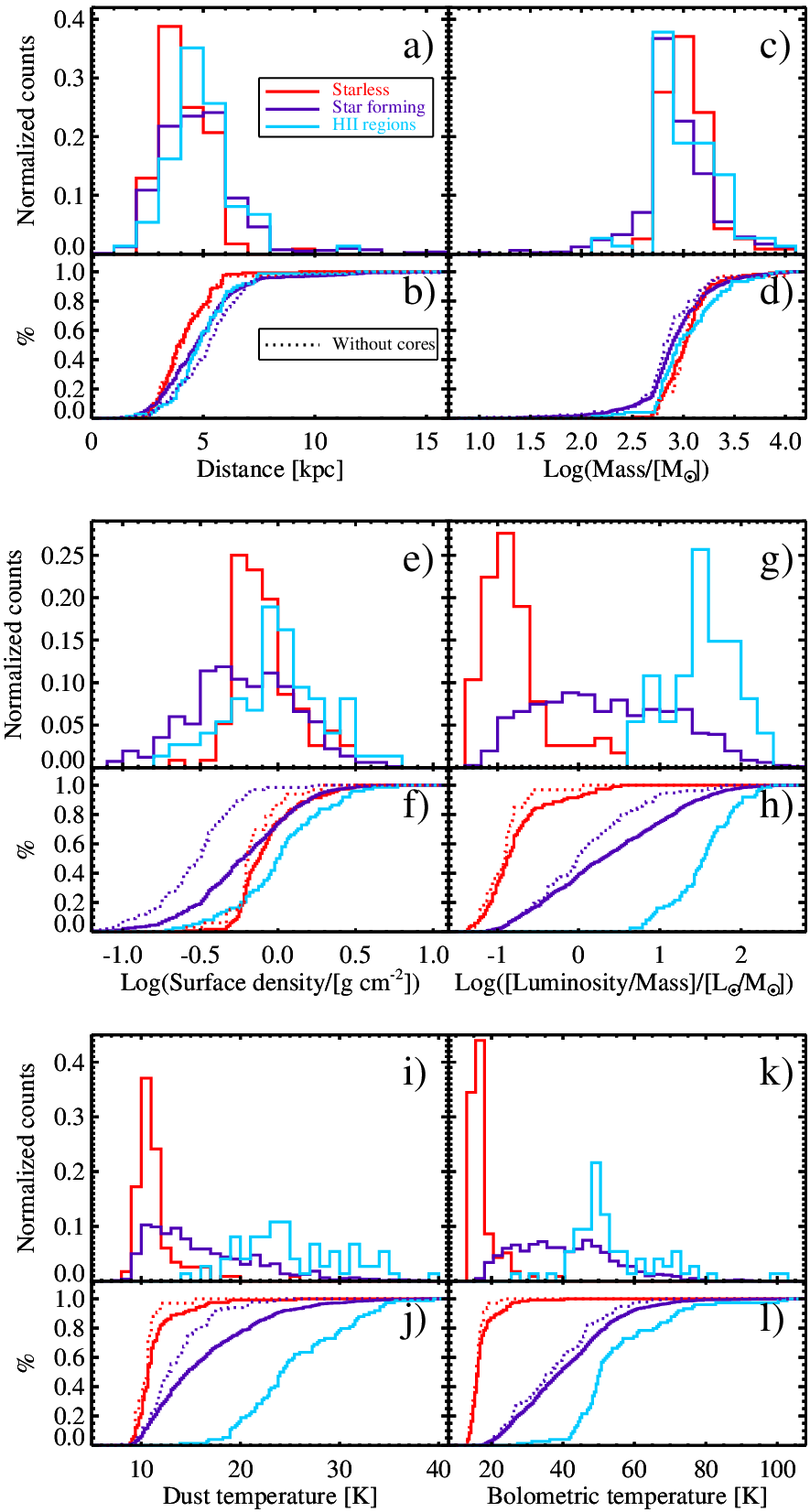} 
   \caption{Distributions of ALMAGAL target physical properties: $a$) heliocentric distance; $c$) mass (common logarithm); $e$) surface density (common logarithm); $g$) bolometric luminosity over mass ratio (common logarithm); $i$) modified black body temperature $k$) bolometric temperature. Panels $b$, $d$, $f$, $h$, $j$, $l$ show the corresponding cumulative distributions. Three different populations are shown separately: in red, quiescent (70-$\mu$m dark) clumps; in blue, star-forming (70-$\mu$m bright) clumps; in cyan, star-forming clumps associated with an ultra-compact H\textsc{ii~} region \citep[notice that, differently from][the counts of the third class of objects are not also included in those of the second class]{eli17,eli21}. All histograms are normalized by their total (\nnce, \nnco, and \nnch\ sources, respectively), therefore they do not mirror the actual number ratios between clump classes, and units on the $y$ axis are arbitrary. Additionally, in the panels containing the cumulative distributions, the red and dotted curves correspond to the subsamples of quiescent and star-forming clumps with no detection of mm cores in ALMAGAL observations (discussed in Sect.~\ref{zerocores}); UCH\textsc{ii~} regions are not shown, as only one of them appears devoid of cores inside.}
   \label{clumphist}
\end{figure}

Figure \ref{clumphist} shows the statistics of clump physical properties, separately for different evolutionary classes (quiescent, star-forming, UCH\textsc{ii}~region candidates). 
The initial requirements for selecting the Hi-GAL-selected subsample included a $M>500~\mathrm{M}_\odot$ constraint, so that the presence of smaller masses in the current list of ALMAGAL target properties is due to part of the RMS-selected targets, as well as to Hi-GAL-selected targets for which an update of distance produced a substantial change in mass.
As already highlighted by \citet{eli17,eli21}, the evolutionary indicators (dust temperature, $L/M$, bolometric temperature) are generally found to be enhanced in star-forming clumps, especially in those associated with compact radio emission.

Remarkably, also after the inclusion of the cross-match with the CORNISH South catalog, a large fraction (\cornishlmten\%) of the targets recognized as UCH\textsc{ii}~region candidates have $L/M$ above $10~\mathrm{L}_\odot/\mathrm{M}_\odot$, confirming the trend highlighted by \citet{urq13} and \citet{ces15}. In particular, \cornishlmthresh\% of them fulfill the threshold of $22.4~\mathrm{L}_\odot/\mathrm{M}_\odot$ suggested by \citet{eli17} as a conservative requirement for compatibility with the presence of a H\textsc{ii}~region.

Finally, a further examination of the classification between quiescent and star-forming clumps, considering the specific positions observed in ALMAGAL, is essential. In many star-forming Hi-GAL clumps, a systematic shift of the position of the emission peak across \textit{Herschel} bands, from 70 to 500~$\mu$m is observed. It can be shown that this migration is intrinsic rather than a result of resolution effects or incorrect astrometric alignment (Elia et al., in prep.). Near-infrared images of these objects often reveal clusters of young stellar objects (YSOs) coinciding with the 70-$\mu$m peak. In contrast, emission at SPIRE wavelengths (250 to 500~$\mu$m) generally originates from a NIR-dark region alongside. Consequently, for these sources, the SED collected from the five Hi-GAL bands reflects contributions from spatially connected but physically distinct components. For the Hi-GAL-selected ALMAGAL targets, the ALMA observations were centered on the 250-$\mu$m peak position. Therefore, in cases where the angular separation between the 70-$\mu$m and 250-$\mu$m peaks approaches or exceeds the ALMAGAL field of view (FOV) radius ($\sim 17.5\arcsec$), the ALMAGAL images may capture the quiescent portion of a clump rather than the star-forming region itself. In the ALMAGAL sample, this separation exceeds 15\arcsec\ for 11~sources, and 18\arcsec\ for three of these.
While these numbers reflect a bias that must be carefully taken into account when analyzing individual sources, their possible influence on overall trends for object classes can be considered statistically negligible.

\subsection{Core physical parameters}\label{corepar} 
The morphologic, photometric, and physical properties of fragments detected in ALMAGAL continuum observations were determined by \citet{col25}. 

The images were produced by combining data from various ALMA configurations \citep[see][] {san25}. Specifically, the compact configurations C-2 and C-3 (collectively referred to as TM2), and the extended configurations C-5 and C-6 (summarized as TM1) were combined with data from the 7m array of the Atacama Compact Array (ACA, referred to as 7M). For this reason, the final products considered by \citet{col25} are called 7M+TM2+TM1 maps.

Heliocentric distances of all cores in a given target are assumed to be the same as the global target distance, assigned as mentioned in Sect.~\ref{clumpparameters}. Based on the distances available at the time of the ALMA proposal, targets were divided into two samples, to be observed with different configurations of the interferometer to preserve some uniformity in spatial resolution: the ``near'' distance for $d \leq 4.66$~kpc (C2+C5 configurations, \nnear\ targets observed), and the ``far'' distance for $d > 4.66$~kpc (C3+C6 configurations, \nfar\ targets observed). The reassessment of distances by Benedettini et al. (in prep.) implied, in a minority of cases, that sources initially observed in the ``far'' configuration were reassigned a ``near'' distance, and vice versa. Details are provided by \citet{mol25}, Benedettini et al. (in prep.), and \citet{col25}.

The fragment\footnote{In this article, we use the terms fragment and core as interchangeable synonyms to indicate the compact sources detected in such images, with a subtle distinction: \textit{fragment} tends to denote the object as it appears in observations and is identified by the detection algorithm, while \textit{core} emphasizes its physical nature.} detection and photometry in the ALMAGAL fields were carried out with the CuTEx algorithm \citep{mol11}, specifically adapted to the ALMAGAL maps, as detailed in \citet{col25}. CuTEx performs background estimation and subtraction to derive final flux density measurements. Consequently, direct comparisons with core flux densities extracted in other similar ALMA surveys \citep[e.g.,][]{san19,svo19,and21,mor23} are not straightforward, as those studies include background emission, leading to systematically higher flux density estimates compared to ALMAGAL.

Here, we briefly summarize the overall morphologic features of the cataloged cores, as presented by \citet{col25}. The angular sizes range from 0.15\arcsec to 1.4\arcsec, with some distinction between the targets observed with the ``near'' configuration (for which the median angular size is $\sim 0.4\arcsec$) and the ``far'' one (median: $\sim 0.2\arcsec$; best: $\sim 0.15\arcsec$), respectively. Coupled with distances, this results in a physical size distribution whose 90\% is contained in the range $800 - 3000$~au, with a median of 1700~au.

The extracted ALMAGAL flux densities were converted into masses through the modified black body formula, assuming an opacity of 0.9~cm$^2$~g$^{-1}$ at 1.3~mm, suitable for dense environments \citep{oss94,san19}, a gas-to-dust ratio of 100 (not included in the aforementioned opacity), and a temperature $T_\textrm{core}$ established according to the following scheme based on the $L/M$ ratio of the parent clump: $T_\textrm{core}=20$~K for $L/M \leq 1~\mathrm{L}_\odot/\mathrm{M}_\odot$; $T_\textrm{core}=35$~K for $1~\mathrm{L}_\odot/\mathrm{M}_\odot < L/M \leq 10~\mathrm{L}_\odot/\mathrm{M}_\odot$; $T_\textrm{core}\propto (L/M)^{0.22}$ for $L/M > 10~\mathrm{L}_\odot/\mathrm{M}_\odot$, respectively \citep[see][for details]{col25}. Core masses are found in the range $0.002 - 345~\mathrm{M}_\odot$, with a mean value of $1.5~\mathrm{M}_\odot$ and a median of $0.4~\mathrm{M}_\odot$. The 90\% flux completeness in the catalog is 0.95~mJy, which translates into 0.13 and 0.37~M$_\odot$ at $d=3.5$ and 6~kpc, respectively.

\citet{col25} investigated whether core flux densities, and consequently core masses, can be overestimated due to contamination by free-free emission. They considered the possible presence of CORNISH and CORNISH South emission (including diffuse emission, not just the compact emission mentioned in Sect.~\ref{clumpparameters}) within the ALMAGAL fields. Through this approach, they identified \cornishcontamination\ targets where at least one core photometry might be affected by such contamination. In Appendix~\ref{nocornish}, we focus on the subsample of ALMAGAL targets confirmed to be free from this contamination, and we demonstrate that trends observed in the relationships between target properties generally remain unchanged, whether the analysis is performed on this subset or the entire sample.

To study distance bias in Hi-GAL clump measurements and clump-core connections, they simulated observing nearby star-forming regions (150–500 pc) at distances up to 7 kpc by progressively degrading their resolution. 

Finally, we summarize here the approach of \citet{bal17} in linking \textit{Herschel} clump properties to those of the cores they contain, whose results are frequently referenced throughout this paper. To study distance bias in Hi-GAL clump measurements and clump-core connections, they simulated observing nearby (150–500 pc) star-forming regions mapped in the \textit{Herschel} Gould Belt survey \citep{and10} at distances up to 7 kpc by progressively degrading their resolution. At each simulated distance, they re-extracted sources and recalculated physical parameters, comparing them to the original data. 
This approach revealed how distance affects \textit{Herschel}-based observables. Similarly, the resolution improvement from \textit{Herschel} clumps to ALMAGAL fragments discussed in the current paper mirrors the relationship between a clump observed in a map artificially degraded by \citet{bal17} to kpc distances and the cores detected in the original Gould Belt map (see also Appendix~\ref{appbaldeschi}).

\section{Clump fragmentation level}\label{ncore}

In this section, we adopt a statistical approach to investigate the level of fragmentation in the ALMAGAL targets and explore its relationship with their photometric, physical, and evolutionary properties.
The data are expected to reveal scattered distributions rather than clear-cut trends, indicating that multiple interacting factors contribute to determining the number of substructures within a fragmented clump. However, thanks to the statistical robustness offered by ALMAGAL, it is possible to identify average trends and to evaluate the potential impact of individual clump parameters.

\subsection{Clump fragmentation vs photometry}
The first question to address is whether it is possible to obtain predictions about the internal fragmentation level of clumps from a macroscopic property, such as their brightness in the far infrared.

\begin{figure}[ht!]
   \centering
   \includegraphics[width=0.47\textwidth]{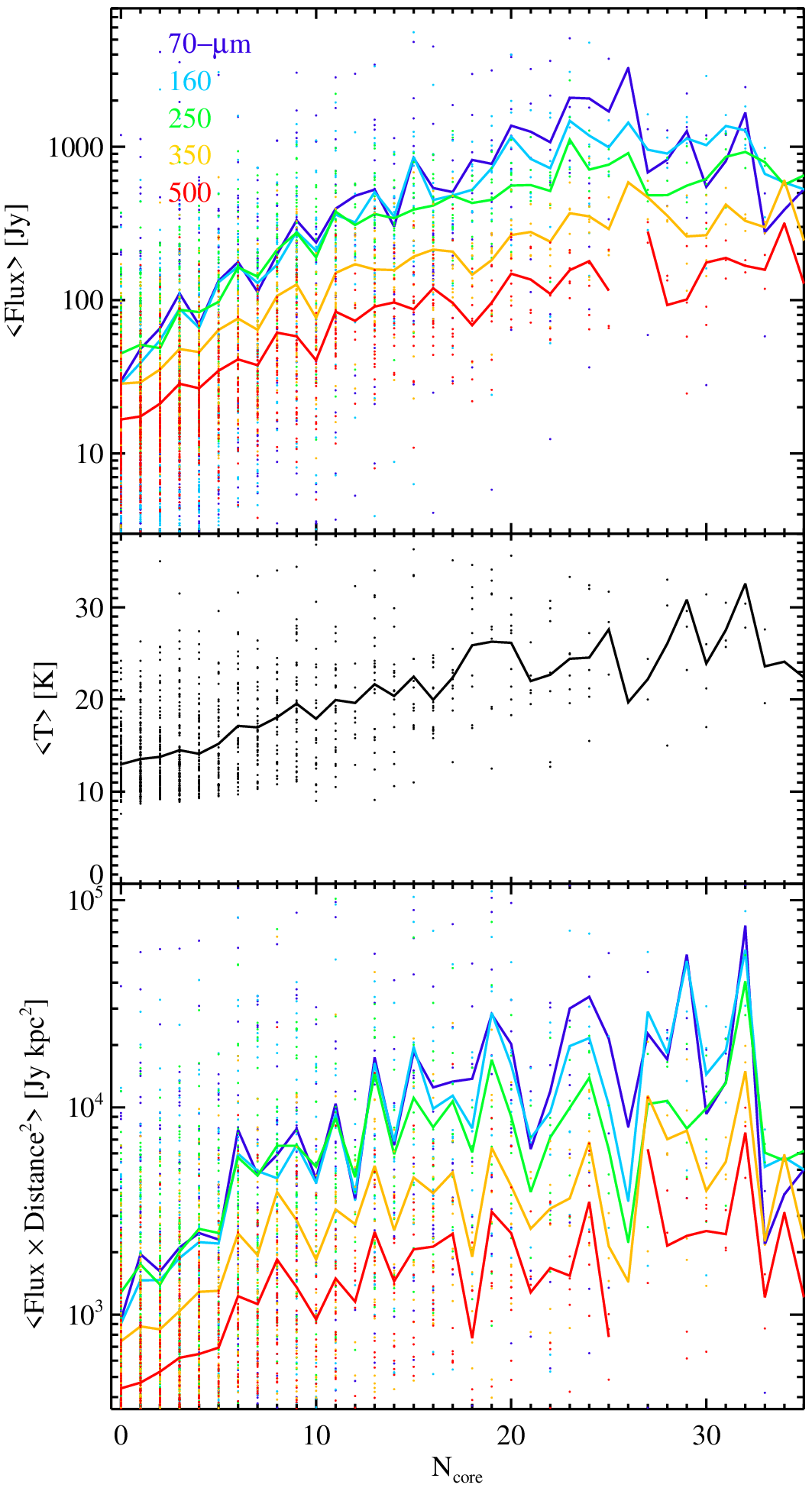}
   \caption{Properties of ALMAGAL targets as a function of the number of cores ($N_\mathrm{core}$) revealed in their interior.
   {\it Top}: fluxes in the five Hi-GAL bands (the core-band correspondence is reported in the legend), and their averages in bins of $\Delta N_\mathrm{core}=1$. 
   {\it Middle}:
   temperature determined using a modified-black-body fit to the Hi-GAL SEDs, and its average. 
   {\it Bottom}: as in the top panel, but for the product of flux densities by the squared distance. Both in the top and bottom panels, a break in the red curve corresponds to a bin populated by sources without a 500~$\mu$m detection.}
              \label{ncoresfluxes}
    \end{figure}

In Figure~\ref{ncoresfluxes}, top, we investigate the relation between the number of fragments detected within the clumps, $N_\mathrm{core}$, and the clump \textit{Herschel} flux densities in the five Hi-GAL bands, averaged in bins of $N_\mathrm{core}$. 
An overall increasing trend appears up to $N_\mathrm{core} \sim 25$, being more pronounced as the wavelength decreases from 500 to 160~$\mu$m. This is the range in which clump temperatures are estimated, and this imbalance among wavelengths is expected to be mirrored in the temperature distribution. In fact, in Figure~\ref{ncoresfluxes}, middle, an increase of the average clump temperature derived from the SED modified black body fit, estimated in bins of $N_\mathrm{core}$, is also seen up to $N_\mathrm{core} \sim 25$. In contrast, the region for $N_\mathrm{core} \gtrsim 25$ is statistically irrelevant, as \citet{col25} showed that, in this range, bins of $\Delta N_\mathrm{core}=1$ get populated by less than ten clumps.

To investigate a possible bias with the distance $d$ on the trends observed in the top panel of Figure~\ref{ncoresfluxes}, in the bottom panel of the same figure, the mean flux densities multiplied by $d^2$ (i.e., a quantity proportional to the monochromatic luminosity) are plotted versus the $N_\mathrm{core}$ parameter. An increasing trend similar to that seen in the top panel (although more scattered and shallower) confirms a general correlation between the number of fragments in the clump and the clump brightness at all wavelengths. A higher number of cores appears to be genuinely correlated with increased fluxes. This correlation cannot be trivially attributed solely to the presence of more emitting objects but instead carries evolutionary implications (i.e., more evolved clumps tend, on average, to host a larger number of cores), as suggested by the middle panel of Figure~\ref{ncoresfluxes}, and further explored in Sections~\ref{fragvsevol} and~\ref{multidiscussion}.

\subsection{Clump fragmentation vs distance}\label{fragvsdist} 

In principle, the number of detected cores in ALMAGAL targets might be underestimated due to both sensitivity limitations and resolution loss at increasing target distances, leading to confusion and loss of lower-mass fragments. Therefore, the pure count of cores detected in a given field turns out to be a quantity more affected by the distance bias than, for instance, the total mass in cores.

\begin{figure*}[p!]
   \centering
   \includegraphics[width=0.99\textwidth]{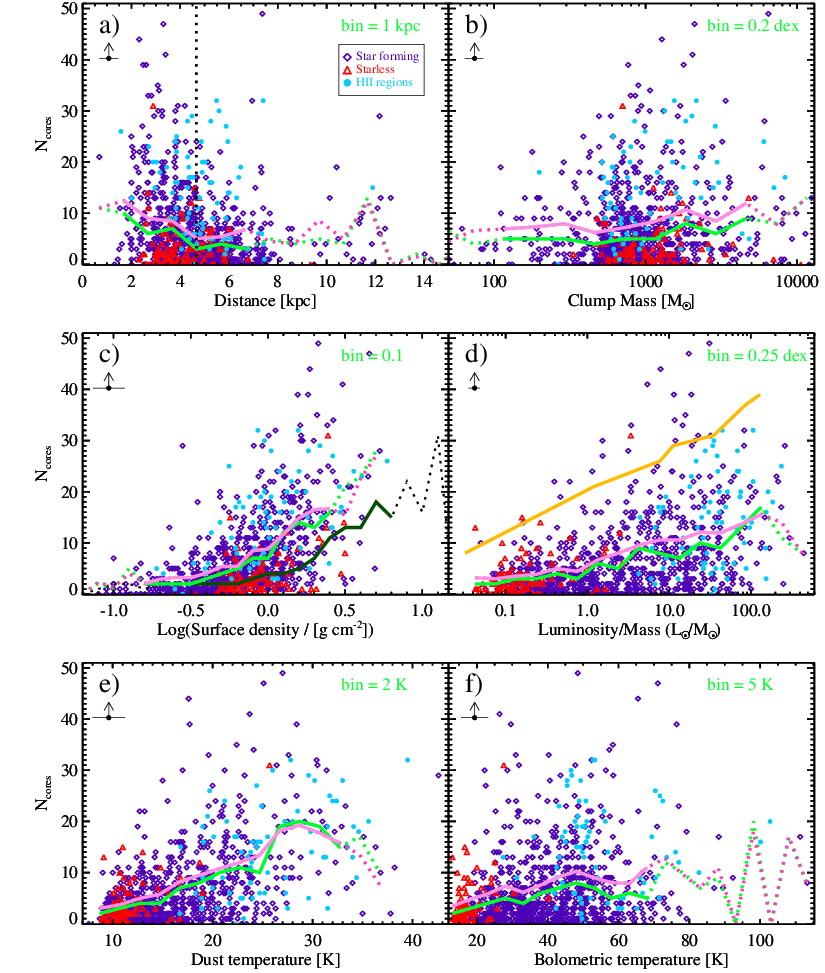}
 \caption{Number of fragments $N_\textrm{core}$ detected in each ALMAGAL target as a function of the target's physical parameters: $a$) heliocentric distance, with the vertical dotted line separating the distances originally assigned to the ``near'' and the ``far'' sample, see text; $b$) mass; $c$) surface density (logarithm); $d$) luminosity over mass ratio, with the orange line representing the prediction of the model of \citet{leb25} with initial conditions set to $M=500~\mathrm{M}_\odot$, $\mathcal{M}=7$, and $\mu=10$ (presented and discussed in Sect.~\ref{simulations}); $e$) modified black body temperature; $f$) bolometric temperature, respectively. Red open triangles are used for quiescent clumps, dark blue open diamonds for the star-forming ones, and light blue filled circles for counterparts of a UCH\textsc{ii}~region. The symbol under each panel label represents the typical error bar associated to data. In this case, the vertical error bar is replaced by an arrow, to indicate that the number of detected cores likely represents an underestimate of the ``actual'' $N_\textrm{core}$. The green line connects the medians of $N_\textrm{core}$ in bins (whose width is specified in green as well); its dotted parts correspond to bins containing low statistics (< 10 values). In particular, in panel $c$ the medians calculated in bins of surface densities based on deconvolved clump sizes is represented as a dark green solid/dotted line. The mean values are also shown, connected by a magenta line.}
   \label{ncoresvsall}
\end{figure*}

In panel~$a$ of Figure~\ref{ncoresvsall}, $N_\mathrm{core}$ is plotted against distance. Despite a large scatter due to the intrinsic level of fragmentation in various clumps, a slightly decreasing trend is observed in the median of $N_\mathrm{core}$ computed in 1-kpc bins of distance, at least in the interval $2< d < 8$~kpc, in which the statistical basis is rich enough (i.e. at least 10 objects per bin), confirming a mild influence of the expected distance bias. 
However, within this range, no discontinuity is observed, both in the overall scatter plot and in the median trend\footnote{In Figure \ref{ncoresvsall}, both the medians and averages of $N_\textrm{core}$ are presented as functions of various clump parameters, to appreciate differences between the two. However, the trends seen across all panels appear qualitatively similar, making it redundant to analyze both. Therefore, the discussion hereafter focuses solely on the behavior of the medians.}, in correspondence of $d=4.66$~kpc, namely the distance used to separate the two subsamples of ALMAGAL targets to be observed with different ALMA baseline configurations. Therefore, the decreasing trend of $N_\mathrm{core}$ with distance appears a more direct consequence of the decrease in mass sensitivity, related to $d^{-2}$, rather than in resolution \citep[cf.][]{col25}.

Finally, it is worth noting that analyzing $N_\mathrm{core}$ as a function of distance-independent clump parameters, as done in Sects.~\ref{fragvssurfd} and~\ref{fragvsevol}, mixes up the ordering of the clumps by distance, thus mitigating the impact of the bias seen in this context.

\subsection{Clump fragmentation vs mass and density}\label{fragvssurfd}

A search for a possible link between the fragmentation level of clumps and their total mass estimated with \textit{Herschel}, namely the mass available for fragmentation, is illustrated in panel~$b$ of Figure~\ref{ncoresvsall}. Confirming the results of \citet{fon18}, \citet{olm23}, and \citet{tra23}, no clear correlation is observed between these quantities, except for a mildly increasing median trend in the range of masses with higher statistics ($500 \lesssim M \lesssim 2000~\mathrm{M}_\odot$).

A more pronounced correlation is observed, instead, between median $N_\textrm{core}$ and clump surface density $\Sigma$ 
(Figure~\ref{ncoresvsall}, panel $c$). Unlike previous scatter plots, there is a noticeable increase in both the maximum and median $N_\textrm{core}$ values with rising surface density\footnote{We note that even when choosing to show, the median trend with respect to the surface density calculated from deconvolved sizes (black dotted line in panel $c$ of Figure~\ref{ncoresvsall}, which is generally higher (see Section~\ref{clumpparameters}), the behavior does not change qualitatively, and all considerations discussed so far remain valid. The same applies also to the trends of the most massive core (Sect.~\ref{mmmcvsmass}) and the core formation efficiency(Sect.~\ref{cfevsmass}), for which we will not revisit this point in detail.}. Additionally, for $\Sigma > 2~\mathrm{g}~\mathrm{cm}^{-2} $ instances of low fragmentation level (say $N_\textrm{core} \leq 3$) are almost absent. These aspects were glimpsed in more limited statistics (11~objects) by \citet{svo19}. More recently, \citet{mor24} highlighted a similar correlation between $N_\textrm{core}$ and $\Sigma$ in a sample of 39 targets. The ALMAGAL data further reinforce this trend, providing strong statistical significance.

On the one hand, the expression of classical Jeans' mass has an inverse square-root dependence on volume density, suggesting that fragmentation into further super-critical substructures is favored in conditions of high density (for an assessment of the equivalence of a clump description based on volume or surface density see Appendix~\ref{otherclumpprops}).

On the other hand, the theoretical prediction by \citet{kru08} proposes that a surface density equal to or larger than 1~g~cm$^{-2}$ is a necessary, though not sufficient, threshold to inhibit further fragmentation (thus promoting high-mass star formation) when a cloud undergoes heating. However, there is no noticeable flattening of the median $N_\textrm{core}$ vs $\Sigma$ trend around this value. Instead, such a flattening might be glimpsed near 2~g~cm$^{-2}$, but in a less statistically significant region of the diagram.

We note that the behavior of $N_\textrm{core}$ versus surface 
density should not be overinterpreted in an evolutionary sense. On the one hand, \citet{eli17} showed that, on average, the surface density is larger in star-forming clumps than in quiescent ones; furthermore, among the star-forming population, the densest sources are, on average, those appearing dark at $\sim 20~\mu$m. However, since a high surface density was imposed as a constraint to select ALMAGAL targets, this is not necessarily true in the objects studied here (see Figure~\ref{clumphist}, $c$). On the other hand, for sources with high $L/M$ ($\gg 10~\mathrm{L}_\odot/\mathrm{M}_\odot$), expected to be the most evolved in the Hi-GAL catalog, a large spread of surface density values is observed, from high values typical of very concentrated envelopes to low ones possibly corresponding to a clump that is being cleaned up by its internal protostellar activity. This aspect is further discussed in Appendix~\ref{otherclumpprops}. In this sense, exploring the $N_\textrm{core}$ number as a function of the parent clump surface density cannot yield clear evolutionary indications as other distance-independent parameters, such as $T$ or $L/M$, can do. This is
confirmed in panel $c$ of Figure~\ref{clumphist} and in panel
$c$ of Figure~\ref{ncoresvsall} by the high degree of scattering shown by the H\textsc{ii} regions along the $x$ direction. 

The differentiation in fragmentation trends as a function of clump surface density and evolutionary stage will be discussed in further detail in Sect.~\ref{multidiscussion}.

\subsection{Clump fragmentation vs evolutionary status}\label{fragvsevol}
The development of fragmentation with time depends on local conditions, and consequently, the final distribution of masses can strongly vary from case to case. The statistics provided by ALMAGAL offer an interestingly varied picture of fragmentation as a function of clump evolutionary indicators.

\begin{figure}[ht!]
   \centering
   \includegraphics[width=0.49\textwidth]{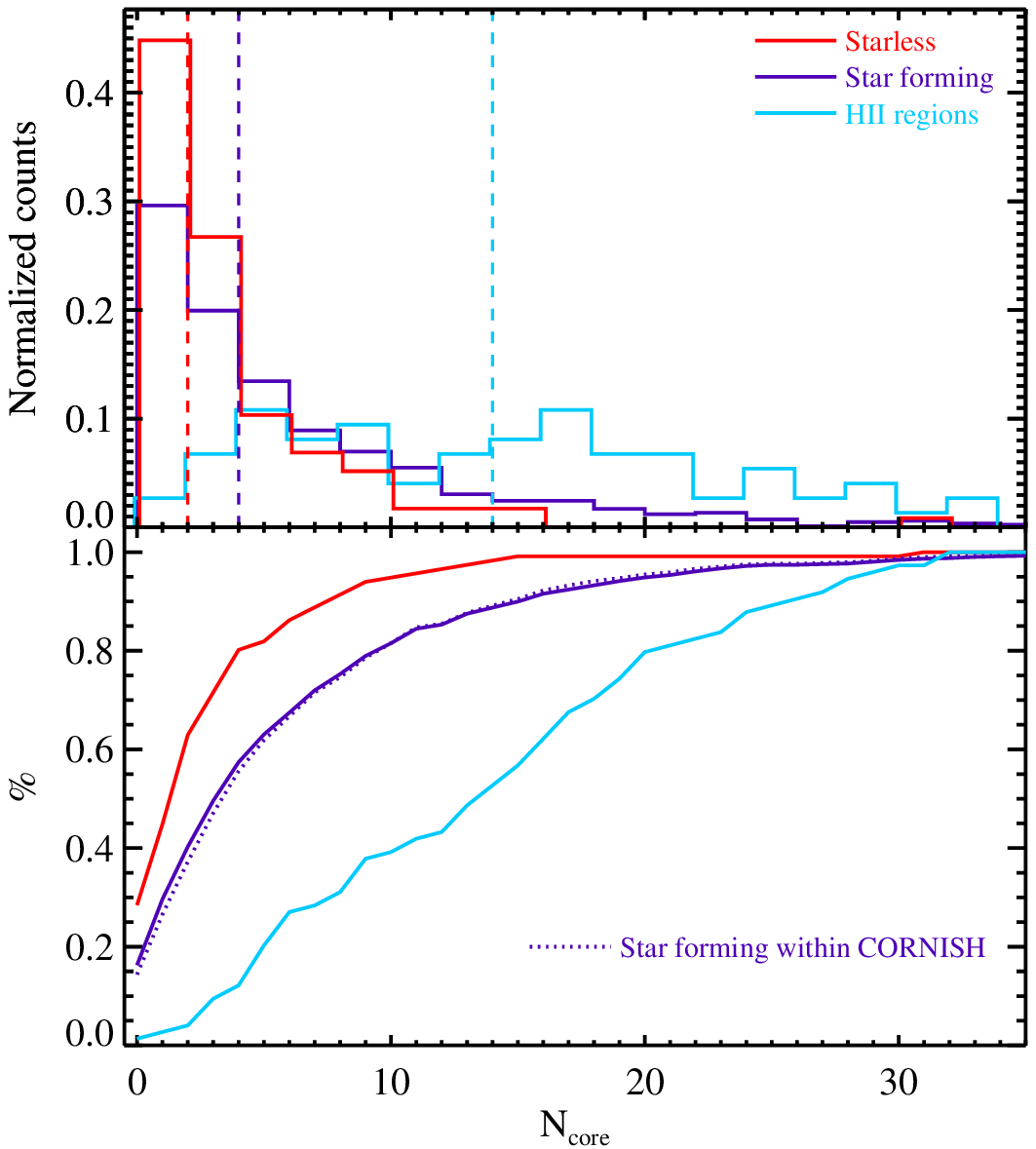}
   \caption{\textit{Top}: statistical distribution of the fragment number $N_\mathrm{core}$ for the ALMAGAL targets, categorized into quiescent (red histogram), star-forming (blue), and CORNISH-based UCH\textsc{ii} regions (light blue).
   The sum of each histogram is normalized to 1, therefore all histograms subtend the same area and do not reflect the actual numerical proportions among the classes. The three vertical dashed lines indicate the medians of these distributions, calculated excluding cases where $N_\mathrm{core}$=0. \textit{Bottom}: Cumulative distributions corresponding to the histograms shown in the top panel. The cumulative distribution of $N_\mathrm{core}$ for star-forming targets located within the regions covered by the CORNISH and CORNISH South surveys is also shown (dotted blue), to highlight the potential impact of contamination from UCH\textsc{ii} regions (see text).}
   \label{ncoresclasses}
\end{figure}

Figure~\ref{ncoresclasses}, top, shows the distribution of $N_\mathrm{core}$ in the ALMAGAL sample, divided into quiescent clumps, star-forming clumps, and UCH\textsc{ii} regions. The UCH\textsc{ii} regions exhibit a flatter distribution, with no strong preference for low $N_\mathrm{core}$ values and a median\footnote{All medians reported in this paragraph are calculated excluding the cases where $N_\mathrm{core}=0$. Results also including $N_\mathrm{core}=0$ are provided in Table~\ref{table_classes}.} of \medncorehii. In contrast, the distributions for the other two classes are qualitatively similar, with the histogram of star-forming clumps showing a stronger right skew than that of quiescent ones. The respective medians are \medncorepro\ and \medncorepre, but a two-sample Kolmogorov-Smirnov (K-S) test is required to determine whether the differences between the data sets are statistically significant. In this case, the K-S statistic is $D=\ksncorep$, which must be compared with the critical value for rejecting, at a 99\% confidence level, the null hypothesis that the two distributions are random samples from the same population. The critical value is given by 
\begin{equation}
    D_{m,n,0.01}=1.63 \times \sqrt{\frac{m+n}{m n}}\,,
    \label{ksequation}
\end{equation}
where $m$ and $n$ are the sizes of the two samples \citep[e.g.,][]{roh15}. For this analysis, the sizes of the quiescent and star-forming clump subsamples are $m=\nncep$, $n=\nncop$, respectively, yielding $D_{\nncep,\nncop,0.01}=\kscriticalncore$. The fact that $D$ is larger than this value indicates a statistically significant difference between the  $N_\mathrm{core}$ distributions for quiescent and star-forming clumps. Repeating the same exercise for a 99.87\% confidence level, i.e. $3~sigma$ increases the factor in Equation~\ref{ksequation}  to 1.92, and consequently increases $D$ to 0.22, which remains below the K-S statistics obtained in this case.

The differences among these distributions become more evident if seen in terms of cumulative functions (Figure~\ref{ncoresclasses}, bottom). The distribution for quiescent clumps appears more distinctly right-skewed compared to that of star-forming clumps. In contrast, UCH\textsc{ii} regions show an almost linear trend over a wide range of $N_\mathrm{core}$, consistent with the nearly flat histogram.

Furthermore, we investigate whether the fact that several ALMAGAL star-forming targets lie outside the areas covered by the CORNISH and CORNISH South surveys can lead to significant contamination by UCH\textsc{ii} regions. Figure~\ref{ncoresclasses}, bottom, shows the cumulative curve for star-forming targets within the coverage of these two surveys: this curve differs only negligibly from the overall cumulative distribution, indicating that any potential bias can be considered negligible both here and throughout the rest of the paper.

\begin{table*}
\caption{Statistics  of core population parameters (namely fragmentation level, mass of the most massive core, and core formation efficiency).}           
\label{table_classes}      
\centering           
\begin{tabular}{lcccccccccccc} 
\hline\hline 
Parameter & \multicolumn{3}{c}{Quiescent} && \multicolumn{3}{c}{Star-forming} && \multicolumn{3}{c}{UCH\textsc{ii}} \\
\cline{2-4}\cline{6-8}\cline{10-12}
 & \# & Median & $Q_1$; $Q_3$  && \#& Median &  $Q_1$; $Q_3$ && \# & Median & $Q_1$; $Q_3$ & \\ 
\hline                   
$N_\mathrm{core}$ & \nnce & \medncorepre & \percncorepre && \nnco & \medncorepro & \percncorepro && \nnch & \medncorehii & \percncorehii \\
 $N_\mathrm{core}$ ($>0$) & \nncep & \medncoreprep & \percncoreprep && \nncop & \medncoreprop & \percncoreprop && \nnchp & \medncorehiip & \percncorehiip \\
 $M_\mathrm{mmc} [\mathrm{M}_\odot]$ &  \nmmmce & \medmmmcpre & \percmmmcpre &&  \nmmmco & \medmmmcpro & \percmmmcpro && \nmmmch & \medmmmchii & \percmmmchii \\ 
CFE [\%] &  \ncfee & \medcfepre & \perccfepre &&  \ncfeo & \medcfepro & \perccfepro && \ncfeh & \medcfehii & \perccfehii\\
\hline 
\end{tabular}
\tablefoot{The statistics (number of objects and quartiles) are grouped based on the evolutionary class of each target. For the fragmentation level, values are provided both for all targets and for the subset with detected cores only.}
\end{table*}

To obtain a further differentiation of clumps across different evolutionary stages, the distribution of $N_\textrm{core}$ as a function of a few descriptors is discussed in the following.
Already in the middle panel of Figure~\ref{ncoresfluxes} an average increase of temperature with $N_\mathrm{core}$ suggested that a larger number of fragments is found in warmer objects. 
 
In panel~$d$ of Figure~\ref{ncoresvsall}, first we analyze the distribution of the number of detected fragments, $N_\textrm{core}$, in each clump as a function of the clump $L/M$ ratio. We observe that cases with few or no fragments appear across the entire range of $L/M$. However, the highest values of $N_\textrm{core}$ at different regimes of $L/M$ tend to increase with this ratio. Specifically, clumps with low $L/M$ ($< 1~\mathrm{L}_\odot/\mathrm{M}_\odot $) reveal a relatively low number ($<10$) of detected cores, whereas higher numbers are increasingly found, in many cases, at larger $L/M$, thus suggesting a possible evolutionary implication. 

This can be summarized in the overall increasing behavior of the median of $N_\textrm{core}$ in bins of $\log(L/M)$, at least for $L/M < 100~\mathrm{L}_\odot/\mathrm{M}_\odot$, in which the statistics are meaningful (more than 10 sources per bin). Noticeably, such a trend was not found by \citet{tra23}, most likely due to a narrower range of $L/M$ investigated, the much smaller statistics, and the coarser angular resolution ($>1 \arcsec$).
A further discussion of the relationship between $N_\mathrm{core}$ and $L/M$ in light of numerical simulations is provided in Sect.~\ref{multidiscussion}.

In panels~$e$ and $f$ of Figure~\ref{ncoresvsall}, $N_\textrm{core}$ is plotted against the clump dust temperature $T$ and bolometric temperature $T_\mathrm{bol}$, respectively. An increasing level of fragmentation corresponds, on average, to the increase of these two evolutionary indicators, confirming what emerged from panel $d$.
In the case of $T$, this may initially seem counterintuitive when considering, for example, the analytical form of the thermal Jeans mass. However, since clump temperature is strongly correlated with other evolutionary parameters \citep{eli17}, the observed increase of $N_\textrm{core}$ as a function of temperature should be regarded as an evolutionary effect rather than a ``static'' one. Specifically, while the clump's average density and temperature reflect global conditions, local fluctuations within the clump can lead to lower Jeans masses, facilitating fragmentation. Over time, as a consequence of the global collapse of the clump, these conditions can be achieved in different regions of the clump, resulting in the formation of new fragments. In practice, barring significant episodes of core coalescence, $N_\textrm{core}$ is expected to increase monotonically with time, or at least to remain constant if strong feedback from newly formed young stellar objects effectively suppresses further fragmentation throughout the clump. Additionally, we cannot rule out the possibility of a selection effect, where warmer cores naturally exhibit higher continuum flux densities.

\subsection{Properties of clumps without core detections}\label{zerocores}
Complementary to discussing the relations of the number of detected cores in ALMAGAL targets is investigating the statistics of parameters, both photometric and physical, of those targets in which \citet{col25} find no compact emission above 5~$\sigma$ in the 7M+TM2+TM1 images.

\begin{figure}[h!]
   \centering
   \includegraphics[width=0.49\textwidth]{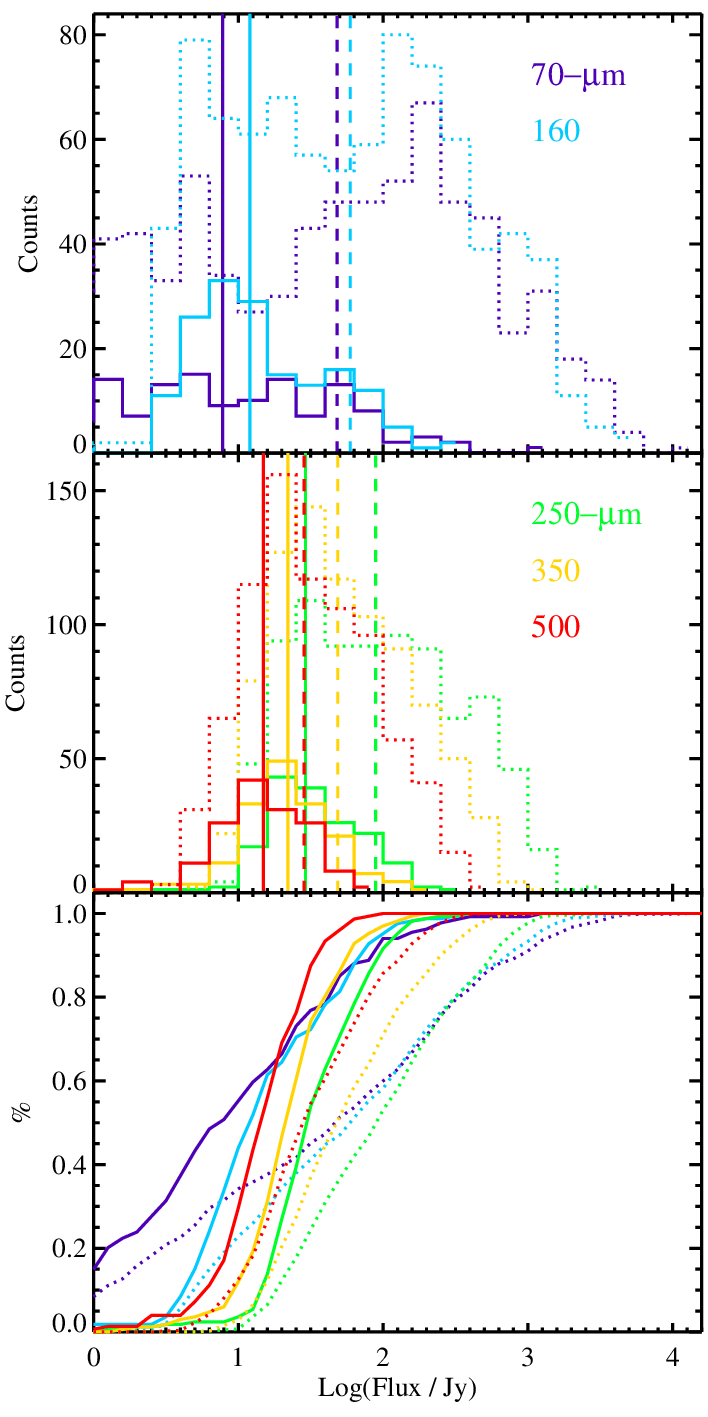}
   \caption{\textit{Top}: histograms of Hi-GAL PACS flux densities (dark blue: 70~$\mu$m; light blue: 160~$\mu$m) for ALMAGAL targets with and without detection of cores (dotted and solid line, respectively). Median values for the two subsamples are shown as vertical dashed and solid lines of the same color, respectively. \textit{Middle}: the same as in the top panel, but for SPIRE flux density distributions (green: 250~$\mu$m; yellow: 350~$\mu$m; red: 500~$\mu$m).  \textit{Bottom}: The cumulative distributions of histograms shown in the top and middle panels, using the same color and line style convention.}
   \label{zerofrag}
\end{figure}

The distributions of \textit{Herschel} flux densities for these sources are shown in Figure~\ref{zerofrag}, top and middle panel. In all bands, they are clearly centered on significantly lower values compared to the distributions of targets with detected cores. This trend is further highlighted in Figure~\ref{zerofrag}, bottom, by showing the cumulative distributions of the histograms displayed in the panels above, and quantified in Table~\ref{zerocoretab}, which reports lower median flux densities for these sources. 

\begin{table}
       \caption{Median photometric and physical parameters of the entire ALMAGAL sample and the subsamples with and witout cores.}   \label{zerocoretab}
    $$
\begin{array}{lccc}
            \hline
             \noalign{\smallskip}
            \mathrm{Parameter}      &  \mathrm{All~clumps} & \mathrm{Without~cores} & \mathrm{With~cores}\\
             \noalign{\smallskip}
             \hline
             \noalign{\smallskip}
             F_{70}~[\mathrm{Jy}]& \medfblue & \medfbluez &\medfbluey\\
             F_{160}~[\mathrm{Jy}]& \medfred & \medfredz & \medfredy\\
             F_{250}~[\mathrm{Jy}]& \medfPSW & \medfPSWz & \medfPSWy \\
             F_{350}~[\mathrm{Jy}]& \medfPMW & \medfPMWz & \medfPMWy \\
             F_{500}~[\mathrm{Jy}]& \medfPLW & \medfPLWz & \medfPLWy \\
             T~[\mathrm{K}] & \medtdust & \medtdustz & \medtdusty\\
             M~[\mathrm{M}_\odot] & \medmass & \medmassz & \medmassy  \\
             \Sigma~[\mathrm{g~cm}^{-2}]  & \medsurfd & \medsurfdz & \medsurfdy \\
        L~[\mathrm{L}_\odot] & \medlum & \medlumz & \medlumy  \\L/M[\mathrm{L}_\odot/\mathrm{M}_\odot] & \medlm & \medlmz & \medlmy\\
              T_\mathrm{bol}~[\mathrm{K}] & \medtbol & \medtbolz & \medtboly  \\
             \noalign{\smallskip}
             \hline
         \end{array}
     $$
\end{table}

Note that the target heliocentric distance can influence the comparison of fluxes between clumps with and without cores. A lower \textit{Herschel} flux in the absence of detected cores may reflect either the intrinsically faint nature of the clump or its large distance, as the sensitivity limit, which constrains the observed $N_\textrm{core}$, follows the same $d^{-2}$ dependence on distance. Rather than distinguishing between these scenarios, here we simply test for a possible link between the non-detection of cores and relatively low flux in one or more \textit{Herschel} bands.

In this respect, however, among the targets without cores, some cases with high flux densities (> 100~Jy in one or more bands) remain. These cases, of course, cannot be explained as unfavorable occurrences of a lower sensitivity. Indeed, we checked this point by comparing the medians for the rms noise for fields without detected cores and for the subsample of those with $F_{250} > 100$~Jy, and for $F_{350} > 100$~Jy, and in both cases we find no differences in the distribution of rms values. These considerations prevent us from extracting a stringent recipe for possible lower limits on the Hi-GAL flux densities, ensuring the presence of bright cores inside the clump.

The ALMAGAL clumps without core detections show, in general, lower values of distance-independent parameters such as $T$, $\Sigma$, $L/M$, and $T_\mathrm{bol}$, as illustrated in Table~\ref{zerocoretab}.
Although distributions remain wide, we can observe that the clumps without core detection can be preferentially found among the less evolved or less dense in the ALMAGAL sample. 

For a deeper insight into this aspect, we also show in detail in Figure~\ref{clumphist} the cumulative distributions of the properties of clumps without core detections, shown separately for quiescent and star-forming ones. First of all, it should be noted that this statistics is based on \nncen\ out of \nnce\ quiescent clumps (\nncenperc\%), and \nncon\ out of \nnco\ star-forming clumps (\nnconperc\%), cf. Table~\ref{table_classes}\footnote{For the UCH\textsc{ii} regions, only one target appear without cores, so this sub-sample is not considered in the present discussion.}. This already provides a first indication of an evolutionary trend, non-detections occur more frequently in quiescent than in star-forming targets. In panels $h$, $j$, and $l$ of Figure~\ref{clumphist}, the cumulative distributions of $T$, $L/M$, and $T_\mathrm{bol}$, respectively, rise more steeply for clumps without cores than for the entire population, both for quiescent and star-forming clumps. This trend is even more evident for $\Sigma$ (panel $f$), confirming what was suggested by the values in Table~\ref{zerocoretab}  for the whole ALMAGAL sample.

In turn, this is consistent with the general picture that emerged for $N_\textrm{core}$ as a function of density and evolutionary state in Sect.~\ref{fragvssurfd} and~\ref{fragvsevol}, respectively.

Finally, panel $b$ shows that, at least for star-forming clumps, distance can be responsible for a loss of sensitivity, as already discussed above, leading to an apparent increase in the number of cases with low $N_\textrm{core}$, while mass does not appear to be systematically affected by this bias (panel $d$). In summary, the absence of detected cores in an ALMAGAL target may result from a joint effect of sensitivity and distance effects, or may be related to the clump density or evolutionary stage, or to a combination of these factors.

It is also necessary, at this stage, to consider two additional potential causes for the lack of core detections. The first is that the ALMAGAL target may in fact correspond to an evolved star or an external galaxy.
A search in SIMBAD revealed only two cases without cores that could be consistent with the former scenario, and none with the latter. The second possibility is that the ALMAGAL pointing, centered on the 250~µm emission peak, is significantly far from the 70~µm peak, leading to missing the region most densely populated of cores within the clump, as discussed in Section~\ref{clumpparameters}.
However, as demonstrated in that section, the number of such cases is not statistically significant enough to produce general trends as those discussed in this section.

Younger, potentially pre-stellar cores, which often exhibit shallower and generally fainter intensity profiles \citep[e.g.,][]{beu02,gia13,gom21} may be more prone to not being detected in the 7M+TM2+TM1 images, for which the higher angular resolution results in worse brightness sensitivity \citep[as highlighted by][]{san25}. A scenario in which only such cores are present, ultimately resulting in observing $N_\mathrm{core}=0$, is expected to occur in ALMAGAL clumps classified as quiescent.
A detailed analysis of the ALMAGAL maps for fields without detected cores is beyond the scope of this paper and will be addressed in a forthcoming study by Coletta et al. (in prep.), through a dedicated analysis using 7M+TM2-only images to identify potential fragments not identified in 7M+TM2+TM1 observations. This approach aligns with the scenario presented here, as clumps with lower densities and/or early evolutionary indicators are likely candidates to host such early-stage cores.

\newcommand{\fmmc}{$f_\mathrm{MMC}$}
\newcommand{\mmmc}{$M_\mathrm{MMC}$}
\section{Mass of the most massive core}\label{mmcmass}

The mass \mmmc\ of the most massive core (MMC) in a clump is a quantity of great interest because it reflects a clump's ability to concentrate matter to build a massive substructure. Being a result of the fragmentation, \mmmc\ is inherently correlated with it and serves as a quantitative descriptor \citep[e.g.,][]{kir16,lin19,san19,and21,mor23}. Another specific quantity that also accounts for the mass of the parental clump is the fraction \fmmc\ of the clump mass contained in its MMC.

These quantities are analyzed below in relation to various clump properties. Despite significant scatter, some average trends emerge. On the one hand, this suggests that certain parameters may indeed influence the observed \mmmc. On the other hand, it is evident that the observed values result from the complex interplay of multiple concomitant factors, making it challenging to disentangle their individual contributions.

\subsection{Mass of the most massive core vs distance}

\begin{figure*}[ht!]
   \centering
   \includegraphics[width=0.99\textwidth]{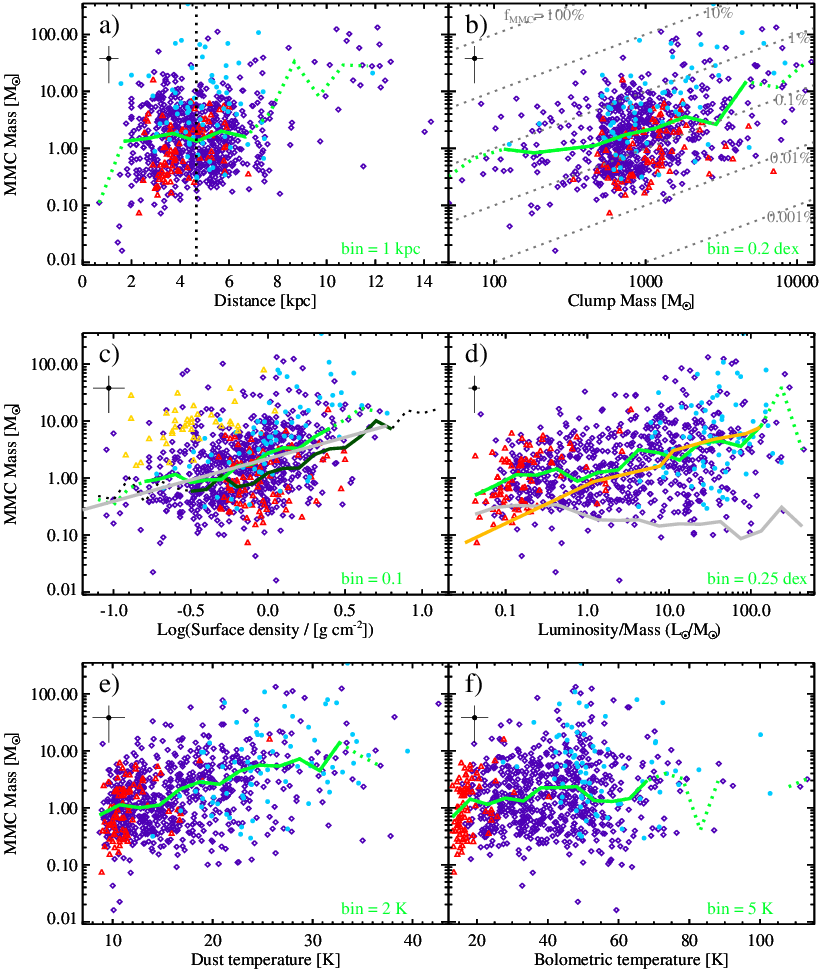}
   \caption{The same as Figure~\ref{ncoresvsall}, but for the MMC mass along the $y$-axis of each panel. The symbol under each panel label represents the typical error bar associated to data.
   The gray dotted lines in panel~$b$ indicate the trends for constant \fmmc\ values.
   In panel~$c$, the solid gray line indicates the power-law fit to the data; also, yellow open triangles represent data of \citet{mor23}. In panel $d$, the gray line represents the median mass of the least massive core in each clump.}
   \label{massmaxvsall}
\end{figure*}

To investigate a possible distance bias in the estimation of the MMC mass, these two quantities are plotted in panel~$a$ of Figure~\ref{massmaxvsall}\footnote{For a plot of all core masses as a function of the heliocentric distance, see \citet{col25}.}.
The points in the diagram exhibit significant scatter, and no overall increase in the median MMC mass can be observed across the $2< d < 7$~kpc range, where robust statistics are available, nor a
break in slope at the separation between the ``near'' and the ``far'' subsamples.

Notably, for $d>10$~kpc we find  $M_\mathrm{MMC} > 5~\mathrm{M}_\odot$. Further fragmentation of the most massive core cannot be excluded in these cases if observed with better spatial resolution.

Despite these minor biases, the subsequent discussion of the MMC mass is not significantly affected. Specifically, the trends observed in the relationships between MMC mass and distance-independent clump parameters (Sects.~\ref{mmmcvsmass}, \ref{mmcvsevol}) appear uncorrelated with distance. For example, as a test, we recalculated all median trends of the other clump parameters shown in panels $b$ to $f$ of Figure~\ref{massmaxvsall} by considering only targets located at $2~\mathrm{kpc} < d < 6~\mathrm{kpc}$. The medians based on this subsample exhibit trends identical to those obtained for the full sample and are, over extended ranges, even indistinguishable from them. Thus, while distance bias may introduce some scatter into these plots, it does not impose a systematic effect.

\subsection{Mass of the most massive core vs clump mass and density}\label{mmmcvsmass}

In panel~$b$ of Figure~\ref{massmaxvsall} the \mmmc\ detected in each clump is plotted against the total clump mass. A large scatter is found, with a Spearman's rank correlation coefficient $\rho_\mathrm{S}=\spmaxmass$. The significance levels of 5\% and 0.03\% (i.e. $3 \sigma$) correspond to critical values $\rho_\mathrm{crit}=0.07$ and 0.13 are, respectively, while \spmaxmass\ would correspond to a $p$-value of $\sim 10^{-24}$, so that one might conclude a direct correlation between the two quantities. However, given the discussion in Section~\ref{spearman} about the significance of Spearman's critical values in the presence of large data sets, and the highly scattered appearance of panel~$b$, we consider the degree of correlation in this plot to be very low. In other ALMA-based studies analyzing smaller sample sizes (then corresponding to a much higher required $\rho_\mathrm{crit}$), \citet{and21} found $\rho_\mathrm{S}=0.54$ for 35 sources ($p=8\times 10^{-4}$), while \citet{san19} $\rho_\mathrm{S}=0.08$ for 12 sources ($p=0.8$), and \citet{mor23} $\rho_\mathrm{S}=0.27$ for 39 sources ($p=0.15$). In all these works, Spearman's correlation coefficients were considered insufficient by the respective authors to support the presence of a clear correlation. 

The lack of correlation is supported by the results of numerical simulations. Based on simulations of turbulence in molecular clouds, \citet{bon04} found that the MMC mass is not correlated with the clump mass, but is determined by how the competitive accretion proceeds in forming a cluster in the clump. More recently, in their simulations \citet{smi23} find that the probability of a low-mass cloud forming a star as massive as those often formed from high-mass clouds is quite low. In this respect, if star formation cannot be completely deterministic, it cannot be purely stochastic either, as suggested by \citet{bon04}. By examining the relationship between maximum stellar mass and cluster mass, \citet{smi23} identify two distinct power-law regimes: one for cluster masses below 500~M$_\odot$ and one above. In the higher mass regime (which is more comparable to the typical clump masses in ALMAGAL), the slope is 0.11, shallower than the slope found in the lower mass range.

To the contrary, \citet{xu24assemble} found a direct correlation ($\rho_\mathrm{S}=0.73$, $p=0.01$) between the MMC and the clump mass for their sample of 11 clumps, selected by means of the following constraints on mass and luminosity (in addition to further constraints on spectral line parameters): $8 \times 10^2 \mathrm{M}_\odot < M < 2 \times 10^4 \mathrm{M}_\odot$, and $1 \times 10^4 \mathrm{L}_\odot < L < 6 \times 10^5 \mathrm{L}_\odot$, respectively. They interpret this correlation as evidence of co-evolution of clump and most massive core, namely in massive star-forming objects gas accretion connects a variety of scales, from filaments to clumps to cores, so that the masses of both the clump and the most massive core are expected to increase with evolution.
In our data, we only see a mildly increasing trend of the median of \mmmc,  
(Figure~\ref{massmaxvsall}, panel~$b$). However, it is not sufficient to state that \mmmc\ is univocally influenced by the total mass of the clump, and/or co-evolving with it. 

\begin{figure}[ht!]
   \centering
   \includegraphics[width=0.49\textwidth]{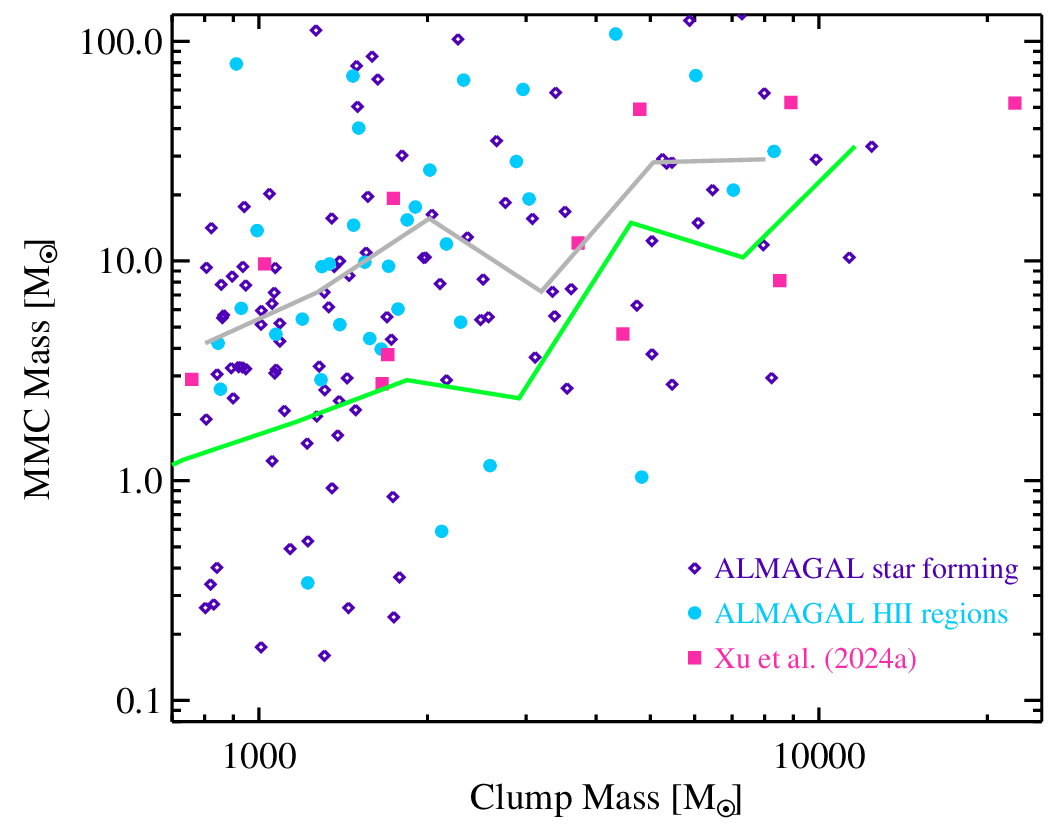}
   \caption{Mass of the most massive core (MMC) vs clump mass for the ALMAGAL star-forming targets (blue open diamonds) and CORNISH(/South) counterparts (light blue filled circles), limited to $8 \times 10^2~\mathrm{M}_\odot < M < 2 \times 10^4~\mathrm{M}_\odot$ and $1 \times 10^4~\mathrm{L}_\odot < L < 6 \times 10^5~\mathrm{L}_\odot$. The green line connects the medians of MMC mass in logarithmic bins of~0.2 for the entire sample (the same as in Figure~\ref{massmaxvsall}, panel~$b$), while the gray line is recalculated only with the points plotted here. Furthermore, magenta-filled squares represent the values of \citet{xu24assemble}.}
   \label{maxmassxu}
\end{figure}

To allow a direct comparison with results of \citet{xu24assemble}, in Figure~\ref{maxmassxu}, we show a separated view of the MMC mass against clump mass, limited to the \ndataxurange\ ALMAGAL sources having $M$ and $L$ fulfilling the constraints imposed by these authors to their sample (see above). The points of \citet{xu24assemble} appear fully embedded in the general scatter of the ALMAGAL subsample ones, whose median shows a behavior similar to that seen for the whole sample. On the one hand, increasing the statistics by one order of magnitude reveals a much larger scatter in our MMC masses, arguing against a scenario of core+clump mass co-evolution. On the other hand, our selection based on mass and luminosity ranges cannot isolate objects with exactly the same characteristics as those in \citet{xu24assemble}, which were additionally filtered using spectral line criteria. Therefore, we cannot rule out the possibility that the narrower \mmmc\ observed by them results from the additional constraints we were unable to apply in this study.

A strong correlation between the MMC and the clump mass was also found by \citet{lin19}, who investigated clump fragmentation using single-dish images at $350~\mu$m. However, it is important to note that the size ratio between the clumps and their fragments (detected at $350~\mu$m at a resolution of 8\arcsec.5) was typically around 2, and the average number of detected fragments was also two on average. This suggests that when the sizes of the structures being compared (clumps and their fragments) are relatively similar, these two quantities still exhibit a correlation. In contrast, when analyzing smaller-scale fragments, which correspond to localized regions within the clump, as it happens in ALMAGAL, the mass of the most massive fragments appears to be more strongly influenced by local conditions within the clump's internal structure, rather than by the clump's total mass, which represents a global parameter.

Following \citet{and21}, panel~$b$ of Figure~\ref{massmaxvsall} also includes lines corresponding to specific values of \fmmc. The vast majority of cases lie in the range $0.01\% <$~\fmmc~$< 1\% $, with mean and median values being \fmmcave\% and \fmmcmed\%, respectively. This distribution differs significantly from the findings of \citet{and21}, where \fmmc\ was reported to range between 3 and 24\%. As noted above, the discrepancy is primarily driven by differences in the numerator, \mmmc, which in most cases exceeds 10~M$_\odot$ in \citet{and21}, while the clump masses in the denominator are comparable to those in our study. This can be attributed, in turn, to several factors (listed in possible order of relevance): 
the coarser angular resolution of their data ($2.8\arcsec-4.7\arcsec$, which corresponds to $0.03-0.07$~pc when combined with distances) compared with that of ALMAGAL ($0.15\arcsec-0.3\arcsec$, corresponding to $0.005-0.01$~pc), an approach to flux evaluation that does not imply background subtraction (see Sect.~\ref{cfesect}), and the adoption of lower core temperatures ($12-30$~K) in \citet{and21}, which systematically increase the derived masses in the flux-to-mass conversion using the gray body relation, further amplifying the \mmmc\ values in their analysis with respect to ours.

Another relevant difference compared to \citet{and21} is that, when examining the two extreme sub-classes of our sample, the quiescent clumps and the H\textsc{ii}~regions, we observe a distinct trend. In the overall scatter shown in panel~$b$, the H\textsc{ii} regions tend to have larger \fmmc\ values than the quiescent clumps, whereas \citet{and21} reported a mild opposite trend in their sample, where IR-dark sources showed higher \fmmc\ values compared to IR-bright sources. The potential evolutionary implications of \mmmc\ will be discussed further in Sect.~\ref{mmcvsevol}.

Conversely, the \mmmc\ shows a slightly clearer upward trend with surface density (Figure~\ref{massmaxvsall}, panel~$c$, $\rho_\mathrm{S}=\spearmmmcsurfd$). Fitting this trend to a power law, the best-fit exponent is found to be $\slopemaxmasssurfd \pm \eslopemaxmasssurfd$ for the entire population (also shown in the panel). A power-law fit to the trend of the average yields an exponent of $\slopemaxmasssurfdave \pm \eslopemaxmasssurfdave$.

The relation between the MMC mass and the surface density supports the idea that high density conditions favor the formation of massive cores \citep[e.g.,][]{mck03,kum20,tok23}. Indeed, clump surface density, rather than mass, is indicated in the literature as a parameter found, both theoretically and empirically, to be critical for the formation of high-mass cores \citep[e.g.,][]{fed23}. In fact, the clump average density is expected to mirror, more directly than the mass, the presence of unresolved strong overdensities in the clump internal structure that can give rise to massive cores. 

\citet{mor23} observed a qualitatively similar relationship between MMC mass and clump mass in a sample of 39 ALMA targets associated with IRDCs. Our findings not only confirm this trend but also extend it with a significantly larger statistical sample and a broader range of evolutionary conditions.

In panel $c$ of Figure~\ref{massmaxvsall} we plot the values of \citet{mor23} as well (yellow open triangles), and we observe that they
lie in the upper-left region of the spread of ALMAGAL values. In particular, their MMC mass lies, in all cases, above 1~M$_\odot$, and the spread in surface density is slightly narrower ($\sim 1.5$ orders of magnitude against $\sim 2$ for ALMAGAL). Again, the position of these points is mostly due to overestimating the core masses, with respect to ALMAGAL, in turn due to the combination of a coarser resolution, leading to extracting larger structures, and of the strategy of including background emission in the core flux extraction procedure (see Sect.~\ref{cfesect} for a more detailed discussion).

Finally, we use the relation plotted in the panel $c$ of Figure~\ref{massmaxvsall} to investigate how predictive the clump surface density is regarding the ability to form massive stars \citep{kru08,kau10,bal17}. Cores with masses of at least 24~M$_\odot$ (i.e., three times the 8~M$_\odot$ threshold for defining a massive star) are exclusively found in clumps with a surface density exceeding \massivesurf~g~cm$^{-2}$, which can be therefore interpreted as a threshold within the ALMAGAL sample, although evolutionary effects should not be neglected (Sect.~\ref{mmcvsevol}). Notice that, although the ALMAGAL sample is biased toward already high surface densities (Sect.~\ref{clumpparameters}), this evidence remains statistically relevant, as there are \nlowsurfd\ targets with $N_\mathrm{core} > 0$ and $\Sigma < 0.3$~g~cm$^{-2}$.

Moreover, it becomes evident once again that any prescription linking clump surface density to compatibility with massive star formation should be regarded as a necessary but not sufficient condition. Indeed, in most cases, a high surface density satisfying the aforementioned criterion of \citet{kru08} for massive star formation is not accompanied by the presence of a high-mass core.

\subsection{Mass of the most massive core vs evolutionary status}\label{mmcvsevol}

\begin{figure}[t!]
   \centering
   \includegraphics[width=0.49\textwidth]{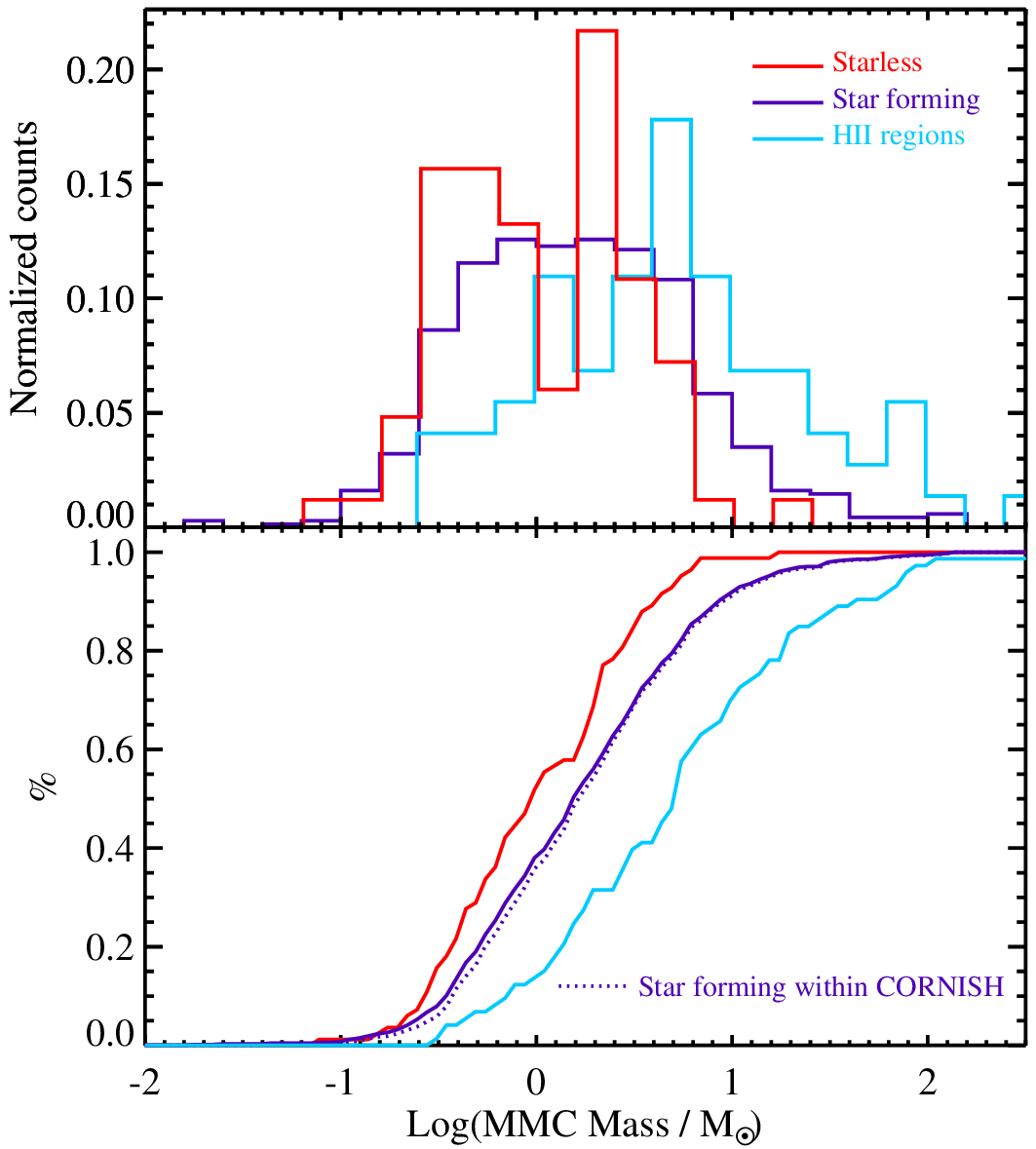}
   \caption{The same as Figure~\ref{ncoresclasses}, but for the mass of the most massive core instead of the number of cores.}
   \label{mmmcclasses}
\end{figure}

Before exploring the behavior of \mmmc\ as a function of specific evolutionary parameters, in Figure~\ref{mmmcclasses} we show the distributions of \mmmc\ separated by evolutionary class. While the top panel shows the distributions as histograms along with their respective medians, the bottom panel presents the corresponding cumulative distributions, which highlight the differences between the sub-classes.

A systematic shift toward higher values in the median and interquartile ranges is observed (see also Table~\ref{table_classes}).
In particular, the MMC masses in clumps associated with radio counterparts are higher than those in quiescent clumps. Specifically, the average masses for these two sub-classes are \medmmmchii~$\mathrm{M}_\odot$ (average: \maxmasshiiave~$\mathrm{M}_\odot$) and \medmmmcpre~$\mathrm{M}_\odot$ (average: \maxmasspreave~$\mathrm{M}_\odot$), respectively.
However, these distributions appear broad, with significant overlap among the classes, making it difficult to define distinct ranges of \mmmc\ for each evolutionary category.

What is clear is that, on the one hand, not all evolved ALMAGAL targets host at least one candidate core capable of forming a massive star (i.e., with a mass of at least 24~M$_\odot$) and, on the other hand, such cores are entirely absent in clumps classified as quiescent. This suggests that the masses of massive cores are not definitely established during the earliest phases of fragmentation (see also below).

Based solely on continuum ALMA photometry, we cannot distinguish, in turn, between the quiescent and star-forming nature of these massive cores, when observed. Further confirmation from ALMAGAL line observations (e.g., Jones et al., in prep.; Benedettini et al., in prep.; Allande et al., in prep.) is required to determine whether these are massive prestellar cores \citep[][and references therein]{xu24assemble}, or if all of them already host active star formation.

When examining the relationships between the MMC mass and clump evolutionary parameters, an increasing trend is observed in the median \mmmc\ in bins of clump $L/M$ and temperature (Figure~\ref{massmaxvsall}, panels~$d$ and $e$, respectively), while a possible similar trend for the bolometric temperature is hard to see (panel$f$).

\citet{col25} already commented on the relation with $L/M$, highlighting that it
is all the more appreciable if compared with the flat trend of the minimum core mass found in the clump. 
This suggests that the mass of the MMC is not determined at the earliest phases of the clump evolution \citep[as supposed, for example, by][for clumps coincident with hubs of filaments]{and21} but the most massive core increases in mass with the advancement of the evolutionary stage.
At the same time, the formation of cores of smaller mass is not inhibited \citep[cf.][]{pil19}. 

The observations show that the minimum $L/M$ where $M_\mathrm{MMC} > 24~\mathrm{M}_\odot$ occurs at $\sim 1~\mathrm{L}_\odot / \mathrm{M}_\odot$, and there are no quiescent clumps with cores exceeding this mass \citep[similarly to what was found by][]{san19,mor23,che24}. This indicates that many quiescent \textit{Herschel} clumps, though meeting the necessary (but not sufficient) conditions in surface density for massive star formation \citep{eli21}, are unlikely to host massive cores at the current evolutionary stage. Furthermore, \citet{bal17} highlighted the potential for misclassifications of such clumps due to distance-related biases. 

These findings imply three possible scenarios. In the first one, quiescent clumps capable of forming massive cores evolve so rapidly that this phase becomes practically elusive, or core masses increase over time, as suggested by the trend in panel~$d$. In the second scenario, newly formed cores are initially insufficiently massive to form high-mass stars but grow through continued accretion, is consistent with the competitive accretion model \citep{bon04,wan10}. This hypothesis, previously put forward by \citet{san17,san19}, \citet{mor21,mor23,mor24}, and \citet{li23}, is here reinforced with a much broader statistical foundation. A third possibility, proposed within the GHC framework by \citet{vaz19}, suggests that the earliest fragments to collapse are the most extreme local density fluctuations, with low total masses; as the mean density of the environment increases, more massive fragments also reach the conditions necessary for collapse.

The relationship between MMC mass and dust temperature is qualitatively similar to the relationship with $L/M$, while, as noticed above, the link with bolometric temperature looks much weaker. It appears that, as with $L/M$, the highest MMC mass values occur before the right tail of the highest values of these indicators. This suggests that for the ALMAGAL targets in the most advanced stages, the MMC might be entering the envelope cleaning-up phase, potentially driven by one or more newly formed high-mass stars.

\begin{figure}[t!]
   \centering
   \includegraphics[width=0.49\textwidth]{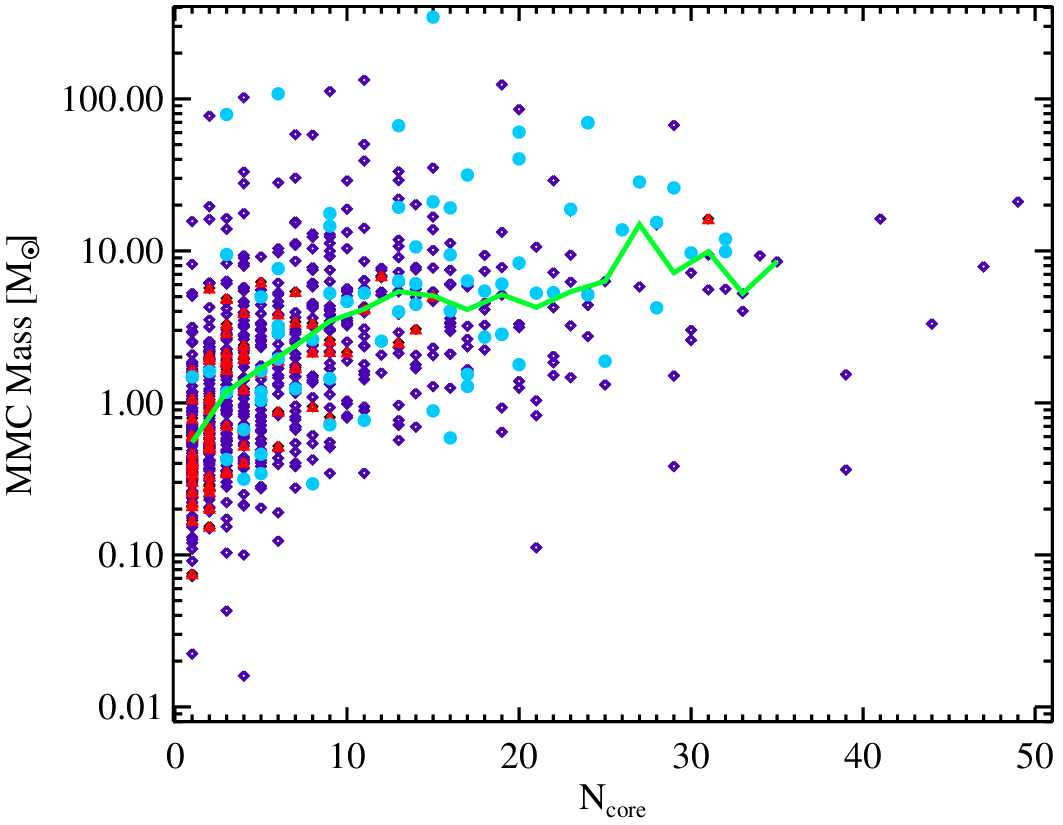}
   \caption{Mass of the most massive core versus the level of fragmentation for ALMAGAL targets. Symbol and color coding for source sub-classes is the same used in Figure~\ref{ncoresvsall} and throughout the paper. The green line represents the median in bins of $\Delta N_\mathrm{core}=2$.}
   \label{maxmassvsncore}
\end{figure}

To assess whether the MMC mass is influenced by the fact the MMC is isolated or, conversely, lies in a cluster with multiple companions, Figure~\ref{maxmassvsncore} displays the MMC mass against $N_\textrm{core}$. 
While the highest MMC masses (\mmmc\ $>10 \mathrm{M}_\odot$) are observed across a wide range of fragmentation levels, appearing largely insensitive to it, the lower envelope of the MMC mass distribution shows a clear trend of increasing with $N_\mathrm{core}$. In particular, MMC masses below $1~\mathrm{M}_\odot$ are almost entirely absent for $N_\mathrm{core} \gtrsim 25$. 

To summarize this evidence, the mass of the most massive core within a clump tends, on average, to increase as the core population becomes more numerous, again meeting the basic trend of the competitive accretion \citet{bon04,wan10} or GHC \citep{vaz19} scenarios. However, it is to be noted that an inverse correlation, albeit based on a significantly smaller statistical sample, has been reported by \citet{pan24}.

\subsection{Maximum core mass vs bolometric luminosity}
The analysis of the relationship between MMC mass and the $L/M$ ratio in panel $d$ of Figure~\ref{massmaxvsall} underscores the importance of investigating whether the increase in core mass is specifically linked to the bolometric luminosity $L$. Our aim is to ascertain whether the growth in the mass of the most massive core within the clump, potentially indicative of a forming star, significantly drives the overall rise in clump luminosity. It is noteworthy that both parameters exhibit the same analytical dependence on distance.

\begin{figure*}[th!]
   \centering
   \includegraphics[width=0.95\textwidth]{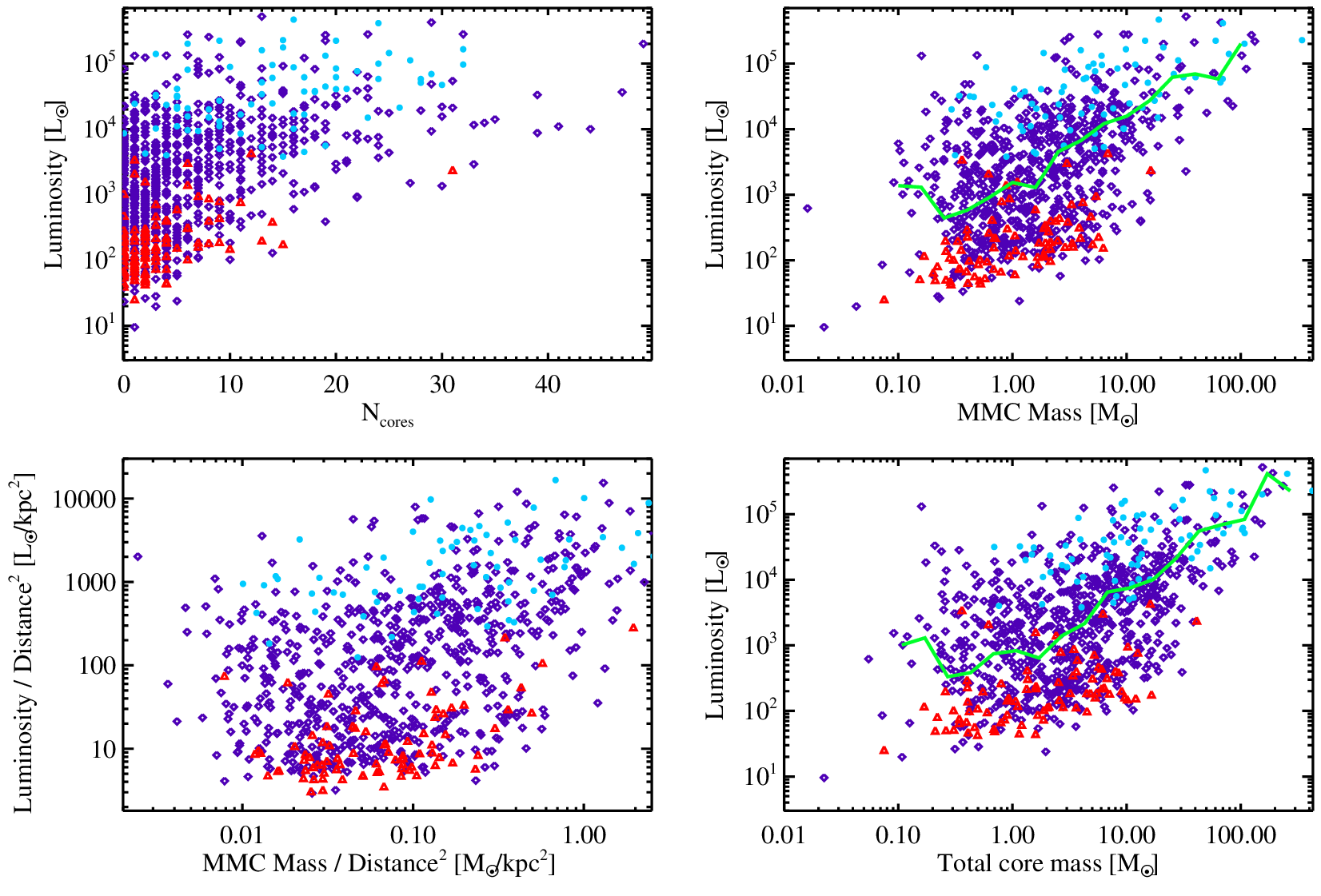}
   \caption{Clump bolometric luminosity $L$ plotted against level of fragmentation (\textit{top left}), mass of the most massive core (\textit{top right}), mass of the most massive core normalized by the square of the distance (with luminosity normalized as well, \textit{bottom left}), and sum of core masses (\textit{bottom right}), respectively. Symbol and color coding for source sub-classes is the same used in Figure~\ref{ncoresvsall} and throughout the paper. In the top-right and bottom-right panels, the green line represents the median in bins of the quantity on the $x$ axis.
   }
   \label{massmaxvslum}
\end{figure*}

For this purpose, we build Figure~\ref{massmaxvslum} based on the following logic: first, we check whether an increased number of fragments can produce, in general, an increment of $L$. This is not necessarily expected, as $N_\mathrm{core}$ can increase due to the fragmentation of larger condensations into smaller and fainter ones. In fact, \citet{pal13} find no apparent correlation between these quantities in a sample of 18~objects. Can, at this point, the MMC mass produce an increment of $L$ and, if so, can we find a similar behavior for the total mass of cores?

In the top-left panel of Figure~\ref{massmaxvslum}, we see that high luminosities ($L > 10^4~\mathrm{L}_\odot$) are found throughout the entire range of $N_\mathrm{core}$. In contrast, the lowest ones tend to increase with $N_\mathrm{core}$, so that in the presence of a large number of fragments (say $N_\mathrm{core} > 30$), only high luminosities (say $L > 10^4~\mathrm{L}_\odot$) are found.

The top-right panel of Figure~\ref{massmaxvslum} illustrates the $L$ vs \mmmc\ relation. A clear increasing trend is observed for the median of $L$, approximable by power-law behavior, with even more distinct trends evident for the quiescent and H\textsc{ii~} region counterparts. Performing a linear fit to the logarithms of the data yields an exponent \lummmcslope\ \citep[approximately three times shallower than that found in][]{pan24}; the Spearman's coefficient is $\rho_\mathrm{S}=\lummmcspear$. To assess whether this trend arises solely from the distance effect spreading $L$ and \mmmc\ along a linear relation, the bottom-left panel of Figure~\ref{massmaxvslum} plots these quantities normalized by the distance, namely $L/d^2$ vs $M_\mathrm{MMC}/d^2$. After normalization, the correlation seen in the top-right panel survives, with a power-law slope \lumdmmcdslope\ and a Spearman's coefficient $\rho_\mathrm{S}=\lumdmmcdspear$. 
In conclusion, while a correlation between the lower luminosity limit and the level of fragmentation is apparent, a mild physical correlation between luminosity and MMC mass is identified, which is not significantly affected by the distance effect.

Using the total mass in cores instead of the MMC mass (Figure~\ref{massmaxvslum}, bottom-right panel) yields a result qualitatively similar to that with MMC mass, with the median luminosity showing a nearly linear increase.

\section{Core formation efficiency}\label{cfesect}
Another parameter used to quantify the fragmentation of a clump is the core formation efficiency (CFE), defined as the total mass of the cores identified in a clump divided by the clump’s total mass. In this case, we adopt the clump mass as estimated by \citet{eli17}, though in principle, it should be determined before fragmentation begins, as the clump could accrete further mass from the parental cloud or lose mass due to star formation activity. Moreover, as seen above and further elaborated in this section, the mass contained in the cores evolves over time, making it more accurate to refer to the ``instantaneous'' CFE. Therefore, hereafter we use the term CFE implicitly implying its instantaneous value.

Based on this definition, the derived CFEs range from \cfemin\% to \cfemax\%, with a median of \cfemed\%. Comparing these results to \citet{and21}, who report a maximum CFE of 33\%, a minimum of 12\%, and a median of 24\%, and \citet{mor23}, who report a maximum of 16\%, a minimum of 0.6\%, and a median of 4.4\%, respectively, reveals a distinctly different regime of values. This discrepancy arises primarily from differences in observational resolution, source extraction strategy, and assumptions in converting photometry to masses.

Firstly, the dust opacity values used by \citet{eli17,eli21} and \citet{col25} to derive clump and core masses, respectively, are different, in order to account for the distinct characteristics of these two environments. As noted in Sect.~\ref{parameters}, taking into account the same gas-to-dust ratio they assumed, the latter adopt 0.9~cm$^2$~g$^{-1}$ at 1.3~mm, whereas the former adopt 10~cm$^2$~g$^{-1}$ at 300~$\mu$m with a dust emissivity exponent $\beta=2$, which scales to $\sim 0.53$~cm$^2$~g$^{-1}$ at 1.3~mm, a factor of~1.7 lower than in \citet{col25}. Consequently, adopting the same value for both lists would roughly double the core masses or halve the clump masses, leading to a significant increase in the CFE.

\begin{figure}[th!]
   \centering
   \includegraphics[width=0.49\textwidth]{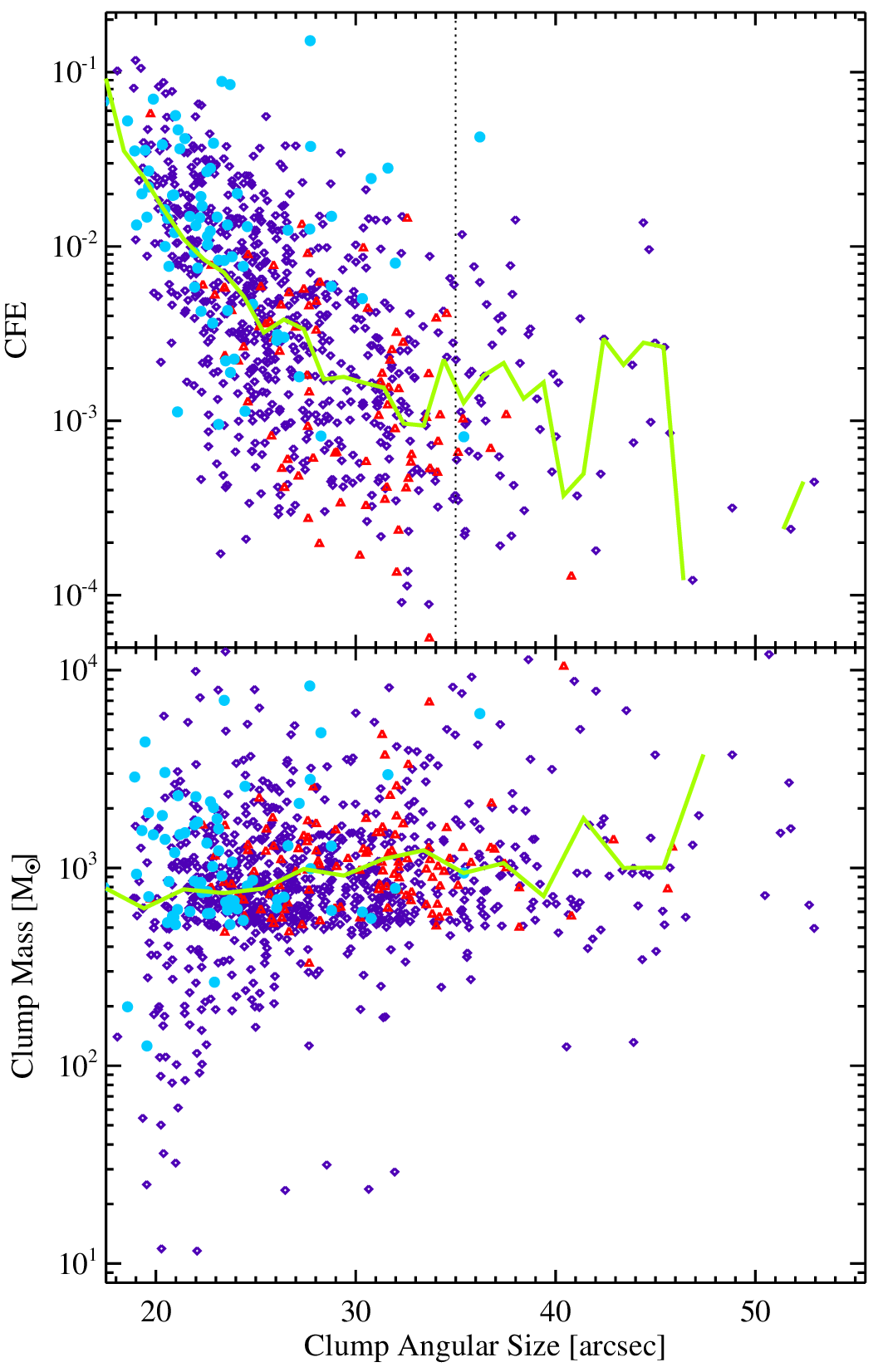}
   \caption{\textit{Top}: core formation efficiency (CFE) against clump angular size, estimated from \textit{Herschel} photometry at 250~$\mu$m. Symbols are defined in Figure~\ref{ncoresvsall}. The green line connects the median of the size estimated in bins of 2\arcsec. The vertical dotted line represents the diameter of the ALMAGAL FOV. \textit{Bottom}: the same as in the top panel, but for the clump mass on the $y$-axis.}
   \label{cfevssize}
\end{figure}

Furthermore, the higher angular resolution of the ALMAGAL images may result in worse brightness sensitivity potentially hindering the detection of fainter and extended structures. Structures lacking prominent peaks may remain undetected when angular resolution improves, as also seen in ALMAGAL when transitioning from 7M+TM2 to 7M+TM2+TM1 maps \citep{san25,col25}. In contrast, in the ASHES surve\footnote{With a resolution 5 to 10 times worse than in ALMAGAL, the images have a better brightness sensitivity that enables the detection of faint and extended structures. These structures are potentially detectable in the 7M+TM2 ALMAGAL images (Coletta et al., in prep).}, such objects can be detected and cataloged, contributing to a higher number of fragments and, consequently, a higher total mass in cores \citep[cf., e.g.,][]{mer15}. Additionally, a higher number of cores in ASHES can be surely due to the larger field coverage, as their maps are mosaics obtained from ten pointings. This difference can be checked in the ten overlapping fields between ALMAGAL and ASHES, where fragment counts vary from 0 to 9 in ALMAGAL (with CFE $ < 1\%$ in all cases) but from 9 to 39 in ASHES \citep[with $0.6\% < \mathrm{CFE} < 16.4\%$,][]{mor23}.
Additionally, the ALMAGAL photometry extraction strategy includes selecting compact sources and subtracting background emission (Sect.~\ref{corepar}), which results in lower flux estimates compared to algorithms like \textsc{Astrodendro} \citep{rob19}. Using background emission values from \citet{col25}, unsubtracted ALMAGAL fluxes are estimated to be, on average, twice as large.
Furthermore, \citet{mor23} use a temperature of 13.8~K to derive core masses, while ALMAGAL adopts a temperature of 20~K or higher (Section~\ref{corepar}), resulting in at least 40\% lower mass.
Lastly, unlike ASHES mosaics, the relatively smaller FOV of ALMAGAL single-pointing maps for the largest clumps may exclude fragments at the edges, underestimating the CFE. 
However, it is to say that, as shown in the top panel of Figure~\ref{cfevssize}, most clumps have circularized sizes under 35\arcsec, i.e. within the ALMAGAL FOV.

A decrease in CFE with the clump size (diameter) is observed (Figure~\ref{cfevssize}, top), while no strong link to clump mass is found (Figure~\ref{cfevssize}, bottom), because the mass also depends on surface density and heliocentric distances, and because of the lower cut adopted on clump masses for selecting most of the ALMAGAL sample (Sect.~\ref{clumpparameters}).

\begin{figure}[th!]
   \centering
   \includegraphics[width=0.49\textwidth]{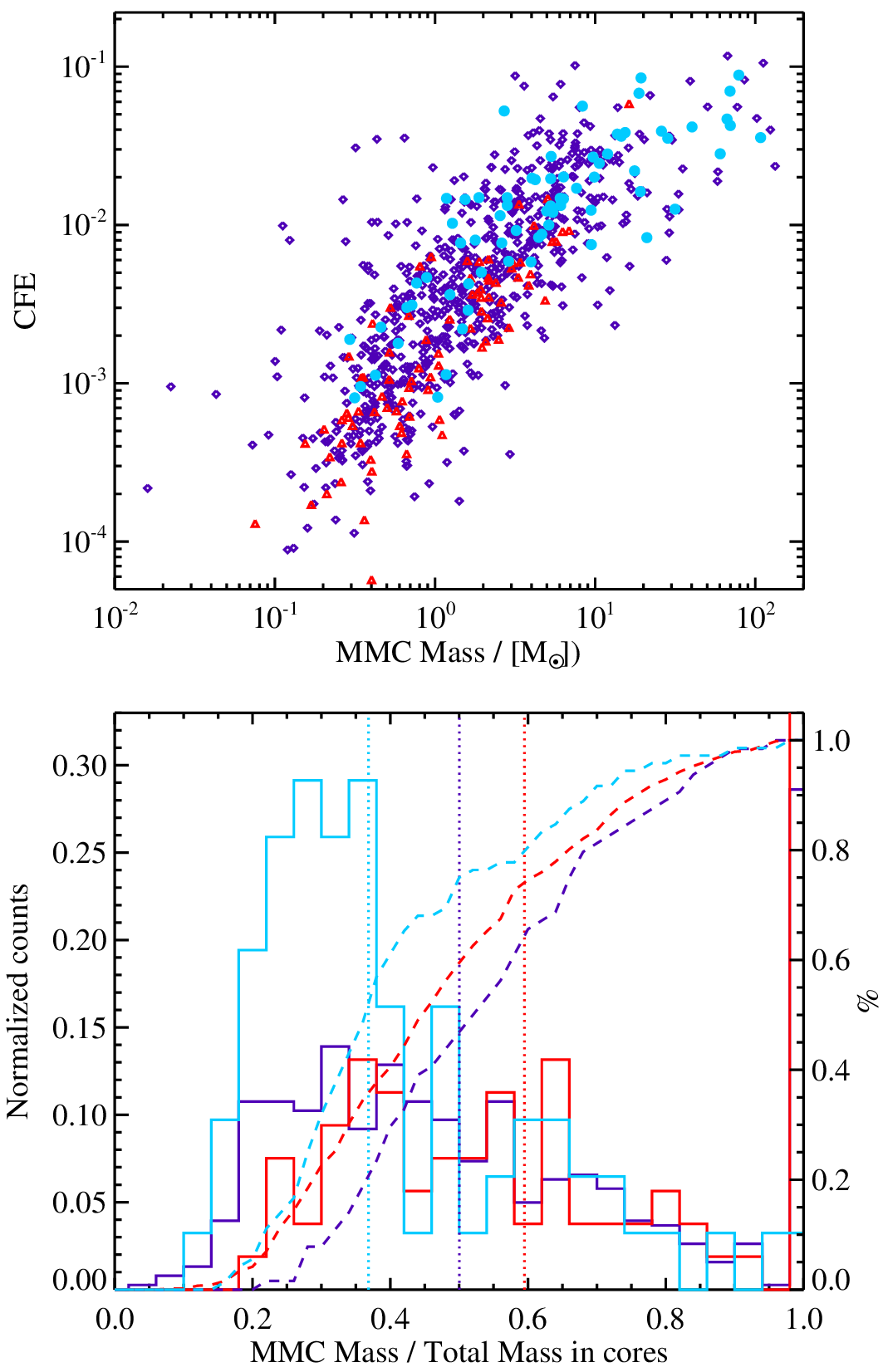}
   \caption{\textit{Top}: Core formation efficiency (CFE) against mass of the most massive core (\mmmc). Symbols are defined in Figure~\ref{ncoresvsall}.
\textit{Bottom}: Distribution of the ratio between the mass of the most massive core and the total mass contained in cores (excluding clumps with $N_\mathrm{core}=0$ or 1), shown separately for quiescent (red histogram), star forming (blue), and UCH\textsc{ii}~regions (cyan). Dashed curves represent the corresponding cumulative distributions, with values referenced to the $y$-axis on the right sight of the plot. Vertical dotted lines mark the median values for each sub-class, using the same color scheme as the histograms.}
   \label{cfevsmaxmass}
\end{figure}

Finally, before examining the relationships between the CFE and clump parameters, it is important to consider the connection between the CFE and the \mmmc\ (Figure~\ref{cfevsmaxmass}, top). 
One might argue that the two quantities are correlated and that discussing both constitutes a circular argument. However, although the mass of the most massive core contributes to the numerator in the definition of CFE, much can depend on the contribution from the other cores. Moreover, CFE also depends on the clump mass in the denominator and is distance-independent, unlike \mmmc.
Nevertheless, the figure reveals a clear, steadily increasing trend, possibly suggesting that the presence of a more massive core generally leads to a higher CFE, irrespective of the total clump mass. This implies that the presence of at least one higher-mass core is associated with increased efficiency of converting clump material into cores, with the most massive core playing a primary role in this process. More likely, as discussed in Sect.~\ref{mmcvsevol} and as will be seen in Sect.~\ref{cfevsevol}, both quantities tend to increase as the clump evolves. In this respect, the trend observed in the upper panel of Figure~\ref{cfevsmaxmass} should primarily reflect an evolutionary effect.

To investigate this aspect, in the lower panel of Figure~\ref{cfevsmaxmass} we show the distributions (also in cumulative form) of the ratio between \mmmc\ and the total core mass, excluding the trivial cases with $N_\mathrm{core}=0$ or 1. Despite the observed spread, the distributions, even more clearly visible through the cumulative curves and median positions, shift toward lower values as the evolutionary stage of the clumps advances. This indicates that, although \mmmc increases on average with the clump age, its relative contribution to the total mass in cores tends to decrease, both because the number of cores grows and because their individual masses also increase.

In conclusion, the correlation observed between \mmmc\ and the CFE in Figure~\ref{cfevsmaxmass}, top, does not appear to be driven solely by the analytical relation between the two quantities, which is also influenced by other parameters. This justifies treating CFE as an independent parameter in the following analysis, exploring its dependence on the clump properties.

\subsection{CFE vs distance} 

\begin{figure*}[ht!]
   \centering
   \includegraphics[width=\textwidth]{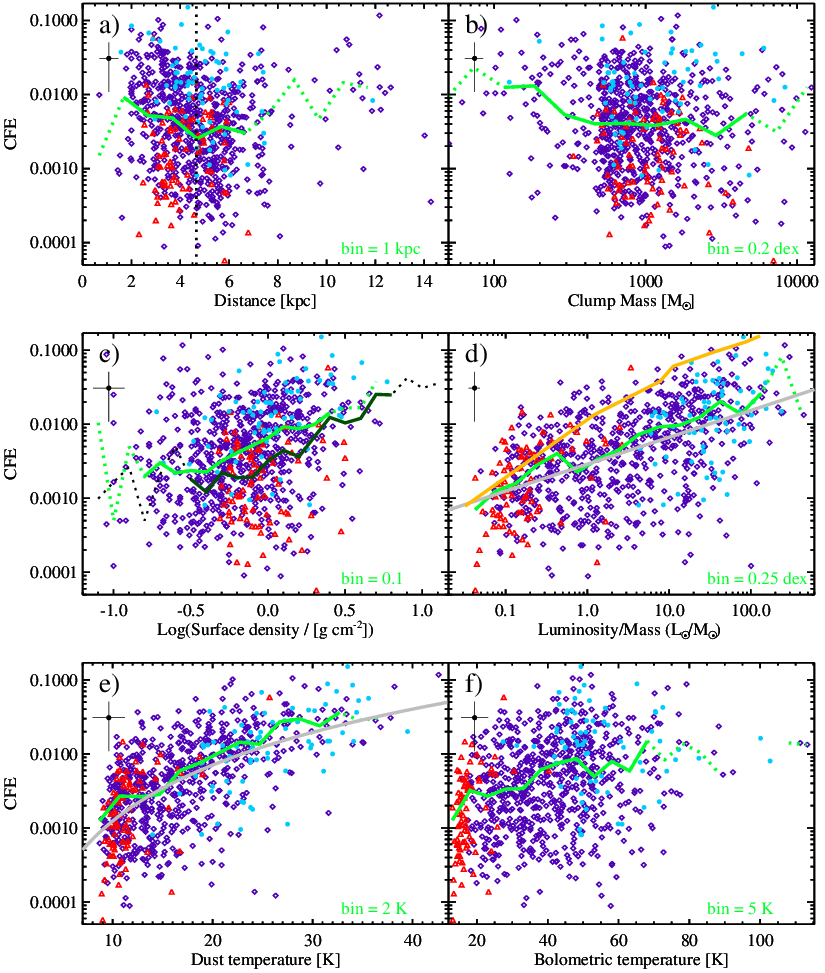}
   \caption{The same as Figure~\ref{ncoresvsall}, but for the CFE along the $y$-axis of each panel. In addition, in 
   panels~$d$ and $e$, the gray solid lines represent the best power-law fit to data.
  }
   \label{cfevsall}
\end{figure*}

A high degree of scatter is seen for CFE plotted against heliocentric distance in panel~$a$ of Figure~\ref{cfevsall}.
The median CFE is seen to slightly decrease with distance in the $2~\mathrm{kpc} \lesssim d \lesssim 6~\mathrm{kpc}$ range. The Spearman's correlation coefficient for $\log(\mathrm{CFE})$ vs.\ $d$ is \spearcfedist\ within this distance range, and \spearcfedistrestr\ for the entire sample, mildly supporting this trend. In the same distance interval, the level of fragmentation is also observed to decrease (Sect.~\ref{fragvsdist}). Therefore, also in this case, if one assumes that the observed trend is genuine, it can be interpreted in light of the progressive lack of low-mass cores that fall below the sensitivity limit as the distance increases.

The behavior of CFE of a region as a function of distance as a pure effect of virtually moving it away (so considering the distance bias due to lack of resolution, but not of sensitivity) was described by \citet{bal17}. They found a constant behavior for CFE at $d \gtrsim 1.5$~kpc, with average values varying from region to region in the range $1-20\%$, again generally higher than CFE we find for ALMAGAL clumps. 
Finally, we observe no evident break in the CFE behavior at the boundary separating the initial ``near'' and ``far'' ALMAGAL sample ($d > 4.66$~kpc). 

We conclude that the observed decrease in the median CFE is mostly attributable to the decrease of sensitivity with distance rather than of resolution.

\subsection{CFE vs clump mass and density}\label{cfevsmass}
As already expected from Figure~\ref{cfevssize}, the plot of the CFE and the clump mass $M$ (Figure~\ref{cfevsall}, panel~$b$) does not show a correlation between the two; despite that, from the analytical point of view, $M$ is present in both quantities. This means that the scatter in the total core mass (numerator of CFE) is predominant in producing the observed large scatter of these two quantities.

A slightly more visible correlation, at least in terms of median values, is seen between the CFE and the clump surface density $\Sigma$ (Figure~\ref{cfevsall}, panel~$c$). Again, the two quantities are inversely related from the analytical point of view (as both depend on the clump mass), so that, in this respect, the CFE should be expected to decrease at increasing mass, then increasing $\Sigma$. In contrast, in the ALMAGAL targets, the CFE increases, on average, with increasing $\Sigma$, i.e. these two quantities appear correlated in a way overpowering the underlying dependence on the mass. This means that other parameters at play, such as the core masses and the clump sizes, are crucial to establish the observed behavior.

The mild ($\rho_\mathrm{S}=\spearcfesurfd$) increase of median CFE (as well as $M_\textrm{MMC}$) with clump density, as also noted by \citet{pal13} and \citet{cse17}, can suggest a growing importance of the role of gravity at increasing densities. 
If this evidence holds despite the scatter in the data, two alternative scenarios are conceivable. In one, an increased CFE can be interpreted as a consequence of high-density conditions. In the other, a high clump density might simply reflect a structure dominated by compact, dense sub-structures, and thus associated with a high CFE.

\subsection{CFE vs evolutionary status}\label{cfevsevol}

\begin{figure}[t!]
   \centering
   \includegraphics[width=0.49\textwidth]{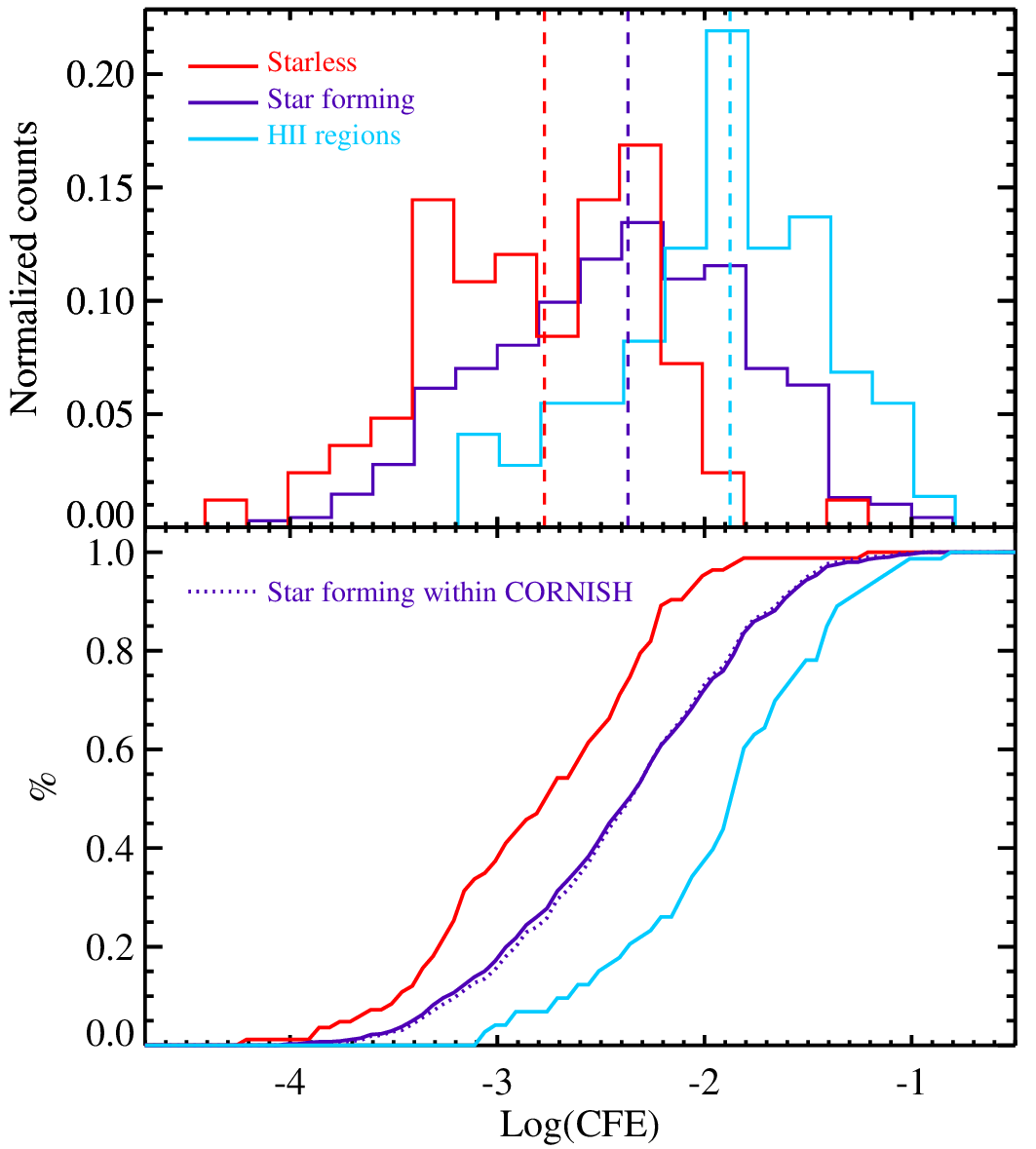}
   \caption{The same as Figure~\ref{ncoresclasses}, but for the core formation efficiency (CFE) instead of the number of cores.}
   \label{cfeclasses}
\end{figure}

The distributions of the CFE for the three evolutionary classes of targets exhibit a progressive shift toward higher values from quiescent to star-forming clumps and then to UCH\textsc{ii}~regions, though with significant overlap (Figure~\ref{cfeclasses}, panel $a$, and Table~\ref{table_classes}). This is even more clear if shown in terms of cumulative distributions (Figure~\ref{cfeclasses}, panel $b$).

The increase in clump formation efficiency (i.e., the ratio of total mass in clumps to the mass of the containing cloud) as a result of feedback-induced triggering by H\textsc{ii}~regions has been proposed by \citet{ede13} and \citet{xu18}. The higher median CFE observed for ALMAGAL targets associated with radio emission may suggest an extension of this mechanism to clump-to-core scales. Additionally, the pronounced left skewness in the $N_\mathrm{core}$ distribution for these sources, compared to other classes (Figure~\ref{ncoresclasses}), would support this interpretation. Also, \citet{zha24} observe the presence of massive and dense cores around H\textsc{ii}~regions, although the origin of their large masses remains unclear. 
However, the large spread in CFE values within this sub-class (spanning more than two orders of magnitude) makes it difficult to generalize this scenario to all cases.

The behavior of the CFE as a function of the evolutionary stage can be also examined by means of panels~$d$, $e$, and $f$ of Figure~\ref{cfevsall}. In all of them, the median CFE exhibits an increasing trend.
In particular, the CFE vs $L/M$ plot (panel~$d$) shows a quite regular increase not only of the median CFE, but also of the highest values of the CFE in bins of $L/M$, whereas the lowest values of CFE (say $< 3\times 10^{-4}$) appear present practically at all $L/M$ regimes. A power-law fit to the entire population of clumps gives a dependence of the form
\begin{equation} \label{cfelmeq}
\mathrm{CFE}=10^{\factorcfelm \pm \efactorcfelm}\;\left[\frac{L/M}{\mathrm{L}_\odot/\mathrm{M}_\odot}\right]^{\slopecfelm \pm \eslopecfelm} \: 
\end{equation}   
(parameter uncertainties are simply derived from least squares fitting to data).
A similar behavior is also seen for 
the dust temperature (panel~$e$). 
\begin{equation} 
\label{cfeteq}
\mathrm{CFE}=10^{\factorcfet \pm \efactorcfet}\;\left[\frac{T}{\mathrm{K}}\right]^{\slopecfet \pm \eslopecfet} \: .
\end{equation}

It is expected that the fraction of clump mass contained in cores increases with time \citep[e.g.,][]{che15}, under the effect of gradual accretion on cores and formation of further cores, unless this is eventually halted by feedback from the first newly formed stars. However, in the range of evolutionary stages covered by the ALMAGAL sample (up to a few $10^5$~years), we do not see, on average, an evident flattening of the CFE trend against clump evolutionary parameters. 

The behavior of the CFE as a function of clump evolution, measured by the $L/M$ ratio, can also be analyzed through comparison with numerical simulations (Sect.~\ref{multidiscussion}).

\section{Discussion}\label{discussion}
\subsection{Ranking correlations between two variables}\label{spearman}
In Sect.~\ref{mmmcvsmass}, the Spearman's coefficient for the \mmmc\ vs clump mass has been discussed to give a quantitative estimate of the possible correlation between the two. We systematically extend this analysis to all relations between clump and corresponding core population properties. In particular, we obtain coefficients $\rho_\mathrm{S}$ by considering the $n=\nok$ ALMAGAL targets with all clump properties defined and $N_\mathrm{core} \geq 1$.

\begin{figure}[t!]
   \centering
   \includegraphics[width=0.49\textwidth]{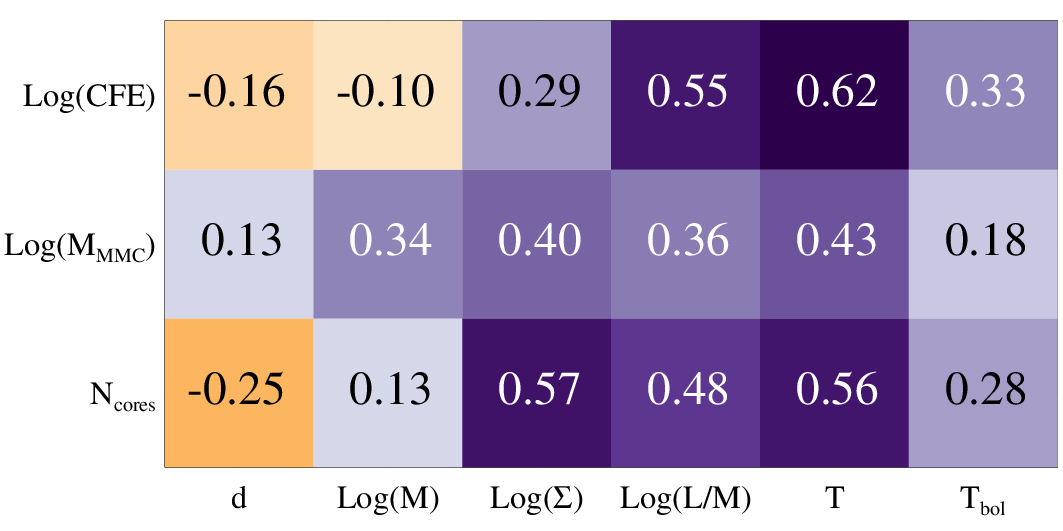}
   \caption{Matrix of Spearman's correlation coefficient $\rho_\mathrm{S}$ for relationships between clump physical parameters ($x$-axis) and core population parameters ($y$-axis). Color scale goes from dark orange ($\rho_\mathrm{S}=-1$) to dark purple ($\rho_\mathrm{S}=1$), through white ($\rho_\mathrm{S}=0$).}
              \label{ranking}
    \end{figure}

Figure~\ref{ranking} contains the matrix of Spearman's coefficients $\rho_\mathrm{S}$ for pairs of parameters: $d$, $\log M$, $\log\Sigma$, $\log(L/M)$, $T$, and $T_\mathrm{bol}$ along the $x$ direction, and $N_\mathrm{core}$, $\log$~\mmmc, and $\log\mathrm{CFE}$ along the $y$-axis.

The matrix of coefficients collects and summarizes the considerations made in Sects.~\ref{ncore}, \ref{mmcmass}, and \ref{cfesect}. It can be immediately seen that none of the reported $\rho_\mathrm{S}$ approaches values corresponding to the highest degree of correlation (+1) or anticorrelation (-1), but the range of the coefficients remains between \minspearman\ and \maxspearman. 

The quantities exhibiting the highest correlation are the surface density (logarithm), the luminosity/mass ratio (logarithm), and the temperature.
In more detail, the largest $\rho_\mathrm{S}$ are found for the relations between $N_\mathrm{core}$ and $\Sigma$ and $T$. Temperature and surface density are quantities most correlated with \mmmc\ as well, but with an inverted role, and with a lower correlation degree in general. Finally, for the CFE the highest correlation is found with $T$ and then with $L/M$, namely the two evolutionary parameters for which, in fact, an analytical relation was tentatively expressed in Equations~\ref{cfeteq} and \ref{cfelmeq}, respectively.
Generally, $T$ consistently shows a stronger correlation with core population parameters compared to $L/M$. The third evolutionary parameter considered, $T_\mathrm{bol}$, instead, shows lower values of the $\rho_\mathrm{S}$ coefficient.
The mass $M$ has notably less influence on fragmentation indicators than the surface density $\Sigma$. Only when combined with \mmmc, which shares the $M$ dependence on the square of the distance, do we observe a stronger correlation. Finally, distance itself shows a weak correlation with all core properties.

A similar analysis was conducted by \citet{tra23}, but using Pearson coefficients, which specifically indicate the reliability of linear correlations. They found direct correlations between \mmmc\ and $\Sigma$, as well as between the CFE and both $L/M$ and $\Sigma$, but no significant correlation between $N_\mathrm{core}$ and other quantities, nor between \mmmc\ and $L/M$. However, a direct comparison with our data is not feasible, as here we use Spearman's coefficients instead of Pearson coefficients to provide a more general assessment of monotonic trends. Additionally, our analysis is based on a much larger statistical sample, which carries implications discussed below.

To determine whether the obtained Spearman's coefficients indicate a true correlation between the two physical quantities, rather than a mere apparent association, the Spearman's test should be invoked.
However, this test may have limitations when applied to data sets with a large number of degrees of freedom, which in our case is $n-2$, with $n=\nok$ being the number of plotted data. With a large sample size, Spearman’s rank correlation coefficient can become overly sensitive, detecting statistically significant correlations that may be weak or practically unimportant. This arises because, with large $n$, even minor deviations from the null hypothesis (no correlation) may yield small $p$-values\footnote{The $p$-value represents the probability of obtaining a Spearman's correlation coefficient at least as extreme as the one observed, under the assumption that there is no true correlation (the null hypothesis) between the variables.}, potentially leading to the so-called Type~I errors, i.e., where trivial relationships are flagged as significant.
However, in our case, this is not straightforward. First, the critical values for the Spearman's distribution decrease as the number $n$ of measurements increases. For instance, in a two-tailed test for a positive association at the 5\% significance level, the critical value is 0.65 for $n=10$, but decreases to 0.36 for $n=30$, 0.20 for $n=100$, and 0.07 for $n=\nok$, respectively (for details on the calculation of the last value, see Appendix~\ref{student}). 
Second, we performed a resampling test by extracting 100 random subsamples of 30 data points (out of \nok) for each pair of variables, calculating the Spearman's coefficient for each subsample, and taking the median of these 100 values. The absolute deviations of these median coefficients from the original ones presented in Figure 11 are small, not exceeding 0.06 in the most extreme cases. However, the 95\% confidence threshold is more stringent for $n$=30 than for $n=\nok$, meaning that for the smaller sample size, only the coefficients for $\Sigma$, $L/M$, and $T$ would exceed it and then be statistically significant.

In conclusion, the interpretation of Spearman's test results depends heavily on the sample size. In this respect, it appears inappropriate to compare coefficients derived from samples of significantly different sizes (Sect.~\ref{mmmcvsmass}). Within the same sample, the coefficients remain useful for comparative rather than absolute analysis, as done in the first part of this section.

\subsection{Multi-parameter analysis}\label{multidiscussion}

The analyses in Sects.~\ref{ncore}, \ref{mmcmass}, and \ref{cfesect} focused on examining the relationship between a single overall descriptor of core populations and individual physical parameters of ALMAGAL targets, such as distance, mass, surface density, and evolutionary descriptors. This approach aimed to isolate the influence of each parameter on the observed characteristics of clump fragmentation. In this section, we take a step further by synthesizing those results and presenting them through multi-dimensional diagrams for a more comprehensive discussion.

\subsubsection{Clump fragmentation level}\label{multincore}

\begin{figure*}[h!]
   \sidecaption
   \includegraphics[width=12cm]{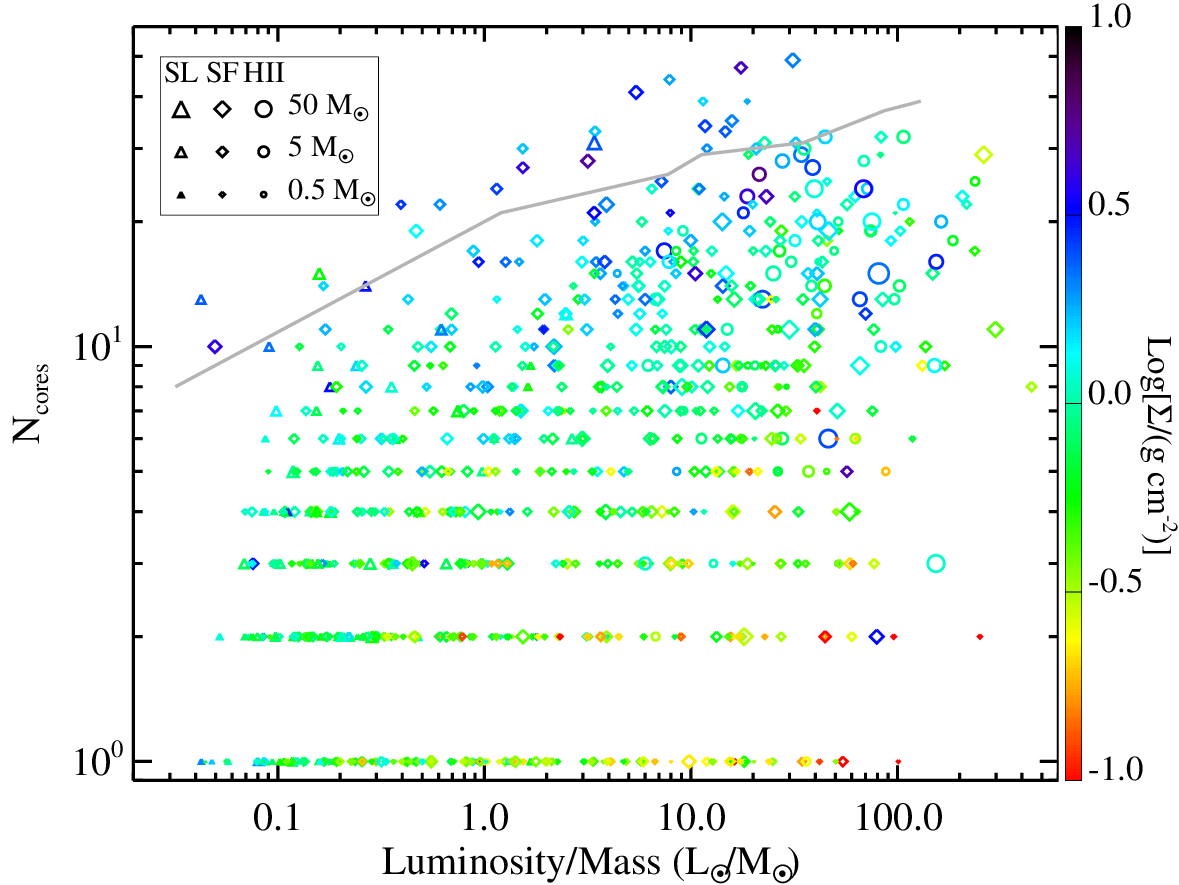}
   \caption{The same as panel $d$ of Figure~\ref{massmaxvsall}, but with different symbol and color coding. The color scale is coded according to the logarithm of the clump surface density in g~cm$^{-2}$ (color bar on the right), and quiescent clumps (SL in the legend in the top-left corner), star-forming ones (SF), and UCH\textsc{ii} counterparts (HII) are displayed by means of open triangles, open diamonds, and filled circles, respectively. The symbol size scales as the logarithm of the mass of the most massive core, as shown in the legend. The gray solid line represents the numerical prediction by \citet{leb25} with initial conditions set to $M=500~\mathrm{M}_\odot$, $\mathcal{M}=7$, and $\mu=10$, also plotted in panel $e$ of Figure~\ref{massmaxvsall} as an orange line.}
   \label{ncoremultiplot}
\end{figure*}

In Figures~\ref{ncoresvsall}, \ref{massmaxvsall}, and \ref{cfevsall}, we highlighted an increasing trend of $N_\mathrm{core}$, \mmmc, and CFE parameters, respectively, both with increasing $L/M$, an evolutionary indicator, and $\Sigma$, which, as explained in Sect.~\ref{fragvssurfd} and Appendix~\ref{otherclumpprops}, is not univocally related to the evolutionary stage of the clump (and, consequently, to $L/M$). 
To better disentangle the interplay of these two parameters against clump fragmentation, in Figure~\ref{ncoremultiplot} we show, similarly to panel~$e$ of Figure~\ref{ncoresvsall}, an $N_\mathrm{core}$ (here in logarithmic scale) vs $L/M$ plot, but with symbol colors and sizes encoded according to $\Sigma$ and \mmmc, respectively.

We highlight again that the clearest trends observable in the $N_\mathrm{core}$ vs $L/M$ plot are the increase of the highest values of $N_\mathrm{core}$ at increasing $L/M$, and the paucity of low fragmentation level (say $N_\mathrm{core}$ < 10) in the (few) cases at $L/M > 100~\mathrm{L_\odot}/\mathrm{M_\odot}$. On the contrary, the correlation between $N_\mathrm{core}$ and $\Sigma$ appears to be tighter, as also testified by the Spearman's coefficient analysis contained in Sect.~\ref{spearman}. 

As for the mass of the most massive core, it appears correlated with the combination of $L/M$ and $\Sigma$ (more strictly with the latter, according to the quantitative analysis in Sect.~\ref{spearman}).
Looking in particular at the \nmassive\ clumps with $M_\mathrm{MMC} > 24~\mathrm{M}_\odot$, i.e. compatible with the formation of a high-mass star, \nmassivelmone\ of them have $L/M > 1~\mathrm{L_\odot}/\mathrm{M_\odot}$, considered by \citet{mol16} as a signature of ongoing star formation, and \nmassivelmten\ have $L/M > 10~\mathrm{L_\odot}/\mathrm{M_\odot}$, that the same authors attribute mostly to the appearance of a first zero-age main sequence star of intermediate/high-mass. 
Among cases with $L/M > 1~\mathrm{L_\odot}/\mathrm{M_\odot}$, \nmassivelmonesigmone\ of them show $\Sigma > 1~\mathrm{g}~\mathrm{cm}^{-2}$ \citep[i.e. the theoretical threshold for compatibility with massive star formation of][]{kru08}, while \nmassivelmonesigmhalf\ fulfill the condition $\Sigma > 0.5~\mathrm{g}~\mathrm{cm}^{-2}$, which is a conservative value, higher than densities empirically found to be consistent with massive star formation \citep[][]{tan13,urq14,tra18}. On the contrary, no cores with mass exceeding $24~\mathrm{M}_\odot$ are found in clumps with both $L/M < 1~\mathrm{L}_\odot/M_\odot$ and $\Sigma < 0.5~\mathrm{g}~\mathrm{cm}^{-2}$.

Among the indications given by the plot in Figure~\ref{ncoremultiplot}, therefore, it emerges that the number of fragments is preferentially determined by density conditions in the clump, although $N_\mathrm{core}$ can also increase with the clump evolution due to further fragmentation, or formation of new condensations \citep[e.g.,][]{zha15}. Certainly, the growth of the fragment masses (or, at least, that of the most massive fragment) generally proceeds with the clump evolution, as would be expected for a clump-fed accretion scenario, and in particular by the GHC theory \citep{vaz19}. As a further argument in favor of the clump-fed model against the core-fed one, we observe that ($i$) neither large-mass cores are seen at early evolutionary stages, ($ii$) nor, as can also be seen in Figure~\ref{maxmassvsncore}, isolated (corresponding to $N_\mathrm{core}=1$ or 2) large-mass cores are found, at any evolutionary stage.
This means, indeed, that the mass reservoir available for the formation of a massive star is not set in the core before it collapses \citep[as required by the core-fed model of][]{mck03}, and also that the mass growth of such a core with time takes place in the presence of and in connection with other cores accreting mass from the same clump \citep{vaz19,pad20}, most likely in a competitive manner \citep{bon01,bon04}.

Of course, as discussed below, superposed on these overall trends there are peculiar conditions in each clump, dictated by magnetic field, turbulence, and protostellar feedback, that can play a role in determining the observed fragmentation level and are likely responsible for the high degree of scattering characterizing all plots contained in Figures from \ref{ncoresvsall} to \ref{cfevsall}. These factors cannot be quantified at this stage, i.e. by means of ALMAGAL continuum photometry alone, but rather need to be further characterized by ALMAGAL spectral line analysis (Jones et al., in prep.; Benedettini et al., in prep.).

Finally, it has to be pointed out that, as highlighted in Sect.~\ref{fragvsdist}, the level of fragmentation is particularly sensitive to the core flux sensitivity limit in the ALMAGAL catalog, so that in principle the $N_\mathrm{core}$ values plotted in various figures might be intended as lower limits for this quantity.

\subsubsection{Comparison with simulations}\label{simulations}

We have the opportunity to compare the ALMAGAL data with predictions from numerical simulations, following an approach similar to \citet{fon18}. We use recent numerical simulations produced by the Rosetta Stone project \citep{leb25}, aimed at investigating the fragmentation of massive clumps ($M \geq 500~\mathrm{M}_\odot$), and subsequent star formation\footnote{An accurate comparison between models and observations would require post-processing of 3-D simulations, with radiative transfer to produce synthetic observations. Such work is currently underway within the Rosetta Stone project \citep{leb25,nuc25}, so here we limit our discussion to general trends, directly comparing data and simulations while noting caveats. \citet{nuc25} also provide a more systematic exploration of the simulation parameter space.}. The first step of this project consists of simulating the clump collapse \citep{leb25} by using the 
RAMSES code \citep{tey02,fro06} and its extension to radiative transfer in the flux-limited diffusion approximation \citep{com11,com14}.
\citet{leb25} simulated the collapse of clumps with 500 and 1000~$\mathrm{M}_\odot$, exploring different regimes of turbulence and magnetization, and obtained predictions of fragmentation and star formation efficiency as a function of the clump evolutionary stage, represented by the $L/M$ ratio. 
We consider, in particular, the case with the following combination of initial parameters, which best matches our data: initial clump size $\sim 0.38$~pc (total size of the cube $\sim 1.5$~pc), mass $M=500~M_\odot$, turbulent Mach number $\mathcal{M}=7$, and mass-to-magnetic flux ratio (in units of the critical mass-to-flux ratio) $\mu=10$; finally, temperature $T=10$~K, which is common to all models. Single forming stars are represented in these models by sink particles \citep{ble14} that form above a number density threshold of $10^9$~cm$^{-3}$. The maximum resolution, in terms of physical scale, from the adaptive mesh refinement is equivalent to $\sim 400$~au.

The predicted number of sink particles (gray line in Figure~\ref{ncoremultiplot}, and orange in Figure~\ref{ncoresvsall} (panel~$d$)) exceeds most ALMAGAL values by a factor $\gtrsim 2$, though it follows a similar increasing trend to that of the highest values of $N_\textrm{core}$.
Other parameter combinations yield stronger fragmentation and poorer agreement, particularly for higher masses ($1000~\mathrm{M}_\odot$) or weaker fields ($\mu=100$).
An anti-correlation between magnetic field and fragmentation degree was tentatively found in observations by, e.g., \citet{cor19}, \citet{ane20}, \citet{pal21}, and \citet{san24}.
Therefore, an opposite scenario, i.e. a model with $\mu<10$, might be expected to better reproduce our data \citep[for example, $\mu=2$ is suggested by][based on observations]{beu18a}.  
However, the simulations produced within the Rosetta Stone framework with $\mu = 3$ have offered complementary insights into how strong magnetization influences fragmentation. \citet{leb25}, who analyzed the number of sink particles with clump age, report that runs with $\mu = 3$ yield fewer sink particles than those with $\mu = 10$, whereas \citet{nuc25} in post-processed versions of the same simulations (at $\sim 7000$~au resolution) find that the $\mu = 3$ and $\mu = 10$ models are not clearly distinguishable in terms of their fragmentation properties, seen as a function of $L/M$.
Post-processing of these simulations at the ALMAGAL resolution is among the next planned steps of the Rosetta Stone project and is expected to help resolve this apparent discrepancy.

Turbulence also plays a role: while it supports large-scale stability, it can enhance small-scale compression \citep{fed13}. Observations indicate that higher turbulence favors massive core formation \citep{olm23}), whereas simulations show that reducing Mach number increases fragmentation \citep{fon18}; thus, values below $\mathcal{M}=7$ may better fit ALMAGAL results.

Stellar feedback, not yet included in these simulations, can contribute to suppress fragmentation \citep{kru07,mye13,kru19,hen20,gee23}, although the relative contributions of these mechanisms remain a subject of active debate \citep[e.g.,][]{dal15,hop20}. As such, it may help explain the observed discrepancy, especially in evolved fragments.

Observational biases must also be considered. 
The number of detected fragments depends on the adopted detection threshold \citep[$5~\sigma$,][]{col25}, which undercounts faint cores. Moreover, while the simulations of \citet{leb25} track sink masses down to $0.001~\textrm{M}_\odot$, the observational catalog is complete only above $0.2~\textrm{M}_\odot$ \citet{col25}, likely missing low-mass fragments. Finally, simulations treat sink particles as discrete entities that do not merge, even when in close proximity. In contrast, in the observations limited angular resolution and projection effects can merge nearby sources, further lowering the observed $N_\textrm{core}$.

The clump collapse simulation of \citet{leb25} also accounts for mass accretion onto sink particles (Figure~\ref{massmaxvsall}, panel~$d$, orange curve). In qualitative terms, the observed increasing trend of \mmmc\ appears similar to that predicted by this model for the mass of the most massive sink particle.

\begin{figure}[h!]
   \centering
   \includegraphics[width=0.5\textwidth]{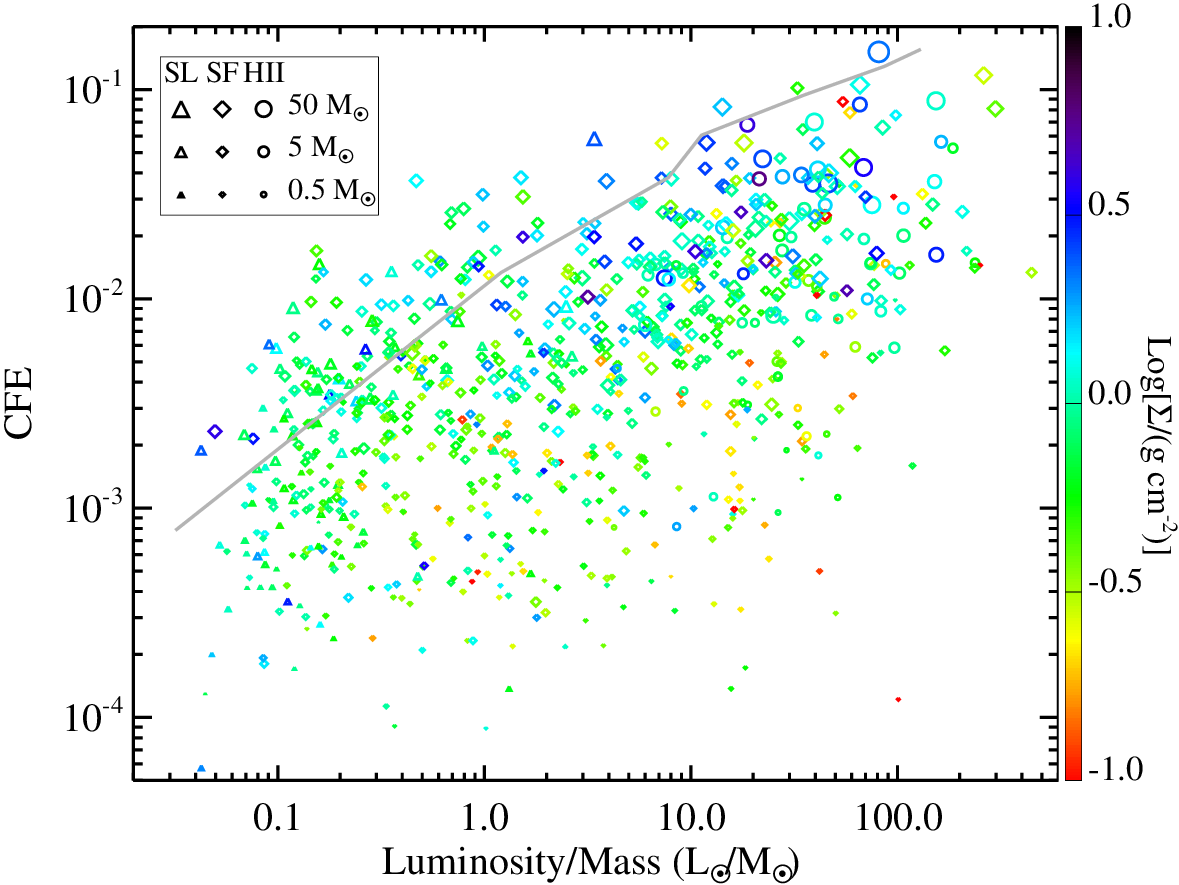}
   \caption{The same as Figure~\ref{ncoremultiplot}, but for the core formation efficiency (CFE) on the $y$-axis.}
   \label{cfemultiplot}
\end{figure}

\subsubsection{Core formation efficiency}
To further examine the CFE (Sect.~\ref{cfesect}), Figure~\ref{cfemultiplot} presents a view of its dependence on clump $L/M$ and $\Sigma$ \citep[seen also by][]{pan24}. As previously noted in Figure~\ref{cfevsall}, the increasing trend of the CFE with $L/M$ is more pronounced than the corresponding trend in $N_\mathrm{core}$ shown in Figures~\ref{ncoresvsall} and~\ref{ncoremultiplot}, and is quantitatively confirmed in Figure~\ref{ranking}. Although there is also a contextual increase in the CFE with $\Sigma$ \citep[cf.][]{kon15}, Figure~\ref{ranking} indicates that this correlation is weaker compared to the CFE-$L/M$ relationship. In other words, the scenario is opposite to what is seen for the number of fragments: compared with $N_\mathrm{core}$, the CFE appears, in general, more sensitive to the clump evolution than to surface density.

Furthermore, the CFE appears strongly correlated with \mmmc, confirming the result of \citet{tra23}, and the combination of high values for both is found prominently at $L/M \gtrsim 10~\mathrm{L}_\odot/\mathrm{M}_\odot$, with less regard for the value of $\Sigma$.

When examining this evidence in the context of discerning between theories of massive star formation, multiple arguments emerge against a core-fed scenario. In this model, cores appear as pre-assembled mass reservoirs; therefore, core masses and the core formation efficiency (CFE) are not expected to increase significantly over time, which contrasts with our observations. If cores are prevented from further accreting mass from the clump, as in the core-fed scenario, those with a mass below a threshold to form massive stars would not fulfill that threshold with time, contrary to what is observed for the highest values of \mmmc.

In contrast, our analysis suggests a more dynamic evolution of the core population within a clump. The number of cores appears to be primarily determined by the density of the clumps, while both CFE and the mass of the most massive core (\mmmc) show a more general and systematic increase as the system evolves. This trend indicates continuous accretion from the clump reservoir, supporting a clump-fed model rather than a static core-fed model \citep[see also][and references therein]{con18,mor24}. Furthermore, the fact that massive cores (\mmmc\ $> 24~\mathrm{M}_\odot$) are only observed in targets where $L/M \gtrsim 1~\mathrm{L}_\odot / \mathrm{M}_\odot$ (Sect.~\ref{mmcvsevol}) implies that massive star formation cannot be predicted solely from the physical parameters at the onset of clump collapse \citep[as also shown by][]{kle00,kle01b,rei05,rei06} and initial fragmentation. Instead, the evolution of core masses appears to be a time-dependent process, with sustained accretion playing a crucial role in the formation of massive stars.
This conclusion gains further significance when considering that the ALMAGAL sample was specifically selected to represent conditions that are expected to be conducive to massive star formation.

A tentative comparison with model predictions can be carried out by displaying in the CFE vs $L/M$ plot (Figures~\ref{cfevsall}, panel~$d$, and~\ref{cfemultiplot}) the sink particle formation efficiency vs $L/M$ curves emerging from the simulation of \citet{leb25} described above in this section. Remarkably, the behavior of these theoretical points is to increase at increasing $L/M$, as well as the median of ALMAGAL data. 

Quantitatively speaking, the model prediction becomes definitely larger than the average of observational CFE for $L/M > 0.3~\mathrm{L}_\odot/\mathrm{M}_\odot$. The considerations to explain this discrepancy can be the same given for the $N_\mathrm{core}$ vs $L/M$ case (Sect.~\ref{ncore}), such as the limitation imposed by sensitivity in detecting cores in observations, compared to the infinite sensitivity of simulations, as well as the absence of stellar feedback in the considered simulations.

In summary, the distribution of core masses in a clump appears to evolve over time, with cores dynamically gaining mass from their surrounding reservoir. This supports a clump-fed scenario of star formation, where the formation and growth of cores are closely related to the ongoing evolution and accretion processes within the clump. A similar conclusion for ALMAGAL is also reached by \citet{col25}, based on the study of the temporal evolution of the core mass function.

\section{Summary and conclusions}\label{summary}

This paper investigates the fragmentation characteristics of clumps as described by their internal core populations, as observed in continuum maps at 1.4~mm of the ALMAGAL survey. Using data from an unprecedented sample of 1007 Galactic plane clumps observed by ALMA at high resolution, this study seeks to correlate fragmentation levels and core masses with the clump photometric and physical parameters. 
Our findings partially confirm earlier results in the literature, but provide greater statistical reliability and uncover previously unexplored aspects of fragmentation.

The main results are:

\begin{enumerate}
     \item The flux densites of the targets at \textit{Herschel} wavelengths increase, on average, as a function of the number of hosted fragments ($N_\mathrm{core}$), particularly for $1 \leq N_\mathrm{core} \leq 25$. The slope of this increase varies systematically with wavelength (decreasing from 70 to 500~$\mu$m), indicating a correlation between $N_\mathrm{core}$ and clump temperature. Clumps without detected cores tend to have lower flux densities, and these are typically less evolved or less dense compared to the whole ALMAGAL sample.
    \item The ALMAGAL sample spans wide ranges in clump properties (physical, evolutionary, and environmental conditions). This diversity leads to significant scatter in the relationships between these parameters and clump fragmentation descriptors. While continuum observations cannot directly quantify the roles of magnetic fields, turbulence, or protostellar feedback, in the future systematic comparisons with numerical simulations based on theoretical models will provide critical insights.
    \item Although the data are widely scattered, average trends emerge. Fragmentation parameters, namely the number of fragments ($N_\mathrm{core}$), the mass of the most massive core (\mmmc), and the core formation efficiency (CFE), show an increasing trend with both clump surface density ($\Sigma$) and evolutionary indicators like dust mean temperature ($T$), luminosity/mass ratio ($L/M$), and bolometric temperature ($T_\mathrm{bol}$). 
    Spearman's coefficient analysis shows $N_\mathrm{core}$ correlates most strongly with $\Sigma$, whereas \mmmc\ and CFE align best with $T$. 
  Therefore, the number of cores appears to be primarily determined by the density of the clumps, while CFE and \mmmc\ show a more general and systematic increase as the system evolves.
    Power-law fits quantify relationships between CFE and $L/M$ and $T$, yielding slopes \slopecfelm\ and \slopecfet, respectively.
    \item In contrast, fragmentation parameters show no clear dependence on clump mass or heliocentric distance, apart from a mild sensitivity-driven decrease in $N_\mathrm{core}$ with distance.
    \item Dividing the ALMAGAL targets into evolutionary classes (quiescent, star-forming, and UCH\textsc{ii} regions), we observe a greater degree of segregation with respect to certain clump parameters, particularly those with evolutionary significance, compared to core population parameters. While the properties of core populations in star-forming clumps and UCH\textsc{ii} regions appear somewhat intermixed, quiescent cores exhibit clear upper limits: $N_\mathrm{core} < 20$, \mmmc~$< 8~\mathrm{M}_\odot$, and CFE~$< 2\%$. The level of fragmentation and the distribution of CFE observed for UCH\textsc{ii} regions may suggest a mechanism of core formation triggering.
\item When comparing the data with the currently available and debated star formation models, the strong dependence of \mmmc\ and CFE on the clump evolution aligns more closely with the family of models categorized under the \textit{clump-fed scenario}, indicating lower compatibility with the \textit{core-fed} model.
\end{enumerate}

 \begin{acknowledgements}
The authors thank the anonymous referee for their constructive and engaged approach, which, with precise suggestions and insightful remark, helped us to improve the overall quality of this article.
The teams at INAF-IAPS and at Heidelberg University acknowledge financial support from the European Research Council via the ERC Synergy Grant ``ECOGAL'' (project ID 855130).
RSK furthermore thanks the German Ministry for Economic Affairs and Climate Action in project ``MAINN'' (funding ID 50OO2206). RSK also thanks the 2024/25 Class of Radcliffe Fellows for highly interesting and stimulating discussions. 
A.S-M. acknowledges support from the RyC2021-032892-I grant funded by MCIN/AEI/10.13039/501100011033 and by the European Union `Next Generation EU'/PRTR, as well as the program Unidad de Excelencia Mar\'ia de Maezto CEZ2020-001058-M, and support from PID2023-144675NB-I00 (MCI-AEI-FEDER, UE).
G.A.F acknowledges support from the Collaborative Research Centre 956, funded by the Deutsche Forschungsgemeinschaft (DFG) project ID 184018867. G.A.F also gratefully acknowledges the DFG for funding through SFB 1601 ``Habitats of massive stars across cosmic time'' (sub-project B1) and from the University of Cologne and its Global Faculty programme.
RK acknowledges financial support via the Heisenberg Research Grant funded by the Deutsche Forschungsgemeinschaft (DFG, German Research Foundation) under grant no.~KU 2849/9, project no.~445783058.
PS was partially supported by a Grant-in-Aid for Scientific Research (KAKENHI Number JP22H01271 and JP23H01221) of JSPS.
CB gratefully acknowledges funding from the National Science Foundation under Award Nos. 2108938, 2206510, and CAREER 2145689, as well as from the National Aeronautics and Space Administration through the Astrophysics Data Analysis Program under Award ``3-D MC: Mapping Circumnuclear Molecular Clouds from X-ray to Radio'', Grant No. 80NSSC22K1125.
LB gratefully acknowledges support by the ANID BASAL project FB210003.
Part of this research was carried out at the Jet Propulsion Laboratory, California Institute of Technology, under a contract with the National Aeronautics and Space Administration (80NM0018D0004).
\end{acknowledgements}

\bibliographystyle{aa}
\bibliography{almagalbib}

\begin{appendix}
\section{Comparison of ALMAGAL target physical parameters}\label{otherclumpprops}

This appendix complements the review of clump physical properties given in Sect.~\ref{parameters}, showing relations and concepts that are useful for the discussion in the main paper.

\begin{figure}[h!]
   \centering
   \includegraphics[width=0.49\textwidth]{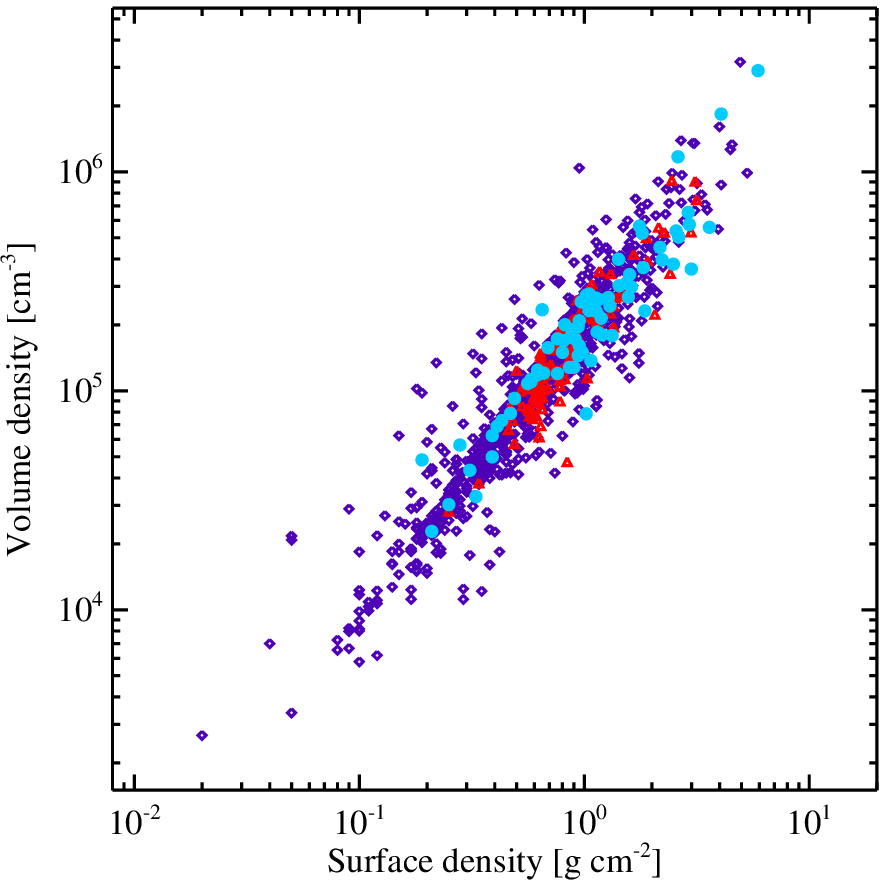}
   \caption{Volume density vs surface density relation for ALMAGAL targets. Colors and symbols are as in Figure~\ref{ncoresvsall}.}
   \label{densities}
\end{figure}

First, we use Figure~\ref{densities} to motivate our choice of discussing surface density $\Sigma$ throughout this paper, rather than using a volume density $\rho$, which could be estimated as the ratio of mass and volume of the sphere with the diameter equal to the circularized (not beam-deconvolved) source size at 250~$\mu$m. In the figure, in most cases, the two quantities appear directly correlated, making it quite redundant to consider these two quantities separately.

\begin{figure}[h!]
   \centering
   \includegraphics[width=0.49\textwidth]{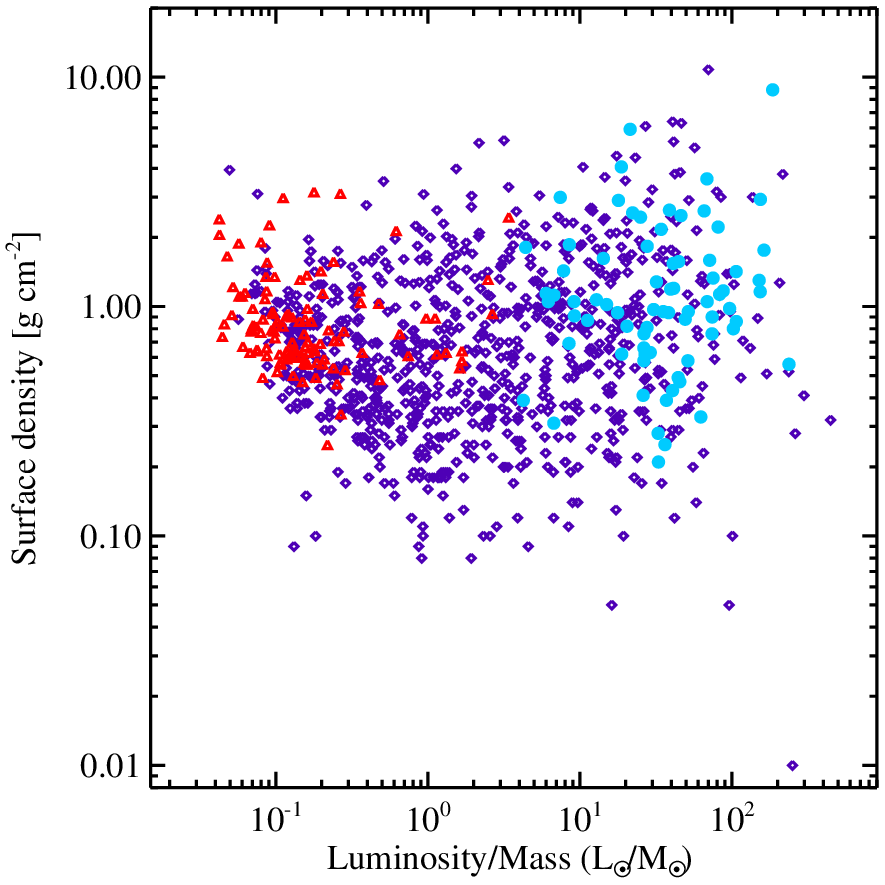}
   \caption{Surface density against $L/M$ ratio for ALMAGAL targets. Colors and symbols are as in Figure~\ref{ncoresvsall}. Notably, values of $\Sigma < 0.1$~g~cm$^{-2}$ are present in the ALMAGAL sample due to $i$) the inclusion of sources selected from the RMS catalog, which were not subject to the physical parameter constraints applied to sources from the Hi-GAL catalog \citep[see][and Sect.~\ref{parameters} of this paper]{mol25}, and $ii$) the use in this work of source sizes that have not been beam-deconvolved when deriving surface densities, differing from the original ALMAGAL target selection criteria.}
   \label{densitylm}
\end{figure}

Another aspect we address here is the relation between the surface density of ALMAGAL targets and their evolutionary stage, expressed by the $L/M$ ratio. A mild general increase of density from the starless to the star-forming phase was highlighted in \citet{eli17}, and summarized in Sect.~\ref{fragvssurfd} of this paper. 
However, the high-pass filters selection criteria applied to both mass and surface density when selecting ALMAGAL targets from the Hi-GAL catalog \citep{mol25} prevent this trend from being observed within the ALMAGAL sample itself, as seen in Figure~\ref{densitylm} in terms of $\Sigma$ vs $L/M$ relation.

A relevant consequence for our analysis is that any trends identified for a given fragmentation parameter as a function of these two quantities (as discussed in Sections~\ref{ncore}, \ref{mmcmass}, and~\ref{cfesect}) are essentially independent of one another.

\section{Testing a prediction for the number of cores vs distance behavior}\label{appbaldeschi}
In this appendix, we test the prediction made by \citet{bal17} about the number of cores detected in a clump as a function of the distance. They found 
\begin{equation}\label{baldist}
N_\mathrm{core}(d)=N_0 \left(d_0/d\right)^\delta,
\end{equation}
where $N_0$ is the number of cores detected at a distance $d_0<d$, and with $\delta$ varying with the \textit{Herschel} band. 

The way Eq.~\ref{baldist} can be applied here is by considering a clump as a structure appearing as a single object (then $N=1$) at its distance $d$, and imagining the increase of resolution from Hi-GAL ($\sim 18\arcsec$ at 250~$\mu$m) to ALMAGAL (using 0\arcsec.3 corresponds to a factor~60), as it was produced by approaching the source by the same factor (i.e. $d/d_0=60$). Adopting a typical exponent $\delta=1.5$ for 250~$\mu$m quoted by \citet{bal17}, one should then expect $\sim 460$ cores inside one clump, which is extremely far from what is actually found in the observations ($N_\mathrm{core}<50$). 

To explain this discrepancy, one has to take into account that:
\begin{itemize}
    \item Both the action of factors contrasting fragmentation (turbulence, magnetic field, feedback) and the decrease in sensitivity with distance \citep[which is not taken into account in the realizations of][]{bal17} surely contribute to decrease in the number of observed cores.
    \item Eq.~\ref{baldist} was empirically determined by probing shorter ranges of ratios between original and ``moved away'' distances ($d/d_0 \lesssim 30$), and, probably, is not fully adequate for this test. 
   \item Connected to the previous point, a limit of this check is to consider that the confusion of the $N$ cores in one clump is achieved at distance $d$, but actually, depending on the source disposition across the ALMAGAL field of view (35\arcsec), it can be observed also at closer distances, making the $d/d_0$ much smaller than 60. That being said, however, to extrapolate $N_\mathrm{core}=49$ (the maximum value achieved in the ALMAGAL catalog), or - say - 10, much lower distance ratios should be invoked ($\sim 13$ and $\sim 5$, respectively).
\end{itemize}

\section{Comparing surface densities of clumps and cores}

The relationship between the clump average surface density ($\Sigma$) and the surface density of the cores within is worth examining. By definition, we expect each clump to contain at least one core whose surface density exceeds the clump’s average.

\begin{figure}[h!]\centering\includegraphics[width=0.49\textwidth]{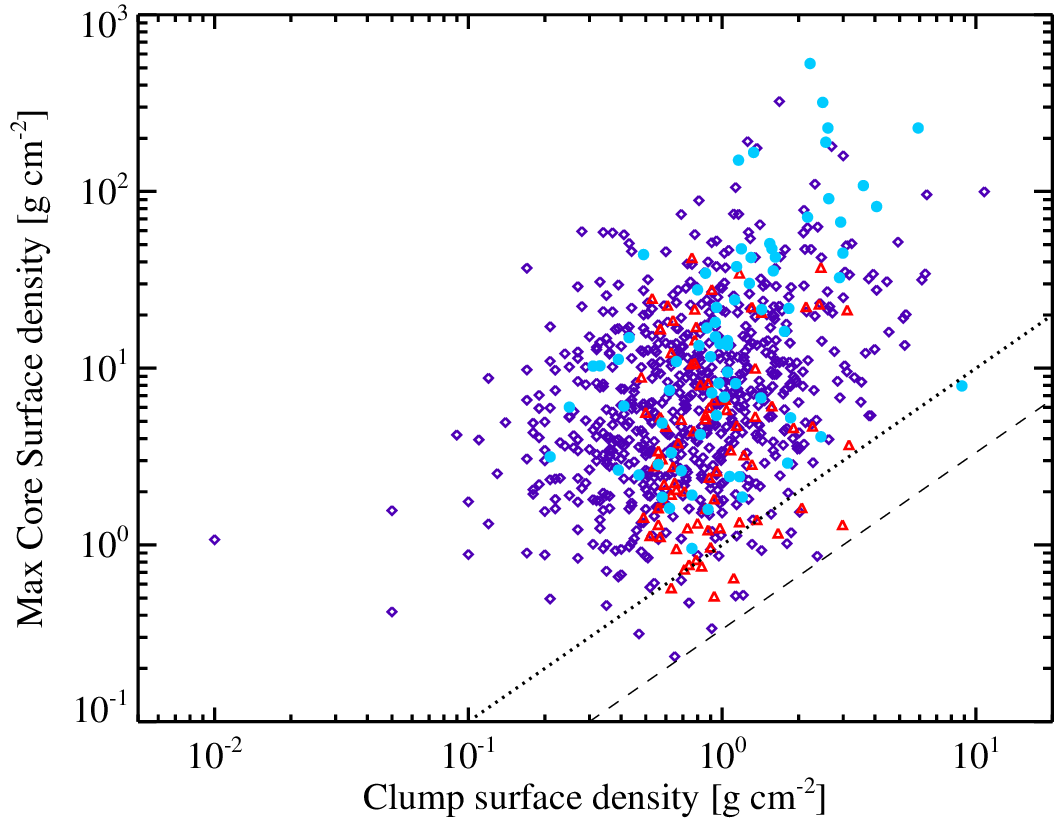}
   \caption{Surface density of the densest core in a clump vs clump average surface density. Symbols are as in Figure~\ref{ncoremultiplot}. The dotted and dashed lines represent the 1:1 and the 1:3 relations, respectively.}
   \label{surfacedensities}
\end{figure} 

To test this, Figure~\ref{surfacedensities} plots the maximum surface density found in each clump (which in \percsamesigma\% of cases corresponds to the surface density of the most massive core) against $\Sigma$. The plot reveals no strong correlation between these quantities. The key observation remains that the maximum core surface density generally exceeds the clump’s average surface density. In the remaining \percsigmalow\% of cases where this condition is not met, the ratio between core and clump surface densities falls within a range of 1/3 to 1 \citep[a similar result can be found in][]{mer15}.

These discrepancies can be attributed to uncertainties in core mass estimates (and, consequently, surface densities) as discussed by \citet{col25}. 
Additionally, the systematic difference in the dust opacity values used for calculating core and clump masses, as pointed out in Section \ref{cfesect}, can account for a factor $\sim 2$, thereby explaining most of the observed discrepancies.

\section{Skimming the ALMAGAL sample for potential free-free emission contamination}\label{nocornish}

\citet{col25} discuss the impact of free-free emission contamination on millimeter continuum flux measurements of cores, as it can lead to an overestimation of the core mass. \citet{col25} exclude from their analysis of the core mass function all cores associated with CORNISH or CORNISH South cm free-free emission. Additionally, they also omit cores within ALMAGAL targets falling outside the coverage areas of these two surveys, where information on the possible presence of radio emission is unavailable.

In this appendix, we assess the potential impact of this contamination on our analysis. First of all, it is important to note that the degree of fragmentation is unaffected by this issue, so our focus here is solely on the mass of the most massive core (\mmmc) and the core formation efficiency (CFE). 
In particular, we examine how the plots in Figures~\ref{massmaxvsall} and~\ref{cfevsall} change after excluding ALMAGAL targets where free-free contamination has been confirmed for at least one core by \citet{col25}, as well as targets outside the coverage of the CORNISH and CORNISH South surveys. For conciseness, we present only four of the six clump properties used to construct the aforementioned figures: $M$, $\Sigma$, $L/M$, and $T$.

\begin{figure*}[th!]
   \centering
   \includegraphics[width=0.99\textwidth]{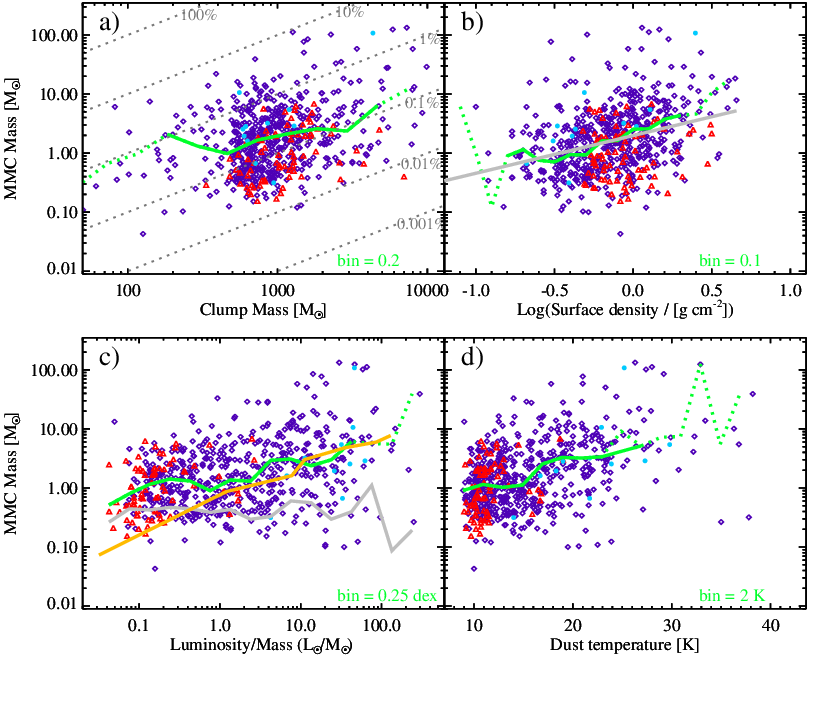}
   \caption{Plots of \mmmc\ versus clump mass ($a$), surface density ($b$), $L/M$ ratio ($c$), and temperature ($d$), after ruling out both ALMAGAL targets with at least one contained core found by \citet{col25} to be contaminated by radio emission from CORNISH or CORNISH South data, and ALMAGAL targets lying outside the areas of the sky covered by these two surveys. Symbols, colors, and lines are as in panels $b$, $c$, $d$, and $e$, respectively, of Figure~\ref{massmaxvsall}.}
   \label{massmaxvsnocornish}
\end{figure*}

In this context, Figure~\ref{massmaxvsnocornish} contains the plots of \mmmc\ vs these four clump parameters, limited to the sources presenting no radio free-free contamination (or risk thereof). Two potential biases must be considered when comparing Figures~\ref{massmaxvsnocornish} and Figure~\ref{massmaxvsall}. First, excluding sources possibly contaminated by free-free emission tends to rule out preferentially more evolved targets, which are characterized by higher values of $L/M$ and $T$, resulting in a relative depopulation of the right side of panels $c$ and $d$ in Figure~\ref{massmaxvsnocornish}. Second, if the continuum emission from the most massive core of a given clump is affected by contamination, its flux density - and consequently \mmmc\ - will be overestimated. However, removing contaminated objects does not necessarily imply the exclusion of the highest values of \mmmc.

That said, the same considerations made in Sect.~\ref{mmcmass} regarding the trends of \mmmc\ versus clump properties appearing in Figure~\ref{massmaxvsall} remain applicable to the plots in this figure, as they exhibit similar qualitative behavior. Noticeably, for this restricted sample, the slope of the power-law fit for \mmmc\ versus $\Sigma$ changes from $\slopemaxmasssurfd \pm \eslopemaxmasssurfd$ (Sect.~\ref{mmmcvsmass}) to $\ncslopemaxmasssurfd \pm \nceslopemaxmasssurfd$, which remain consistent within the error bars.
Furthermore, out of \nmassive\ clumps with $M_\mathrm{MMC} > 24~\mathrm{M}_\odot$ (Sect.~\ref{multincore}), only \nmassivenocorn\ do not present free-free contamination.

\begin{figure*}[ph!]
   \centering
   \includegraphics[width=0.99\textwidth]{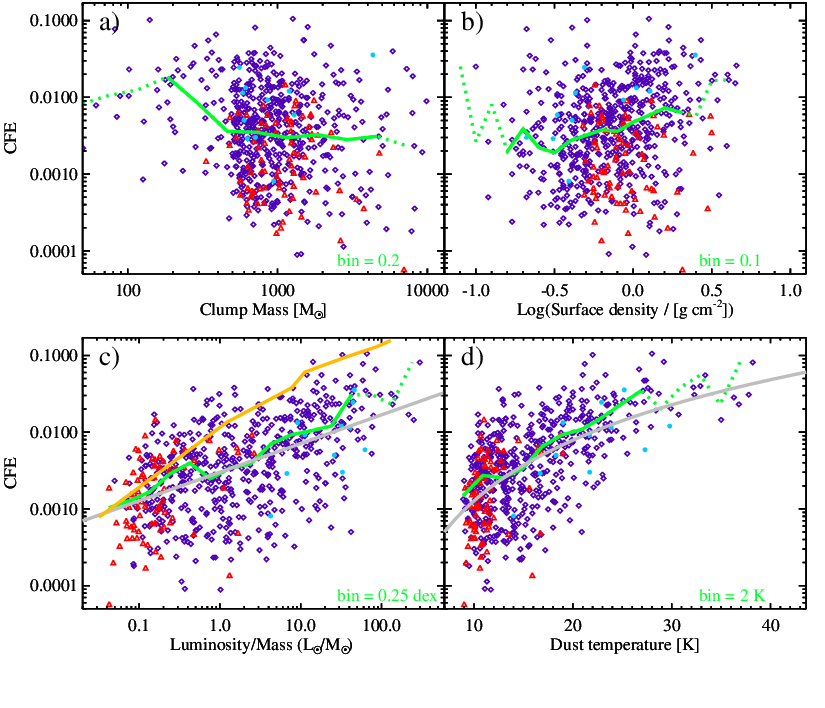}
   \caption{The same as Figure~\ref{massmaxvsnocornish}, but for the CFE on the $y$-axis. Symbols, colors, ad lines are as in panels $b$, $c$, $d$, and $e$, respectively, of Figure~\ref{cfevsall}.}
   \label{cfevsnocornish}
\end{figure*}

A similar behavior can be seen also for CFE, by comparing Figures~\ref{cfevsnocornish} and~\ref{cfevsall}, as the monotonic nature of observed median trends remains unchanged across all plots after ruling out possible cases of free-free contamination. In particular, in more quantitative terms, in Sect.~\ref{cfevsevol} we fitted a power law to the plot of CFE versus $L/M$ and $T$, obtaining the exponents $\slopecfelm \pm \eslopecfelm$ and $\slopecfet \pm \eslopecfet$, respectively. Repeating this procedure for the plots in panels $c$ and $d$ of Figure~\ref{cfevsnocornish}, we obtain a power-law dependency as $\ncslopecfelm \pm \nceslopecfelm$ and $\ncslopecfet \pm \nceslopecfet$, respectively, which are fully consistent with the previous ones, within the error bars.

In conclusion, whereas an analysis based on core masses should ideally exclude instances of possible mass overestimation due to free-free contamination, in practice the general indications taken from the subsample, selected by applying stricter constraints, are consistent with those found in the full sample. Therefore, we can confidently affirm the validity of conclusions drawn in the current paper for the entire ALMAGAL sample.

\section{Calculating the Spearman's critical value for 5\% significance level and $n=\nok$ data points}\label{student}
The critical values of Spearman's rank correlation coefficients ($\rho_\mathrm{S}$) are typically tabulated for sample sizes up to $n=100$ \citep[e.g.,][]{zar99}. As $n$ further increases, these critical values tend to exhibit asymptotic behavior. For large $n$, the distribution of Spearman's approximates that of Student's $t$ distribution, according to the following relationship\citep[e.g.,][]{ram89}:
\begin{equation}\label{studentspearman}
    t=\rho_\mathrm{S}\sqrt{\frac{n-2}{1-\rho^2_\mathrm{S}}}\, .
\end{equation}

For $n=\nok$, and a significance level of 5\% (corresponding to 2.5\% in each tail for a two-sided test, i.e., $p=0.025$), the critical $t$ value is $t_\mathrm{crit}=1.963$. 
Applying the inverse of the equation above,
\begin{equation}
\rho_\mathrm{S}=\frac{t}{\sqrt{t^2+n-2}}\,,
\end{equation}
yields a critical $\rho_\mathrm{S}$ value of $\rho_\mathrm{crit} \simeq 0.07$, i.e., the value reported in Sect.~\ref{spearman}.

Following the same procedure, but at a significance level of 99.7\% (i.e., 3-$\sigma$), the critical $t$-value is $t_\mathrm{crit}=3.89$, yielding $\rho_\mathrm{crit} \simeq 0.13$.
\end{appendix}

\end{document}